\pdfoutput=1

\documentclass[twocolumn]{aastex61}
\accepted{to The Astrophysical Journal on January 15, 2017} 

\shorttitle{Solar Energetic Particle transport near a Heliospheric Current Sheet}
\shortauthors{Battarbee et al.}

\usepackage{graphicx}
\usepackage{amsmath}

\begin{document}

\title{Solar Energetic Particle transport near a Heliospheric Current Sheet}

\correspondingauthor{Markus Battarbee}
\email{mbattarbee@uclan.ac.uk}

\author[0000-0001-7055-551X]{Markus Battarbee}
\affiliation{Jeremiah Horrocks Institute, University of Central Lancashire, PR1 2HE, UK}

\author{Silvia Dalla}
\affiliation{Jeremiah Horrocks Institute, University of Central Lancashire, PR1 2HE, UK}

\author{Mike S. Marsh}
\affiliation{Met Office, Exeter, EX1 3PB, UK}



\begin{abstract}

Solar Energetic Particles (SEPs), a major component of space weather, propagate through the interplanetary medium strongly guided by the Interplanetary Magnetic Field (IMF).
In this work, we analyse the implications a flat Heliospheric Current Sheet (HCS) has on proton propagation from SEP release sites to the Earth.
We simulate proton propagation by integrating fully 3-D trajectories near an analytically defined flat current sheet, collecting comprehensive statistics into histograms, fluence maps and virtual observer time profiles within an energy range of 1--800 MeV.
We show that protons experience significant current sheet drift to distant longitudes, causing time profiles to exhibit multiple components, which are a potential source of confusing interpretation of observations. We find that variation of current sheet thickness within a realistic parameter range has little effect on particle propagation.
We show that IMF configuration strongly affects deceleration of protons. We show that in our model, the presence of a flat equatorial HCS in the inner heliosphere limits the crossing of protons into the opposite hemisphere.

\end{abstract}


%

\section{Introduction} \label{sec:intro}

A significant component of space weather is the flux of Solar Energetic Particles (SEPs), accelerated during energy release events such as flares and Coronal Mass Ejections (CMEs) at the Sun. These high-energy charged particles can, after propagating to the Earth, disrupt satellite communications and impact astronaut health and safety \citep{Turner2000}. Charged particles propagating through interplanetary space are guided and deflected by the solar wind's magnetic field and its spatial and temporal variations. Modern efforts in modeling space weather effects include performing numerical simulations to solve particle fluences at the Earth based on parent active region and observer locations (see, e.g.,  \citealt{Chollet2010} and \citealt{Marsh2015}). The most common approach is to use a transport equation (see, e.g., \citealt{Roelof1969}, \citealt{Aran2005}, and \citealt{Luhmann2007}), where particles are effectively bound to the Interplanetary Magnetic Field (IMF) lines, described as the Parker spiral \citep{Parker1958}.

Recent research (\citealt{Marsh2013}, \citealt{Dalla2013}, \citealt{Dalla2015}) has shown that particle drifts, which are not modeled by a classical transport equation, play a significant role in SEP propagation to the Earth. They have been shown to be significant for protons and especially for heavier elements \citep{Dalla2017}. Other significant factors include field-line meandering \citep{Laitinen2016} and cross-field diffusion \citep{Zhang2003,He2011}. One significant characteristic of the IMF which has not been previously modeled in the context of SEP propagation is the Heliospheric Current Sheet (HCS), providing the boundary between the two hemispheres of the Solar dipole field. The presence of a current sheet changes motion of charged particles due imposing two distinct regions of drifts and the break-down of guiding centre motion at the sheet \citep{Speiser1965}.

The HCS is a vast area of space where the magnetic fields associated with the northern and the southern hemispheres of the solar magnetic field transition between outward and inward-directed polarities. Due to the varying and complicated distribution of mean magnetic flux direction on the solar surface, and the tilt of the solar magnetic axis with respect to the rotation axis, the HCS consists of a complex 3D structure, especially at greater heliocentric distances. The HCS has been the topic of much research, but mainly from the point of view of very energetic particles called galactic cosmic rays (GCRs), propagating inwards from the outer boundary of the heliosphere (references include, but are not limited to, \citealt{Jokipii1977}, \citealt{Burger1985}, \citealt{Kota2001}, \citealt{Pei2012}, \citealt{Strauss2012}, and \citealt{Guo2014}). The role of the HCS in SEP propagation has previously been briefly investigated in \cite{KuboYukiNagatsumaTsutomu2009}.

In this paper, we present a first analysis of how the presence of the HCS affects the propagation of SEPs from the Sun to the Earth. We consider a flat current sheet and assess effects of current sheet thickness and different dipole configurations on SEP propagation for protons of different energies. We also present SEP time profiles at virtual observers, providing a basis of comparison with real observations.


\section{Heliospheric Current Sheet Model} \label{sec:HCS}

In this work, we model the HCS as a flat plane separating two hemispheres of opposite polarity, with each hemisphere based on a simple analytical magnetic field model.\added{ We model purely radial outflow of solar wind plasma, which, combined with solar rotation and flux freeze-in, results in a non-radial magnetic field.} The IMF is described through spherical heliocentric coordinates as a scaled Parker spiral magnetic field ${\bf B}$
\begin{align}
  {\bf B} &= S(\theta) {\bf B}_\mathrm{Parker}, \label{eq:SParker}
\end{align}
where the Parker field is defined as
\begin{align}
  B_{r,\mathrm{Parker}} &= B_0 \frac{r_0^2}{r^2}  \label{eq:Br}\\
  B_{\theta,\mathrm{Parker}} &= 0 \label{eq:Btheta}\\
  B_{\phi,\mathrm{Parker}} &= -\frac{B_0 r_0^2 \Omega_\odot}{u_\mathrm{sw}} \frac{\sin \theta}{r}. \label{eq:Bphi}
\end{align}
Here $B_0$ is the field strength at $1\,r_0$, normalized to provide a field strength of $B(1\,\mathrm{au})=3.85\,\mathrm{nT}$, $\Omega_\odot = 2.87\times10^{-6}\,\mathrm{rad}\,\mathrm{s}^{-1}$ is the average solar rotation rate, $u_\mathrm{sw}=500 \,\mathrm{km}\,\mathrm{s}^{-1}$ is the radial solar wind speed and $S(\theta)$ is a scaling function providing the change of polarity in a gradual fashion and describing current sheet thickness. Due to $S$ being only a function of colatitude $\theta$, the analytical field remains divergence-free. \replaced{Effects caused by a deformed, non-planar current sheet, visible through observations as sector boundaries, are postponed to further studies.}{This simplified HCS model, where the current sheet is completely flat, is thus symmetric in respect to the heliocentric coordinate $\phi$. It is an approximation which is strictly valid only within the inner heliosphere and during solar minimum. Modeling of a non-planar current sheet is postponed to further studies.}

As the field magnitude, and thus the HCS profile, depends solely on $\theta$, thus varying along a direction perpendicular to the solar wind flow, there is no compression of the current sheet and thus no driven reconnection. Therefore, the current sheet modeled in this work does not contain additional electric fields beyond the regular motional electric field 
\begin{align}
{\bf E} = -\frac{{\bf u}_\mathrm{sw}}{c}\times{\bf B}, \label{eq:motionalfield}
\end{align}
where $c$ is the speed of light. \replaced{Within the inner heliosphere this is an acceptable approximation.}{This electric field causes particles to experience ${\bf E}\times{\bf B}$ drift, analogous with corotation of field lines.} In the case of a wavy HCS (see, e.g., \citealt{Strauss2012}, \citealt{Pei2012}, and \citealt{Burger2012}), especially with greater heliocentric distance, an assumed radial solar wind flow will no longer be wholly in the current sheet plane, requiring more detailed analysis of possible reconnection.

Observations estimate the HCS thickness to be in the region of between 5000 and \mbox{40000 km} at \mbox{1 au} (see, e.g., \citealt{Eastwood2002} and \citealt{Winterhalter1994}).\added{ We examine effects of a gradual transition between hemispheres, and the effects of current sheet thickness. Although energetic protons may have Larmor radii much larger than the listed current sheet thicknesses, effects such as beamed injection and adiabatic focusing may cause the perpendicular velocity component of particles to be quite small, resulting in smaller than expected Larmor radii, thus warranting this approach.} Thus, we define the HCS thickness shape function $S$ to be a function of colatitude through latitude \mbox{$\delta=90^\circ - \theta$}, as
\begin{align}
S(\theta) = A \left(-1 + 2\,\mathcal{S}\,(\tfrac{1}{2} + \frac{2\delta}{l_\mathrm{HCS}}) \right) \label{eq:shapefunction}
\end{align}
where $A$ is a configuration parameter with values $+1$ or $-1$, $l_\mathrm{HCS}$ is the thickness of the HCS, and $\mathcal{S}$ is the \emph{Smootherstep} function \citep{ebert2003texturing} which maps the parameter range $[0,1]$ to the values $[0,1]$ as $\mathcal{S}(x)=6x^5-15x^4+10x^3$, resulting in a smooth transition with nil first and second-order derivatives at the endpoints. Closer to the Sun, this parametrization results in smaller current sheet thicknesses. The parameter $A$ defines the polarity of the dipolar field according to cosmic ray physics standard notation, i.e. a configuration of $A+$ ($A=+1$) has an outwards-pointing field in the northern hemisphere, and a configuration of $A-$ ($A=-1$) has an inwards-pointing field in the northern hemisphere, with the direction of the field in the southern hemisphere reversed. We additionally assess the validity of implementing a HCS with zero thickness, using a shape function $S_\mathrm{H}$, which implements the Heaviside step function H as
\begin{align}
S_\mathrm{H}(\theta) = A \left(-1 + 2\,\mathrm{H}(\delta) \right). \label{eq:shapefunctionH}
\end{align}

Protons propagating within the fields given by equations (\ref{eq:SParker})--(\ref{eq:shapefunctionH}) will experience drifts due to the electric field, and the gradients and curvature of the magnetic field. A full analytical treatise of particle drifts in a Parker spiral, far from the HCS, can be found in \cite{Dalla2013}, where a better-suited field-aligned frame of reference $(\hat{\bf e}_l,\hat{\bf e}_{\phi'},\hat{\bf e}_{\theta'})$ is introduced. Within this system, $\hat{\bf e}_l$ is directed along the Parker spiral magnetic field line, outwards from the Sun. $\hat{\bf e}_{\theta'}$ is antiparallel to the standard spherical coordinate vector $\hat{\bf e}_{\theta}$, and $\hat{\bf e}_{\phi'}$ completes the coordinate system. Below, we summarize the non-relativistic forms of the main drifts, the electric field, $\nabla {\bf B}$, and curvature drifts, for the simple case of a unipolar IMF ($S(\theta)\equiv 1$), as
\begin{align}
{\bf v}_{E} &= \frac{u_\mathrm{sw}r}{(r^2+a^2)^{1/2}} \hat{\bf e}_{\phi'} \\
{\bf v}_{\nabla} &= \frac{\mu c}{q} \frac{r \cot \theta}{r^2+a^2} \hat{\bf e}_{\phi'}
                  - \frac{\mu c}{q} \frac{r^2+2a^2}{(r^2+a^2)^{3/2}} \hat{\bf e}_{\theta'}  \\
{\bf v}_{c} &= -\frac{m c}{q B} v_\parallel^2 \frac{r \cot \theta}{r^2+a^2} \hat{\bf e}_{\phi'}
               -\frac{m c}{q B} v_\parallel^2 \frac{r^2+2a^2}{(r^2+a^2)^{3/2}} \hat{\bf e}_{\theta'}, 
\end{align}
where $a$ is a function of colatitude $a = u_\mathrm{sw} (\Omega_\odot \sin \theta)^{-1}$ and $\mu$ is the particle magnetic moment $\mu = m v_\perp^2 (2 B)^{-1}$. Here $m$ and $q$ are the particle mass and charge, and $v_\parallel$ and $v_\perp$ are the components of velocity parallel and perpendicular to the magnetic field, respectively. See \citep{Dalla2013} for the more general relativistic expressions.

The analytical forms show that for near-equatorial latitudes, the term aligned with $\hat{\bf e}_{\theta'}$ dominates both curvature and gradient drifts. For both these drifts, when considering the two polarity configurations of the Solar dipolar field, we find that for $A+$, both hemispheres cause drift of positively charged particles towards the equator, and for $A-$, away from it, the patterns well known from GCR studies. Thus, for the $A+$ configuration, the equator is a stable position, and for the $A-$ configuration, a labile position.

Inclusion of the HCS, for example defined through a shape function $S(\theta)$, will cause additional drifts due to change of magnetic field as a function of $\theta$. The first drift, valid for both smooth and step-mode current sheet profiles, is the current sheet drift, described commonly as Speiser motion \citep{Speiser1965}. With $B$ approaching zero, the guiding centre approximation of particle motion breaks down. Particles oscillate between the two magnetic field polarities by performing partial gyromotion in each side, then crossing the sheet, and performing gyromotion of opposite chirality on the other side. For particles with positive charge, this motion is in a western direction for $A+$ and an eastern direction for $A-$. For a step-mode field transition and an isotropic distrubution, this was found to lead to an average velocity of $\langle v_\mathrm{S}\rangle = 0.463v$ \citep{Burger1985}.

If the gyroradius of particles is smaller than the characteristic length scale describing the rate of change for ${\bf B}$ due to the shape function $S(\theta)$, a second drift is found at the current sheet, taking the form of classical gradient drift, and defined as
\begin{align*}
{\bf v}_\mathrm{g} = \frac{cm}{2q} \frac{v_\perp^2}{B^3} {\bf B}\times(\nabla B).
\end{align*}
If this drift is present, then $\nabla B$ would be aligned with $\theta'$, leading to the gradient drift being aligned with $\phi'$. The direction of this gradient drift would be opposite to that of current sheet drift (or Speiser motion). 


\section{Simulations}

In our simulations, we investigate SEP trajectories in the fixed frame (spacecraft frame) in the presence of a flat HCS using a numerical test particle model \citep{Dalla2005,Kelly2012} with modifications suited to heliospheric problems introduced in \cite{Marsh2013}. Instead of using the focused transport equation (see, e.g., \citealt{Roelof1969}), we solve the full three-dimensional differential equations of motion for each particle. In our model, drifts are not introduced into the relevant equations analytically, but instead arise naturally from the Lorentz and electric force due to the fields given by equations \ref{eq:SParker}--\ref{eq:shapefunctionH} acting on particles during each step of their motion. 

We simulate the propagation of energetic protons, injected instantaneously at time $t=0$ from a heliocentric distance of $2\,R_\odot$. Protons are injected from a region with angular extent $6^\circ \times 6^\circ$, centered at the heliographic equator. For select studies, the injection latitude was varied in order to assess latitudinal drifts. Particles have initial pitch-angles pointing in a random direction within a hemisphere pointing outwards from the Sun along the Parker spiral. The relativistic differential equations of particle motion and acceleration are solved using a self-optimizing Bulirsch-Stoer method \citep{press1996numerical}. Particles are propagated in the prescribed magnetic and electric fields, where the motional electric field is solved using a solar wind speed of $u_\mathrm{sw}=500\,\mathrm{km}\,\mathrm{s}^{-1}$. In order to model the effects of turbulence and wave-particle effects, particles experience large-angle scattering in the solar wind frame, with Poisson-distributed scattering intervals, and a constant rigidity-independent mean free path of \mbox{1 au}, in agreement with an assumed low level of scattering. 

We inject $N=10^5$ particles and trace their propagation within the heliosphere for 100 hours. Snapshots of particle profiles are provided every 60 minutes. A collection sphere is placed at a heliocentric distance of \mbox{1 au}, over which crossings are tracked, allowing the generation of fluence maps, histograms, and virtual observer time profiles. Fluence maps were generated with tiles of angular extent $1^\circ \times 1^\circ$ over the full length of the simulation, whereas time profile generation used $6^\circ \times 6^\circ$ windows and 30 minute time binning.

\begin{figure}[!ht]
\centering
\includegraphics[width=0.45\textwidth]{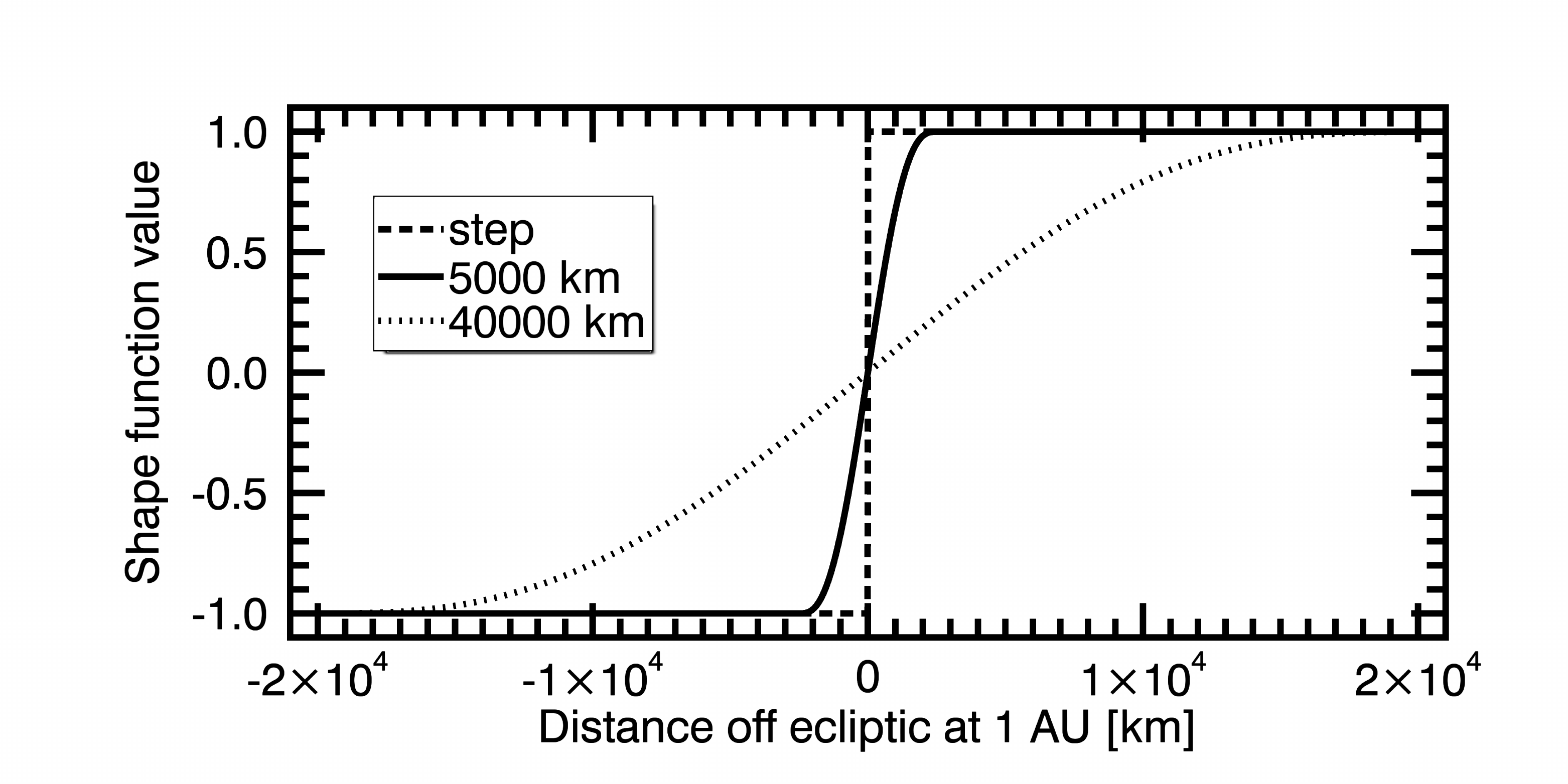}\\
\includegraphics[width=0.45\textwidth]{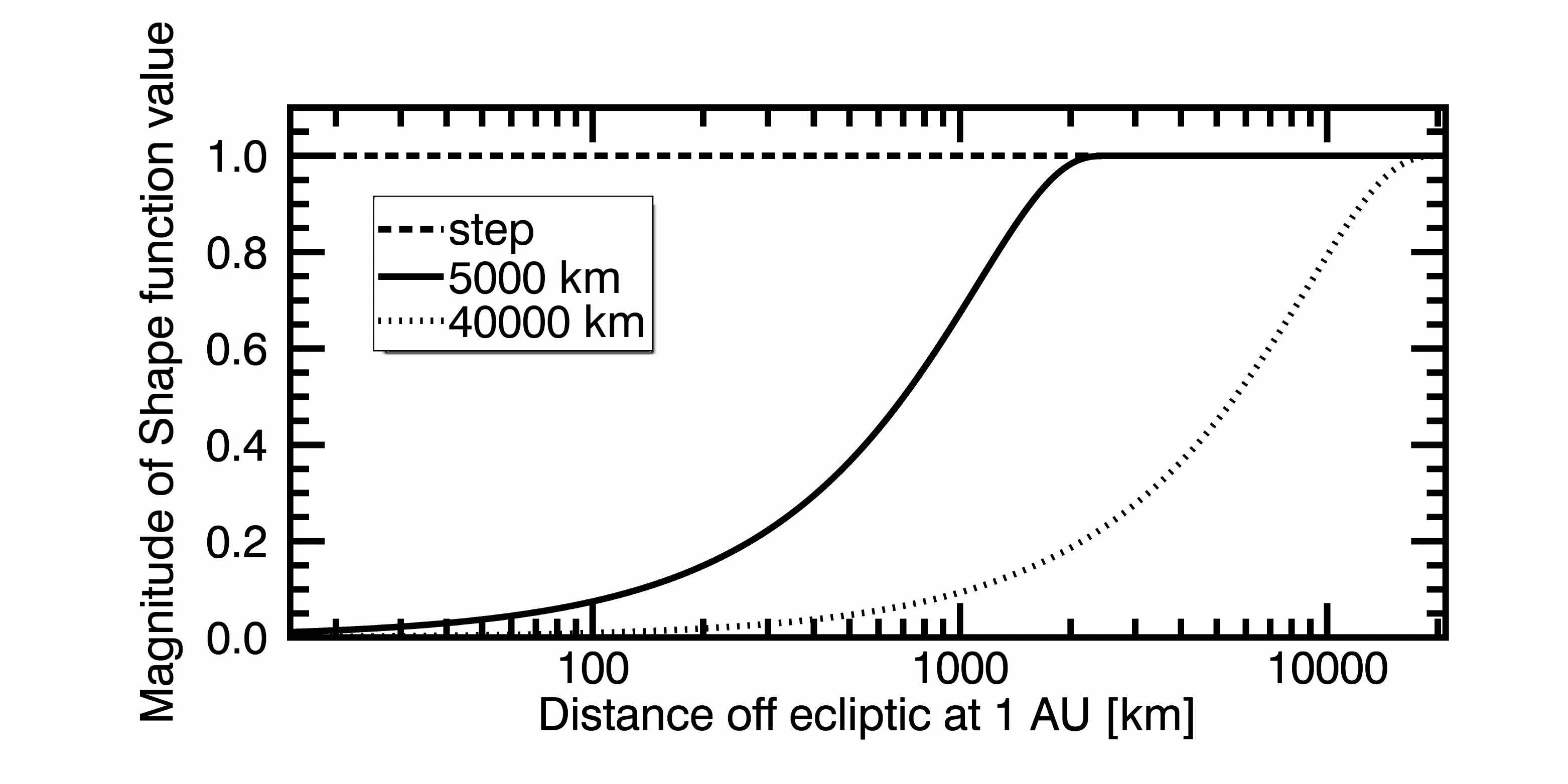}
\caption{Shape function $S(\theta)$ as seen at a heliocentric distance of \mbox{1 au}. The dashed, solid, and dotted lines correspond with HCS thicknesses of \mbox{0 km}, \mbox{5000 km}, and \mbox{40000 km}, respectively. The shape function is displayed with both linear (top) and logarithmic (bottom) distance from the heliographic equator, where the logarithmic plot shows only the positive half of the function.}\label{fig:shapefunction}
\end{figure}

We chose eight different magnetic field configurations for use in our simulations. As reference cases, we simulated particle propagation in both inwards- and outwards-pointing unipolar fields ($S(\theta)=\pm 1$). For heliospheric current sheets we used three different thicknesses, by varying the parameter $l_\mathrm{HCS}$ in equation (\ref{eq:shapefunction}). Each current sheet thickness was simulated for both $A+$ and $A-$ configurations, as described in section \ref{sec:HCS}. The current sheet was simulated with thicknesses of \mbox{0 km}, \mbox{5000 km}, and \mbox{40000 km} at \mbox{1 au}. The first case was in fact modeled as a Heaviside step function using the shape function (\ref{eq:shapefunctionH}). A plot of $S(\theta)$ at \mbox{1 au} for various current sheet thicknesses is shown in Figure \ref{fig:shapefunction}. The shown thicknesses of \mbox{5000 km} and \mbox{40000 km} at \mbox{1 au} correspond with angular extents of $0.0019^\circ$ and $0.015^\circ$, respectively.

In order to simulate the infinitesimally thin current sheet, henceforth referred to as the Heaviside case, we could not use the regular Bulirsch-Stoer algorithm as it could not automatically optimise particle propagation over the step function. Instead, we used an adaptive-step leapfrog Boris-push method \citep{Boris1970}, which is a solver commonly used in Particle-In-Cell (PIC) codes.

For each run, protons are injected as either monoenergetic populations with initial energies of \mbox{1 MeV}, \mbox{10 MeV}, \mbox{40 MeV}, \mbox{100 MeV}, \mbox{400 MeV}, or \mbox{800 MeV}, or as a power-law between \mbox{10 MeV} and \mbox{400 MeV} with a spectral index of $\gamma=-1.1$. 


\section{Results} \label{sec:results}

\begin{figure*}[!htp]
\centering
\includegraphics[width=0.245\textwidth]{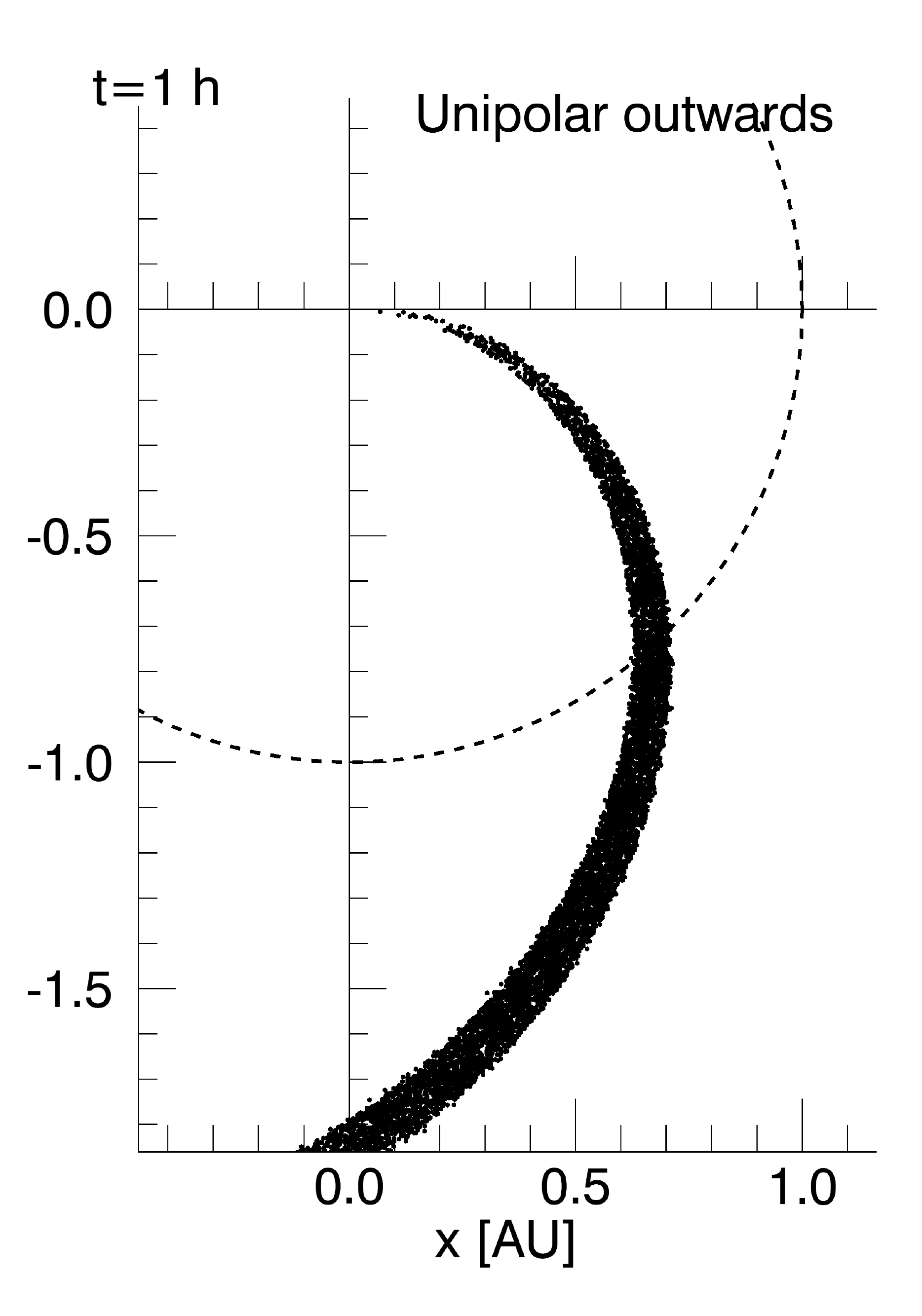}
\includegraphics[width=0.245\textwidth]{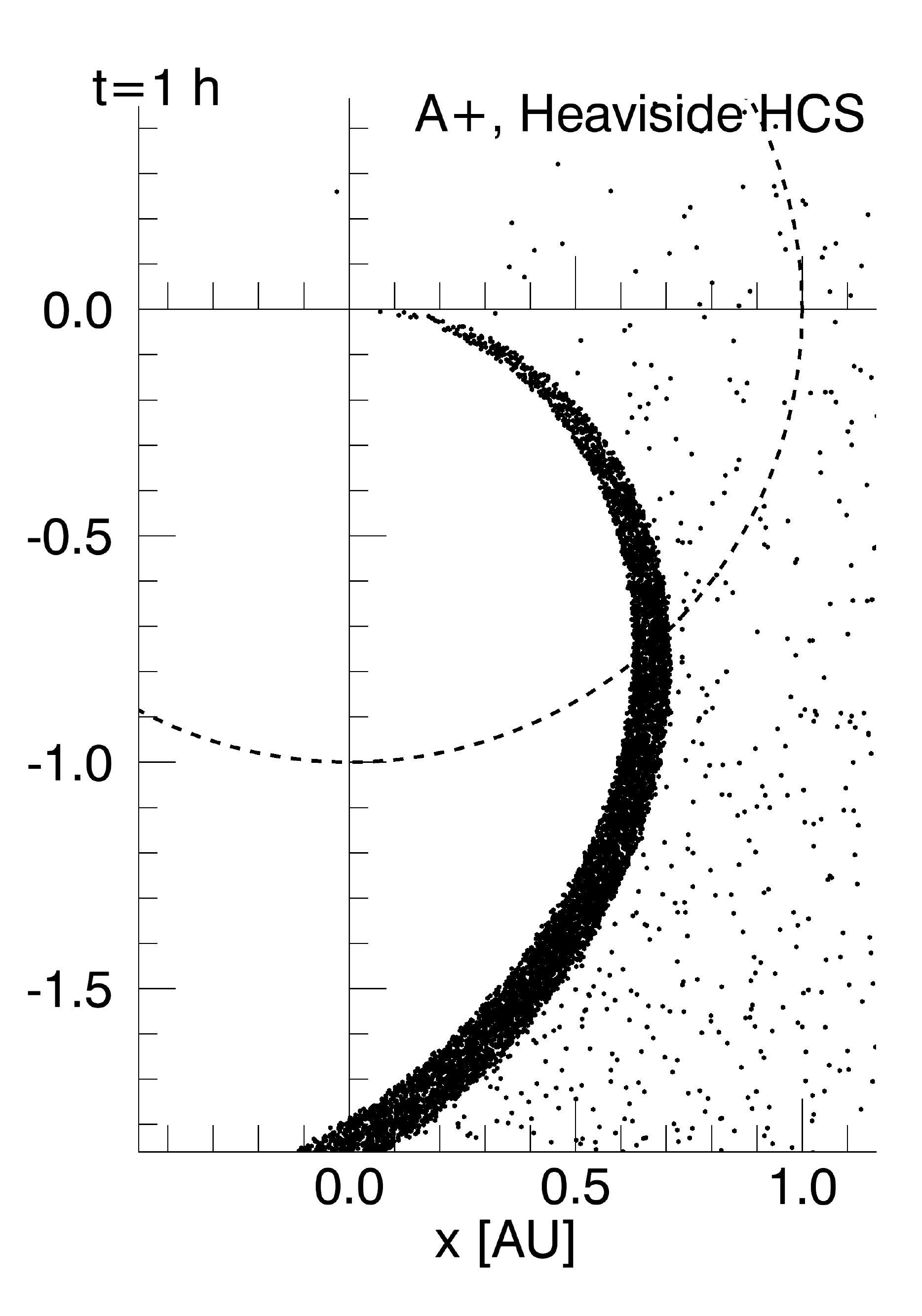}
\includegraphics[width=0.245\textwidth]{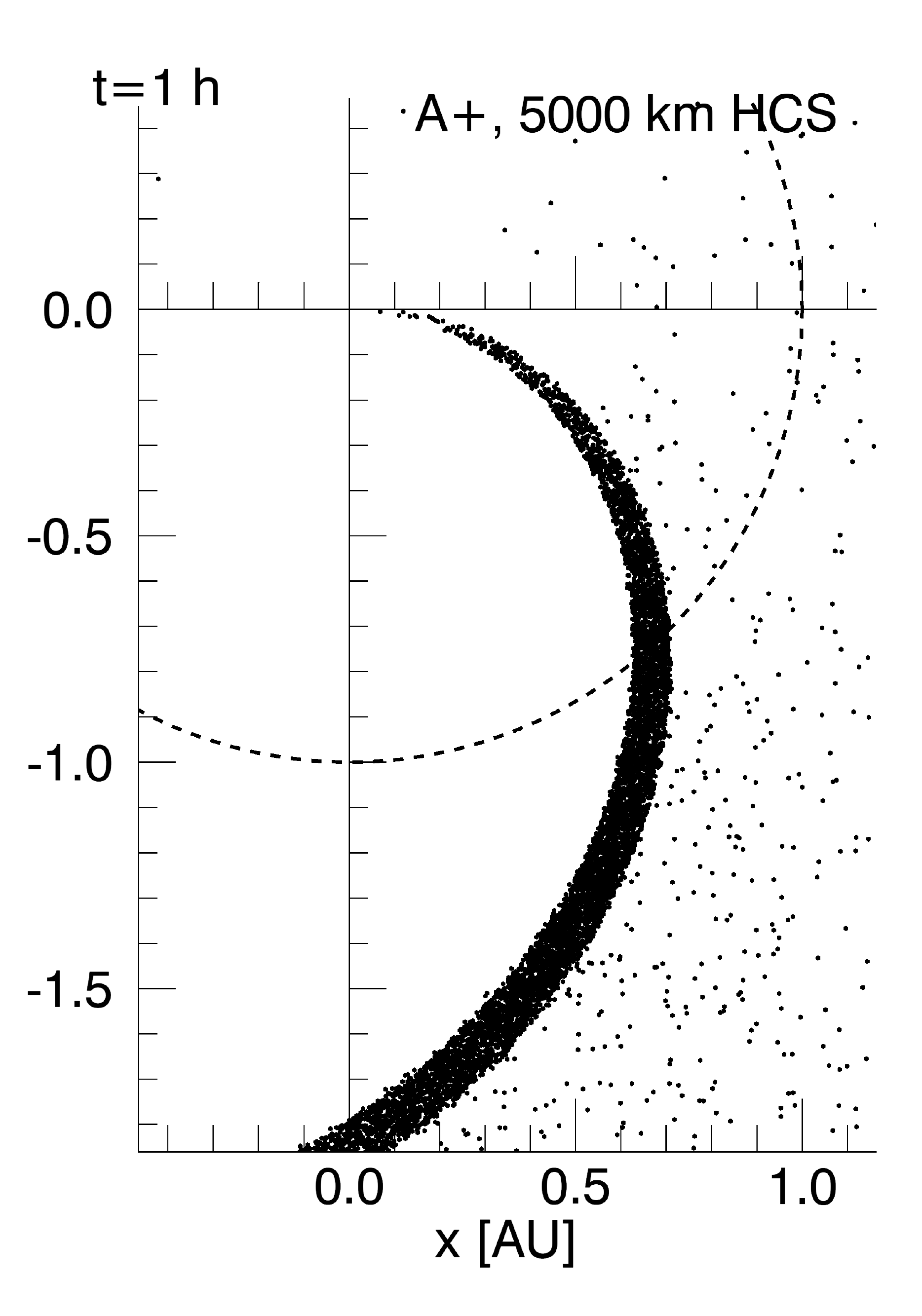}
\includegraphics[width=0.245\textwidth]{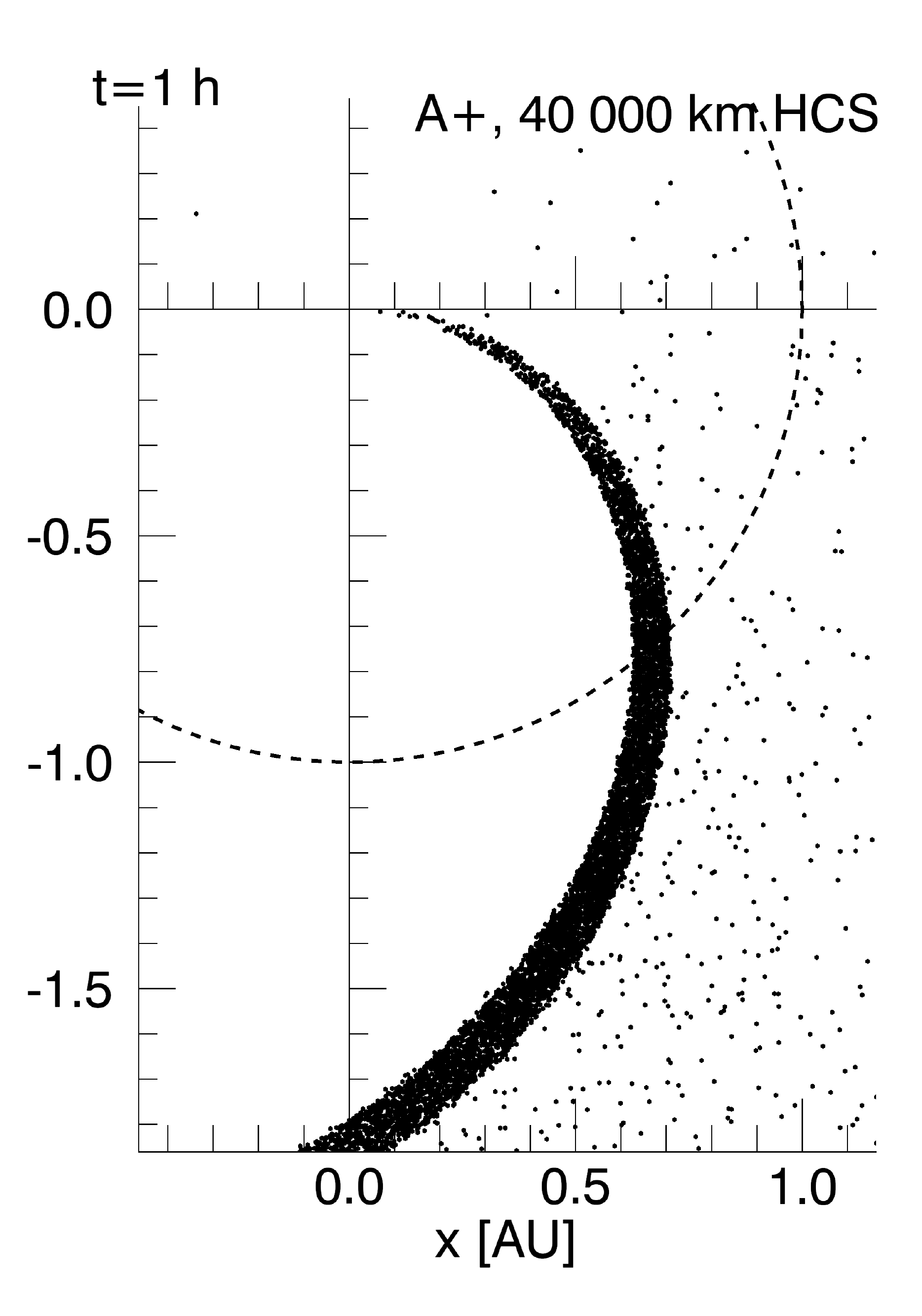}
\includegraphics[width=0.245\textwidth]{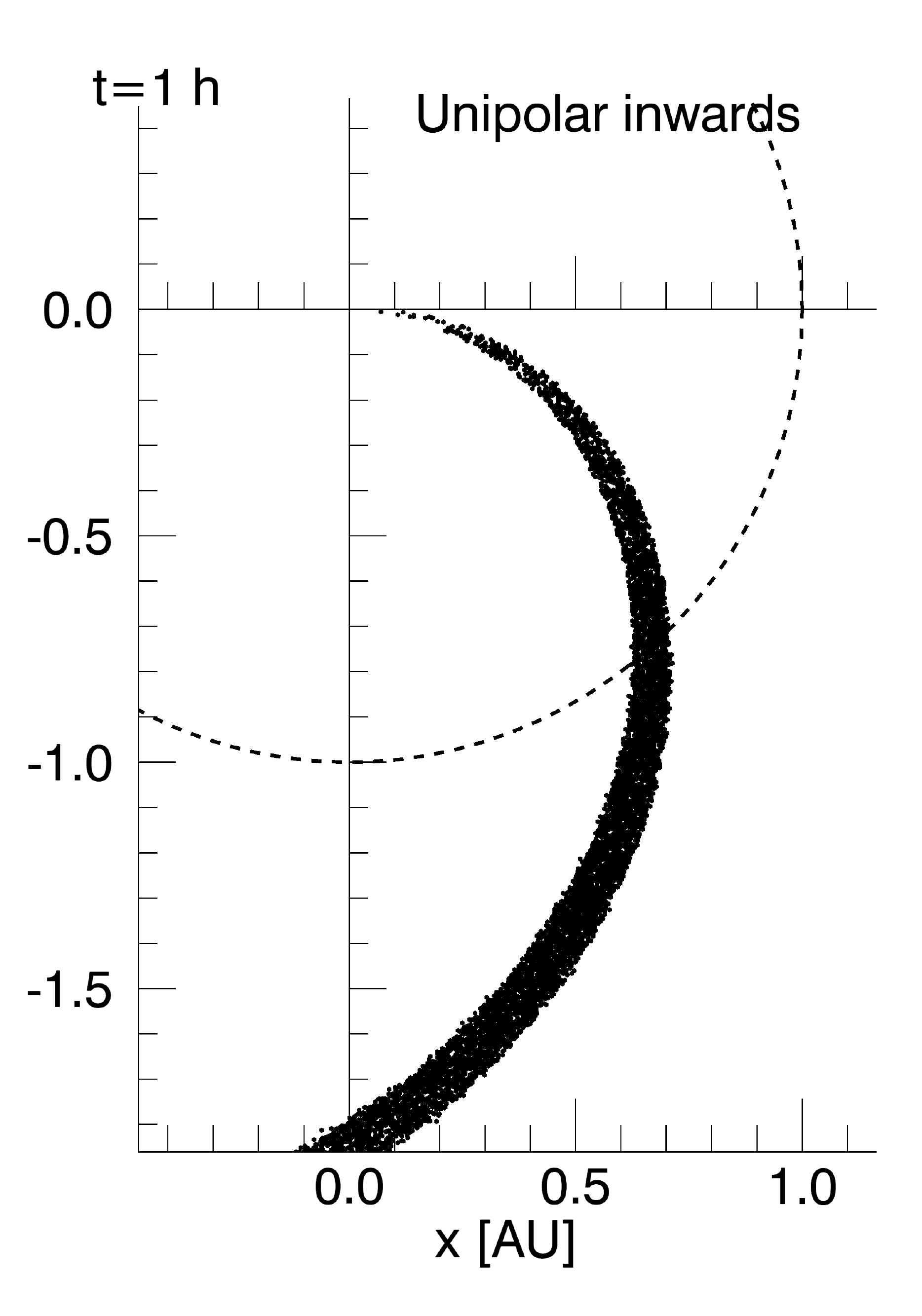}
\includegraphics[width=0.245\textwidth]{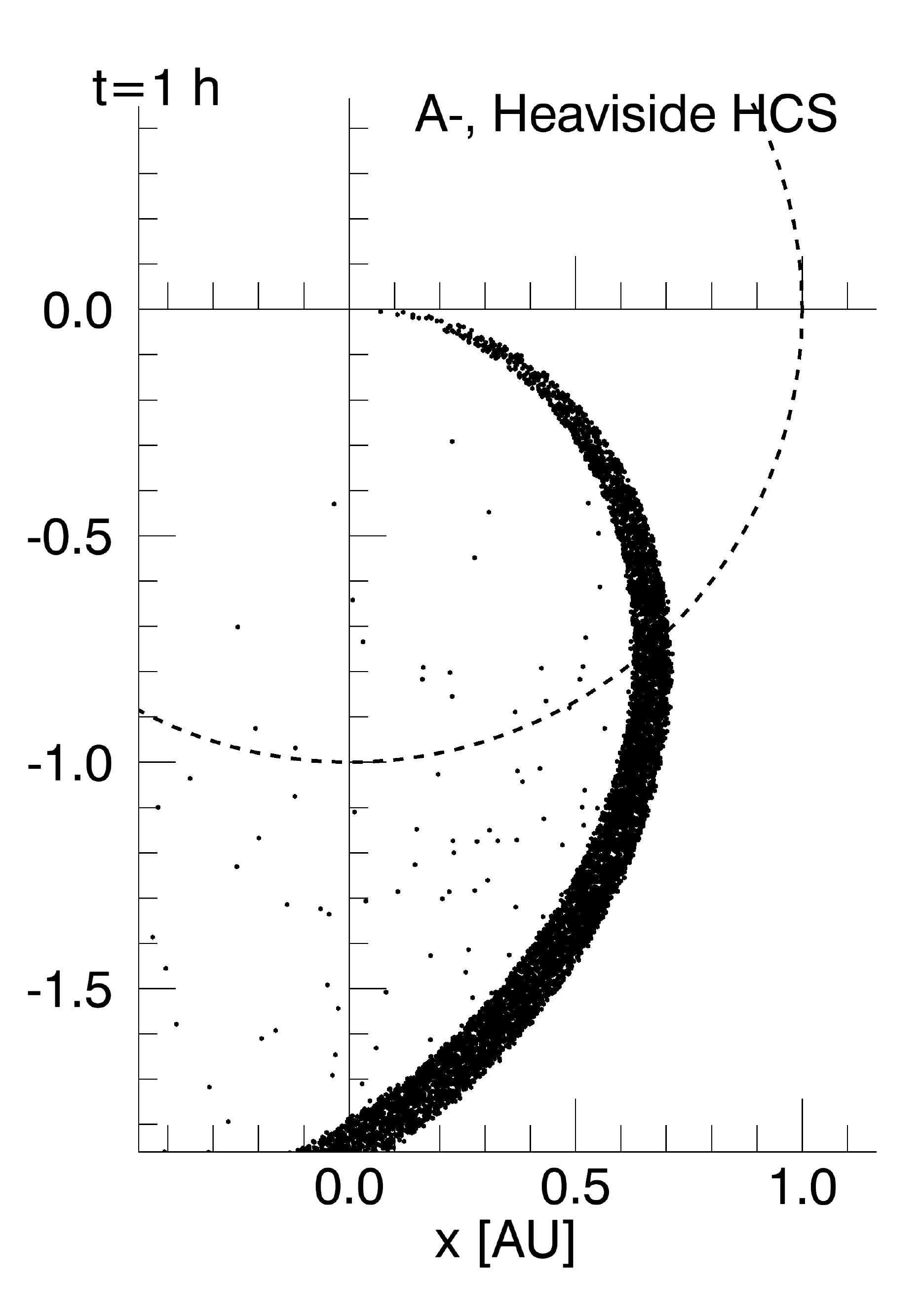}
\includegraphics[width=0.245\textwidth]{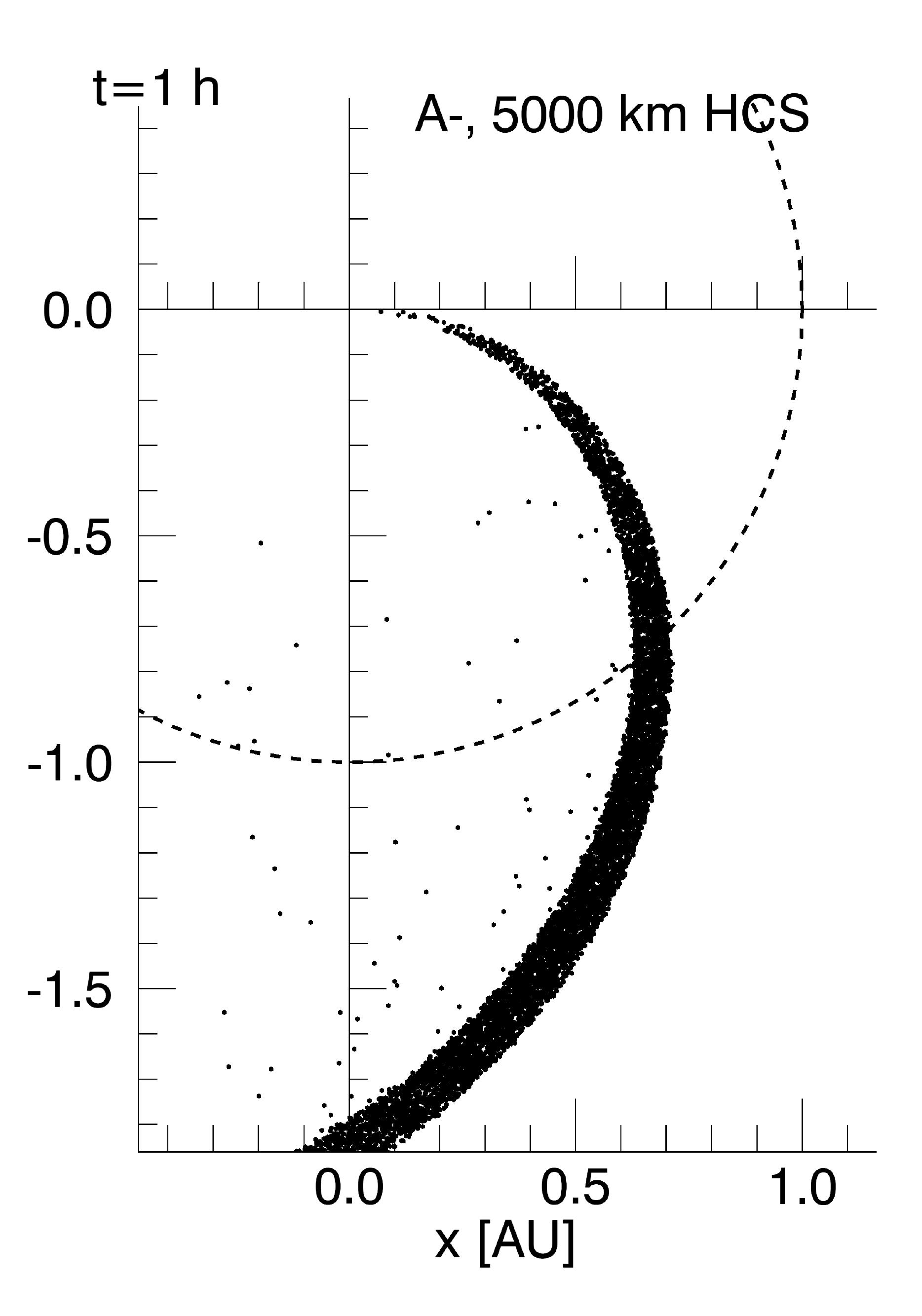}
\includegraphics[width=0.245\textwidth]{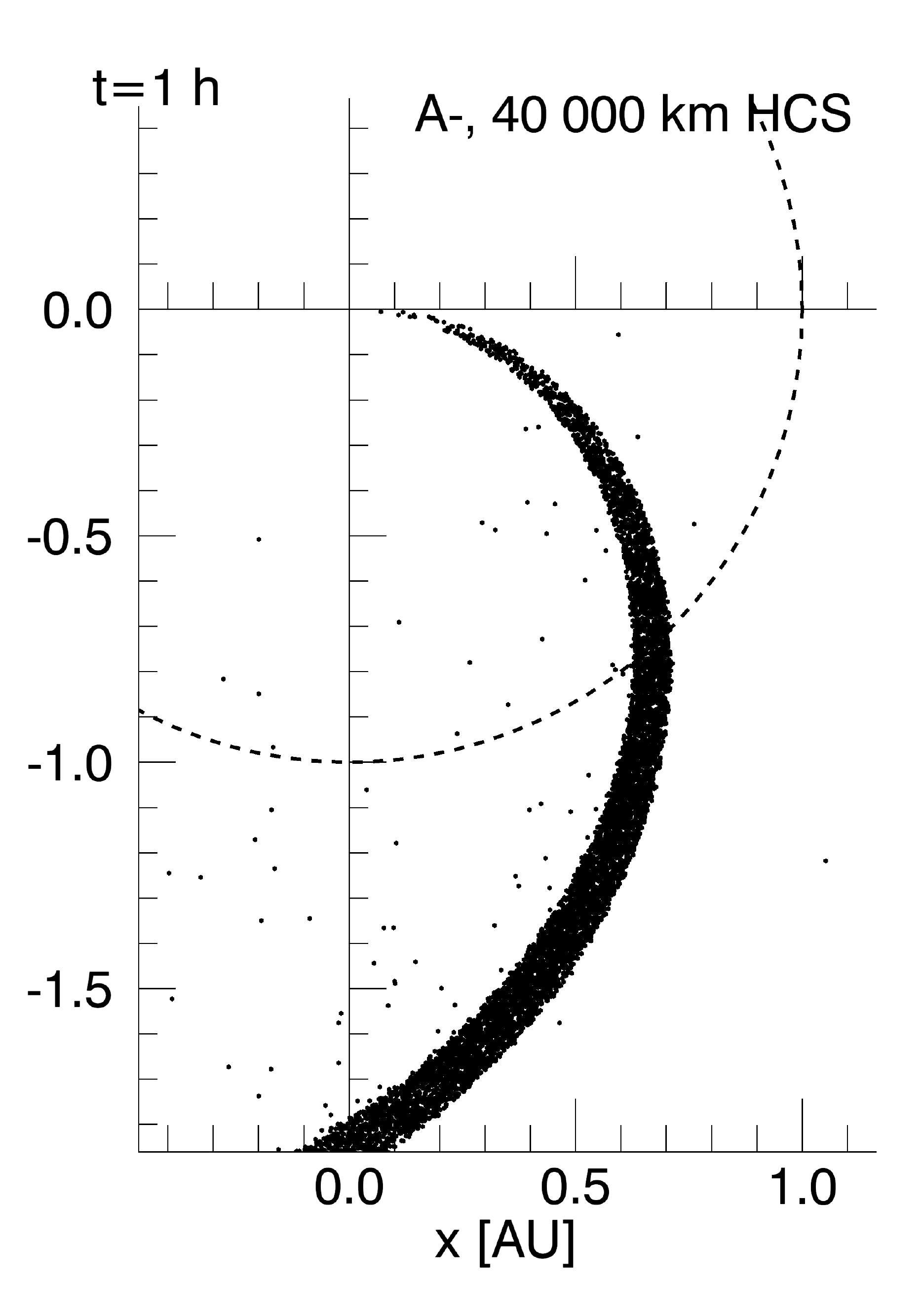}
\caption{Projection of protons injected at \mbox{100 MeV}, after \mbox{1 hr} of simulation, onto the $x-y$ plane for eight different magnetic field configurations. Top row, from left: outwards-pointing unipolar field, followed by $A+$ configurations with current sheet thickness parameters corresponding with \mbox{1 au} thicknesses of \mbox{0 km} (Heaviside step), \mbox{5000 km}, and \mbox{40000 km}. Bottom row, from left: inwards-pointing unipolar field, followed by $A-$ configurations with current sheet thickness parameters corresponding with \mbox{1 au} thicknesses of \mbox{0 km} (Heaviside step), \mbox{5000 km}, and \mbox{40000 km}. A distance of \mbox{1 au} is displayed with a dashed circle. The proton spreads show that current sheet drift (top row: to the right, bottom row: to the left) is noticeable for all current sheets.}\label{fig:thickness}
\end{figure*}

Our first step was to perform qualitative assessment of apparent HCS drifts as a function of sheet thickness. In Figure \ref{fig:thickness} we show comparisons between all eight simulated IMF configurations. We plot the distribution of protons injected at \mbox{100 MeV} after \mbox{1 hour} of propagation flattened to the $x-y$ plane (the equatorial plane of the Sun, with the x-axis pointing in the direction of $0^\circ$ longitude). Injection was centered at $(0^\circ, 0^\circ)$. The unipolar cases (leftmost column) show that within \mbox{1 hr}, little drift has taken place. The three $A+$ panels (top row) show that the presence of a current sheet generates significant current sheet drift to the right (west), and the three $A-$ panels (bottom row) show current sheet drift to the left (east). Gradient drift associated with the variation of $B$ over the thickness of the current sheet is found to be negligible. 

We also performed a check to verify that the protons which appear to have drifted are indeed drifting protons, not a projection effect due to the $x-y$ plot. Protons experiencing current sheet drift were confirmed to be located in the vicinity of the HCS. Plots performed for other proton energies show comparable results, with increase in proton energy resulting in greater deviation from the well-connected field lines. At later stages of the simulation, up to \mbox{100 hrs}, the distribution of protons in the inner heliosphere remains characteristically comparable with the \mbox{1 hr} case, although corotation causes an \replaced{eastward}{westward} transition of all protons, and the general propagation of protons outwards from the Sun causes the proton counts close to the Sun to decrease.

\begin{figure}[!ht]
\centering
\includegraphics[width=0.45\textwidth]{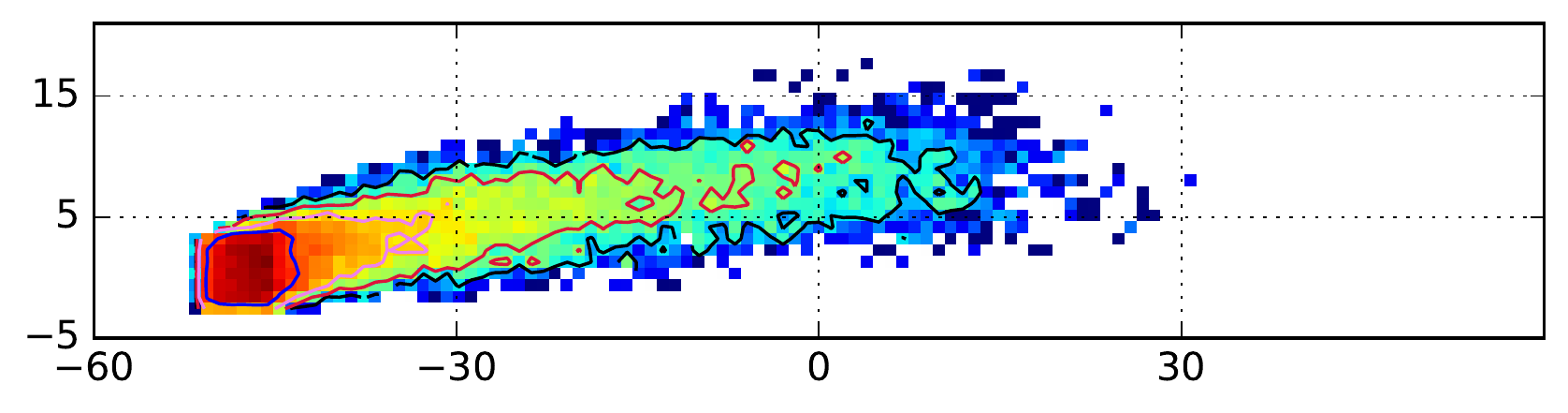}\\
\includegraphics[width=0.45\textwidth]{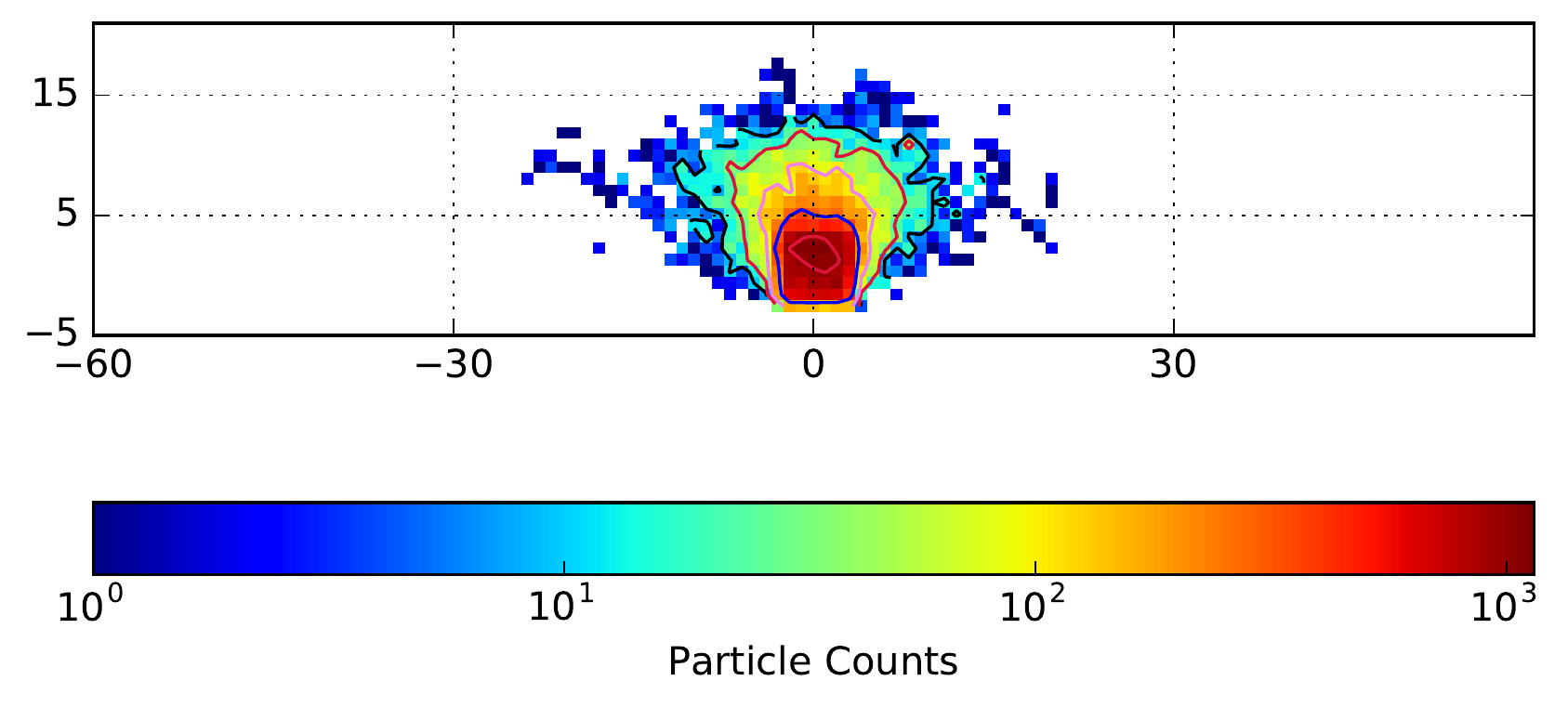}
\caption{Map of protons, injected at \mbox{100 MeV}, crossing the \mbox{1 au} sphere using $1^\circ \times 1^\circ$ binning for a unipolar inward-pointing magnetic field. Fluence colours and contours are on a logarithmic scale, with two contours per decade. Top panel: Proton crossing coordinates (in degrees) relative to injection site, showing how the ${\bf E}\times{\bf B}$ drift and latitudinal drifts both work concurrently. Bottom panel: Proton crossing coordinates relative to the best-connected fieldline with the effects of ${\bf E}\times{\bf B}$ drift removed.}\label{fig:map_correction}
\end{figure}

In order to assess the magnitude of proton drifts, we gathered all proton crossings across the \mbox{1 au} sphere, saving the time of crossing, the longitude and the latitude of each proton. In Figure \ref{fig:map_correction} we show a map of 100 MeV proton crossing counts, for a unipolar inwards-pointing magnetic field, relative to the injection coordinates, using $1^\circ \times 1^\circ$ binning, adding up all counts over the \mbox{100 h} duration of the simulation (top panel). We also show a comparative picture where we have removed the effects of corotation (bottom panel). Corotation, also described as the ${\bf E}\times{\bf B}$ drift, is caused by the field lines along which the particles propagate being frozen into the\added{ radially outflowing} solar wind plasma\added{, resulting in the intersection points at \mbox{1 au} being rotated westwards}. We also added a longitudinal offset to the bottom panel, so coordinates are shown in relation to the best-connected field line. Henceforth, we will utilise these corrections.

The proton distributions in Figure \ref{fig:map_correction} show that the effect of corotation is significant, which is unsurprising considering the \mbox{100 hr} extent of the simulation. The strongest fluence is found at the well-connected fieldline. A drift in latitude (upwards for this polarity) is seen, as well as one in longitude, moving protons away from the well-connected field lines (see also \citealt{Marsh2013}).

We now refer to \cite{Dalla2013} as a theoretical basis of drift analysis. The strongest drifts in longitude (gradient and curvature) are found to be proportional to a function $g(\theta)$, which approaches zero at the equator. This explains protons displaying significant longitudinal drift only after having drifted to higher latitudes. In this field configuration, gradient drift ($\propto v_\perp^2$) pushes protons to the west whereas curvature drift ($\propto v_\parallel^2$) causes drift towards the east. Both gradient and curvature drifts are in the same latitudinal direction, which for this field configuration is to higher colatitudes. The polarisation drift is of smaller magnitude, and thus, ignored in this work.


%
%
%

\begin{figure*}[p] 
\centering
\includegraphics[width=0.45\textwidth]{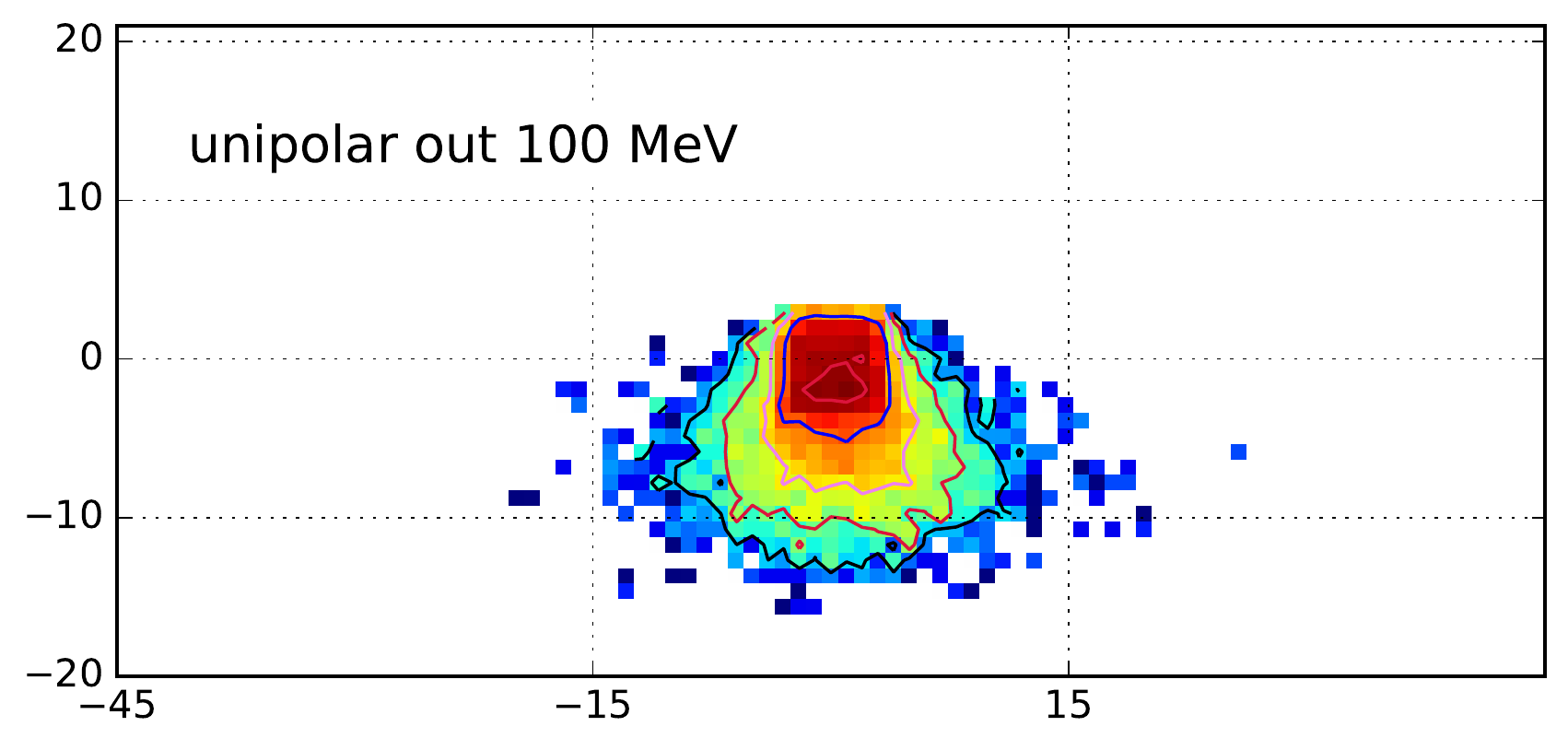}
\includegraphics[width=0.45\textwidth]{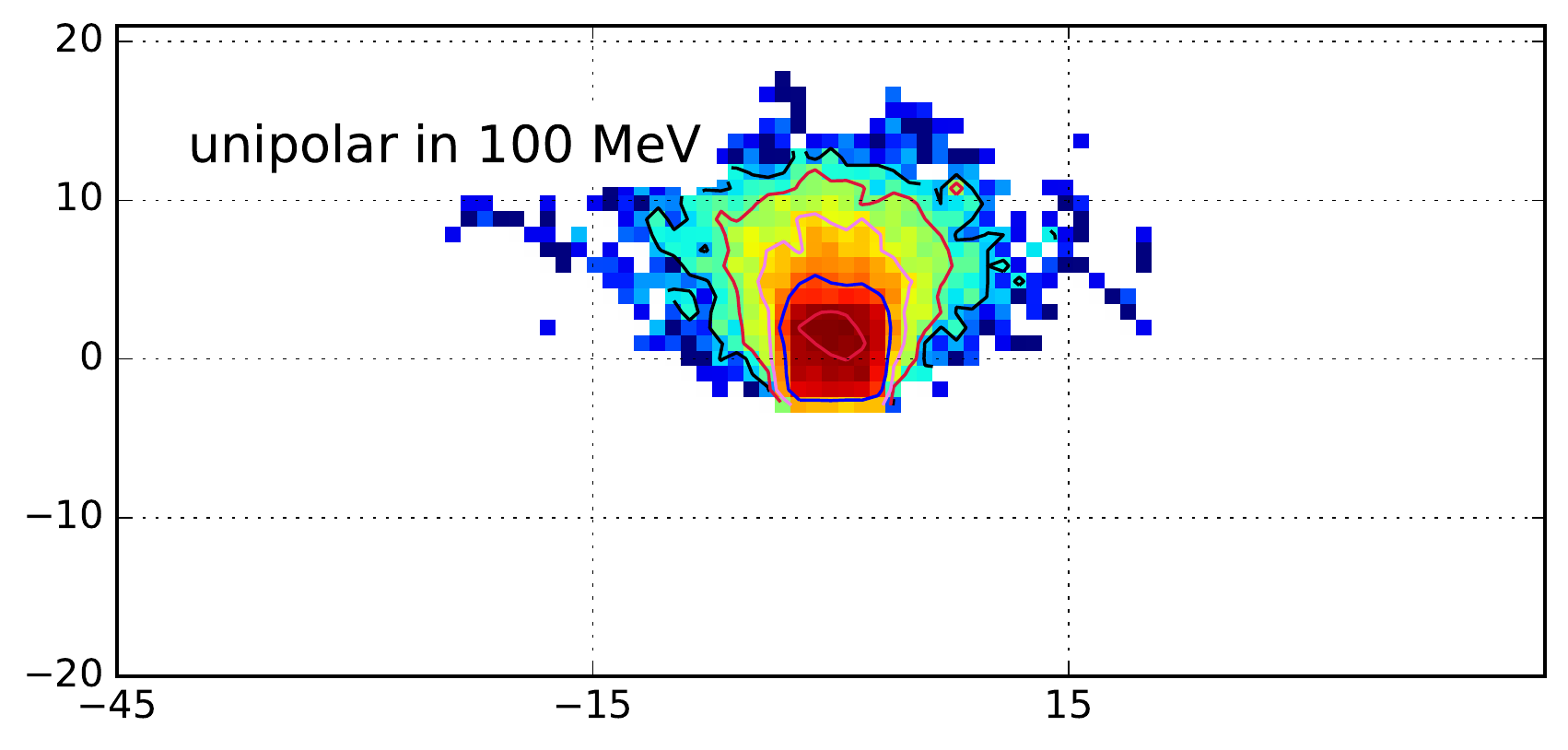}
\includegraphics[width=0.45\textwidth]{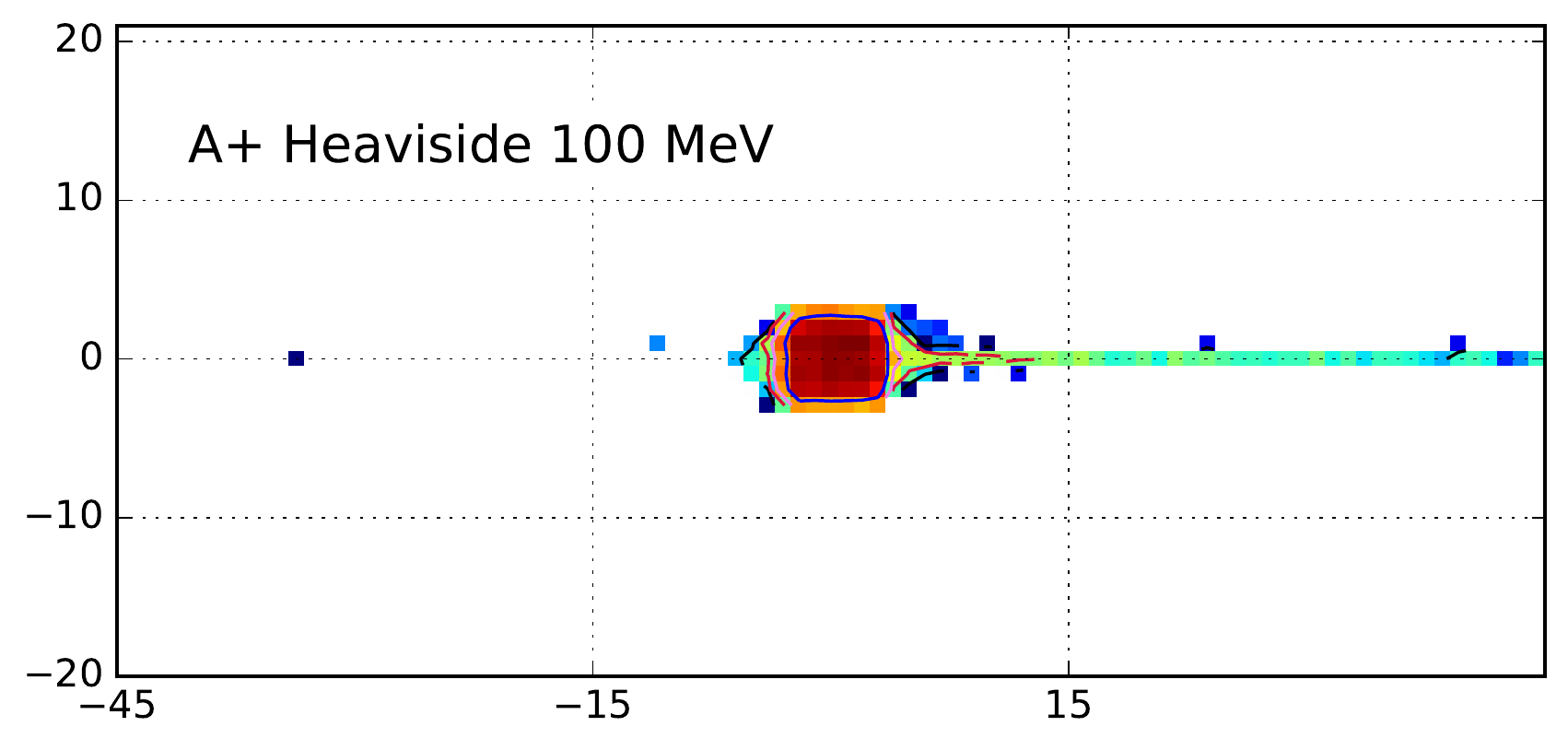}
\includegraphics[width=0.45\textwidth]{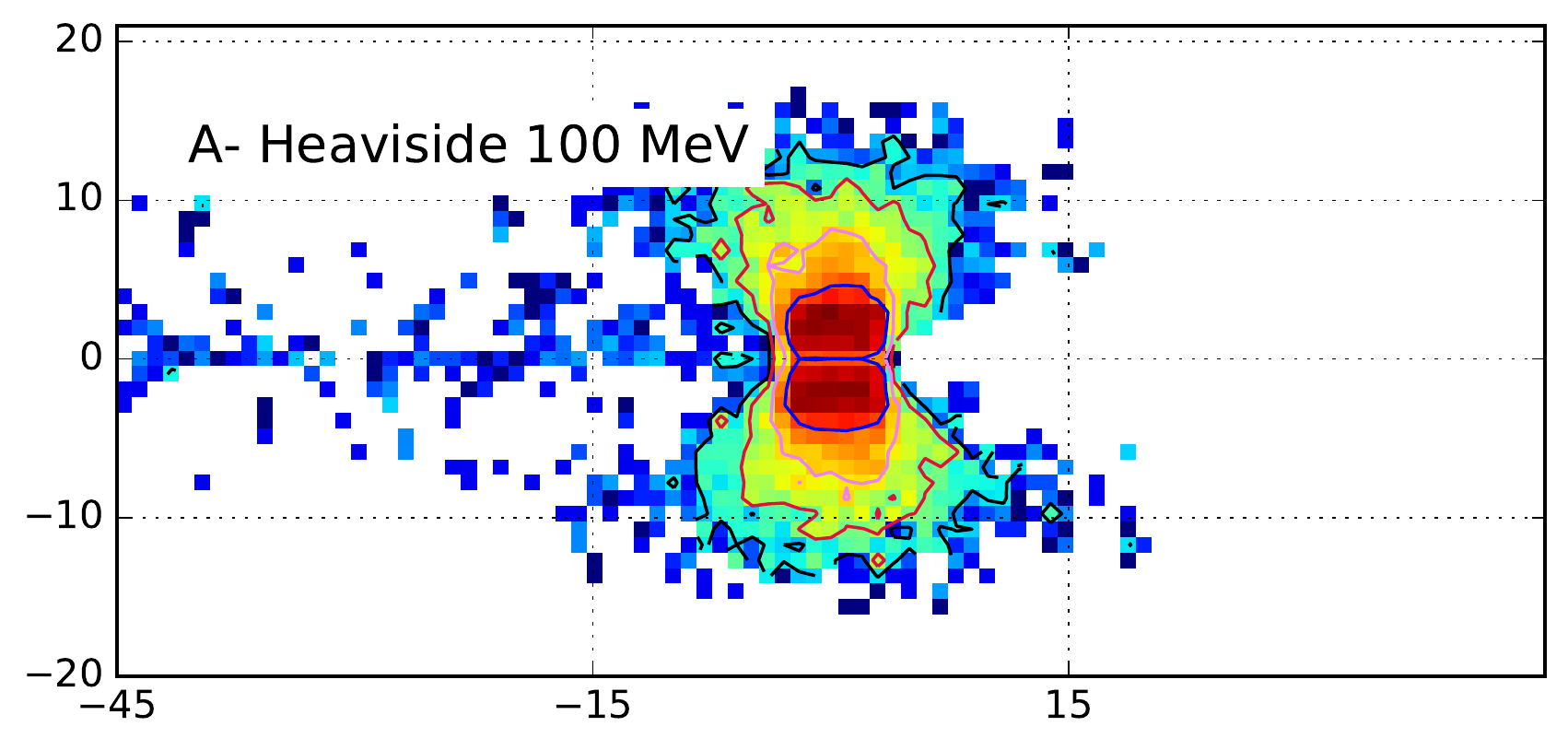}
\includegraphics[width=0.45\textwidth]{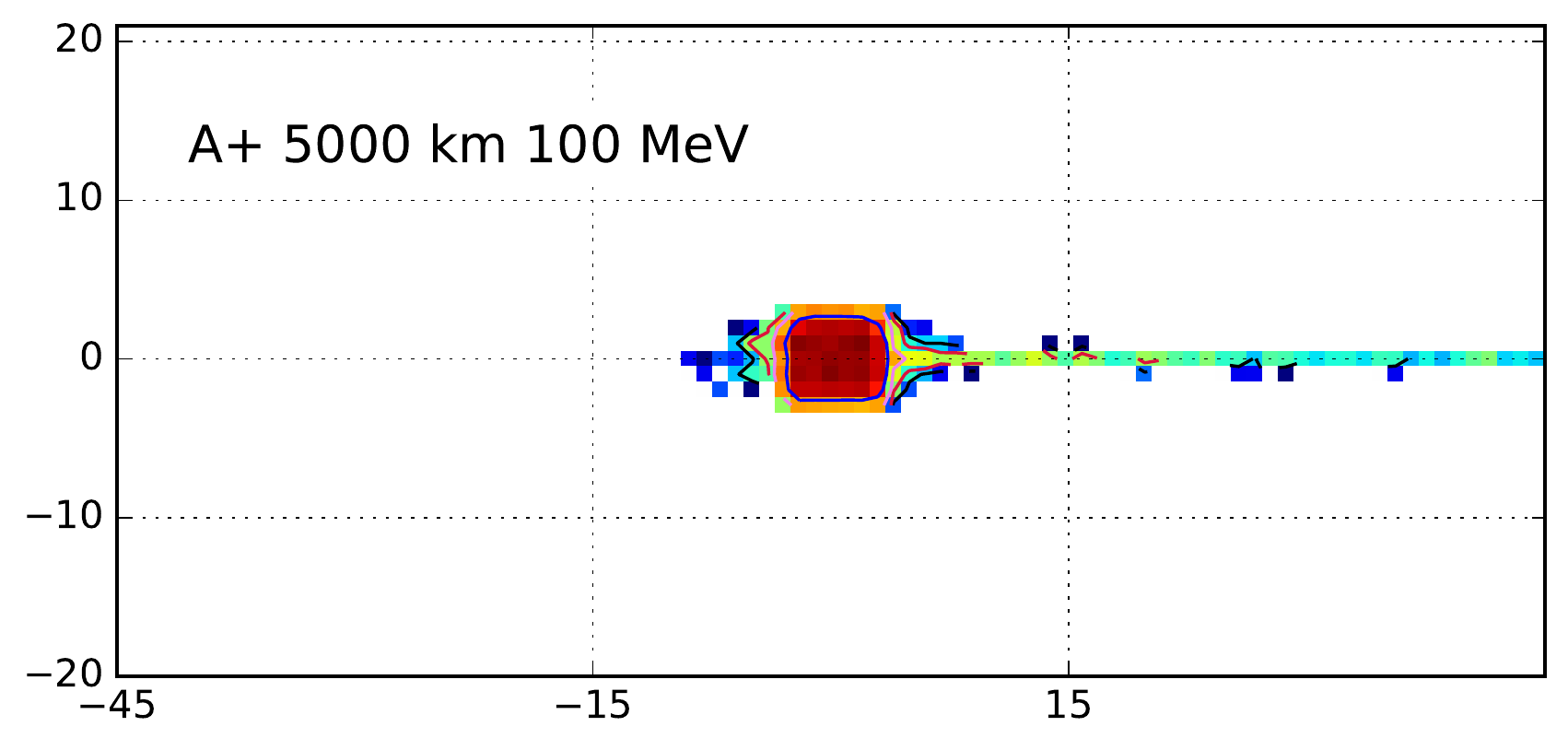}
\includegraphics[width=0.45\textwidth]{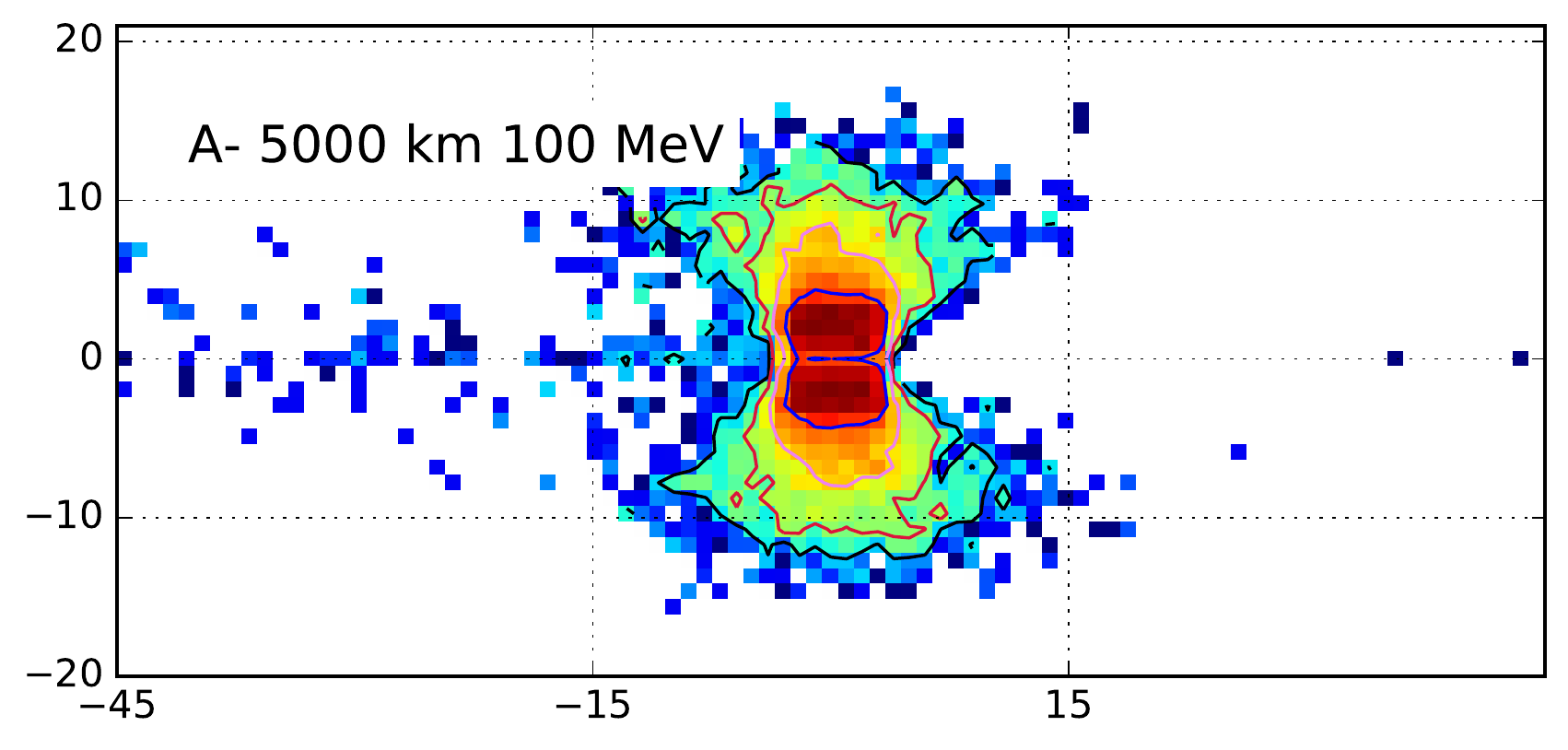}
\includegraphics[width=0.45\textwidth]{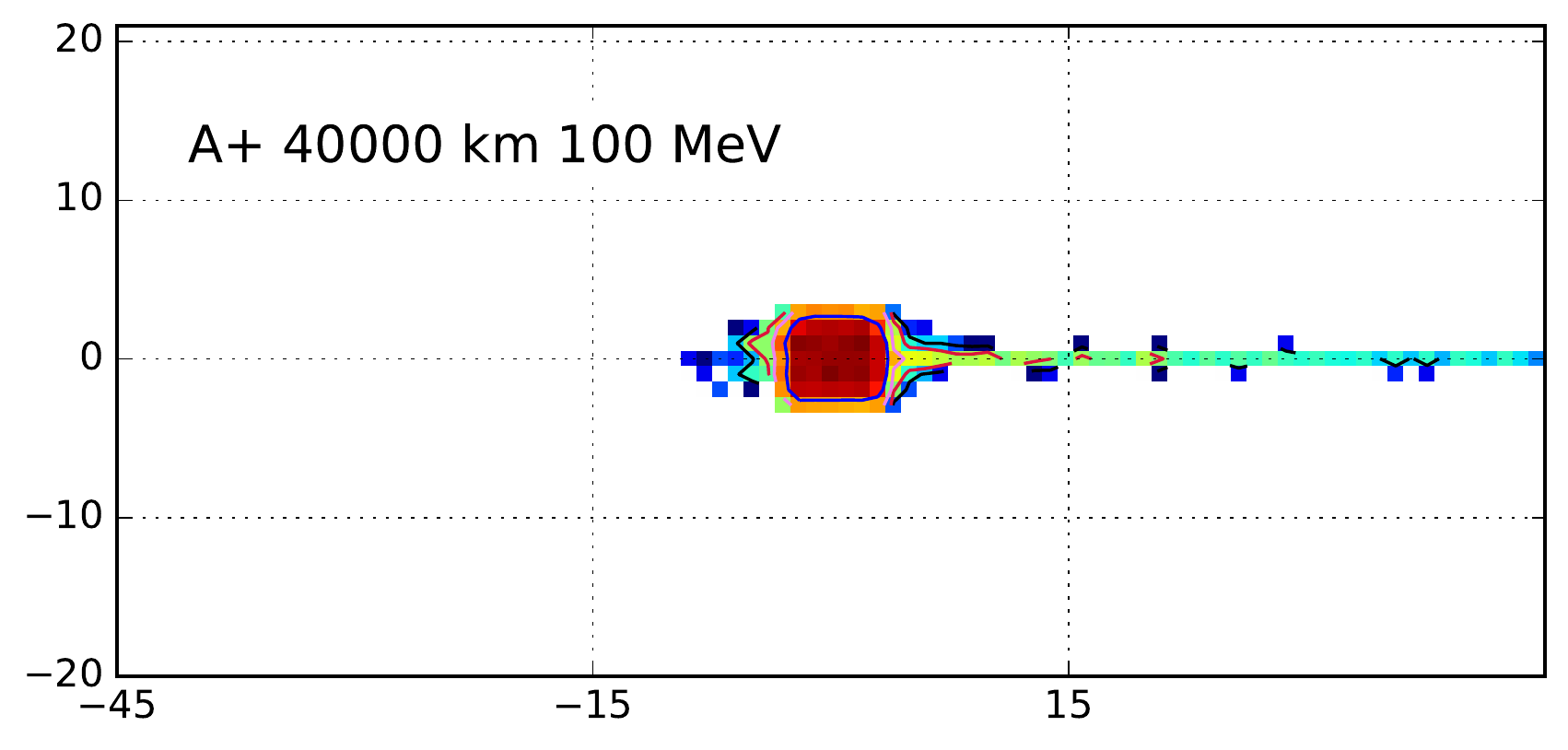}
\includegraphics[width=0.45\textwidth]{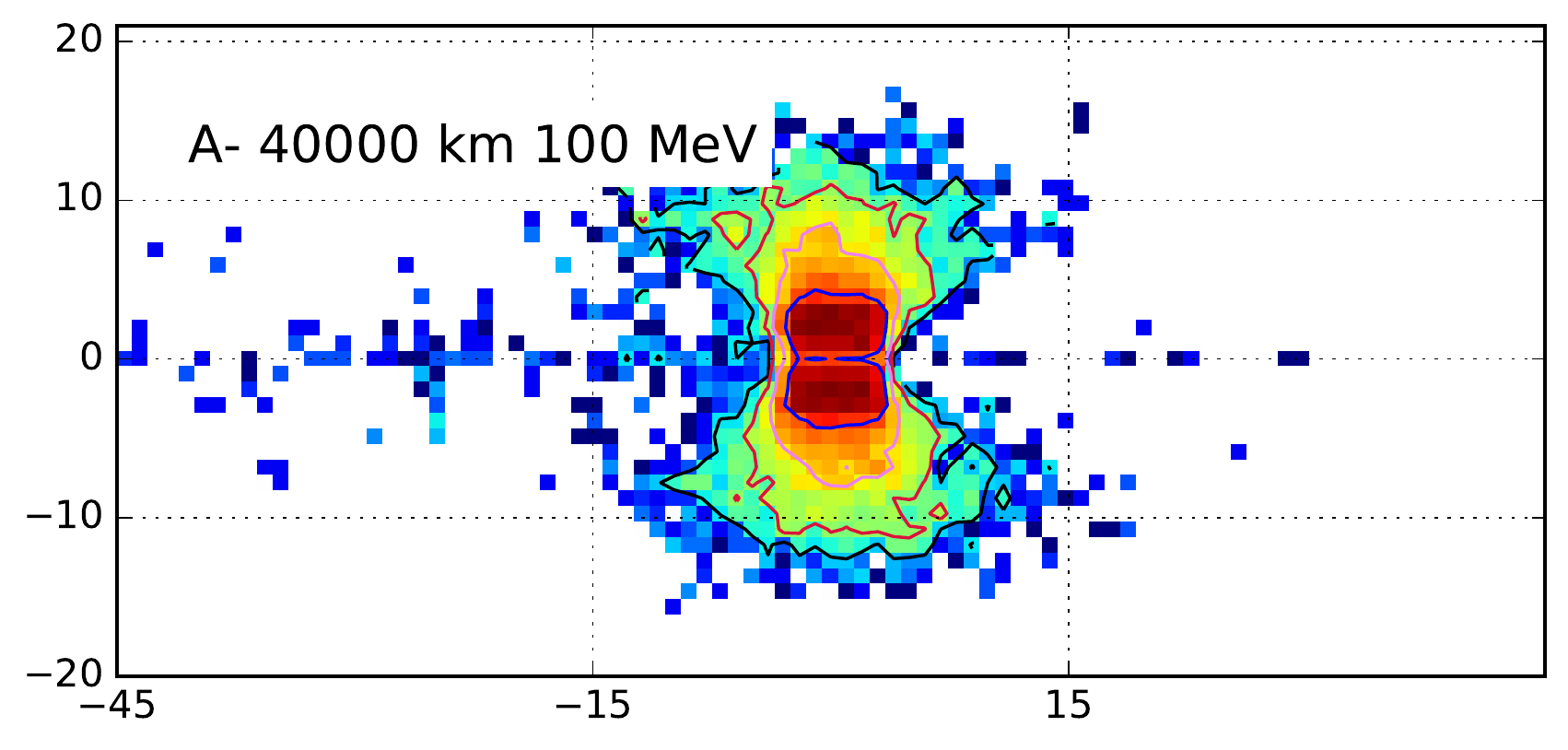}
\caption{Map of crossings of protons, injected at \mbox{100 MeV}, across the \mbox{1 au} sphere, over a time of \mbox{100 hr}, relative to best-connected fieldline at injection time, with the effects of corotation removed. Fluence colours and contours are on a logarithmic scale, with two contours per decade. Top row: Unipolar field, pointing outwards (left) and inwards (right). Second to fourth rows: HCS thickness scaled to \mbox{0 km}, \mbox{5000 km}, and \mbox{40 000 km} at \mbox{1 au}, respectively, with with an $A+$ (left column) or $A-$ (right colum) field configuration.}\label{fig:maps_grid}
\end{figure*}

In the presence of the heliospheric current sheet, the latitudinal drifts in each hemisphere play a significant role to how protons propagate (see, e.g., \citealt{Jokipii1977}). For protons injected at and near the HCS, as in our simulations, we find the dynamics presented to differ significantly from the unipolar case. In Figure \ref{fig:maps_grid}, we plot fluence maps of \mbox{1 au} crossings of protons, injected at \mbox{100 MeV}, for all eight simulated IMF configurations, in the same format as in the lower panel of Figure \ref{fig:map_correction}. The latitudinal drifts in the $A+$ configuration are found to efficiently trap protons close to the current sheet, where they experience current sheet drift. For the $A-$ configuration, curvature and gradient drifts push protons away from the HCS, but Speiser motion nevertheless allows some protons to propagate along the current sheet, until they are ejected and drift away from it. As our model does not include an intrinsic electric field at the sheet, ejection happens due to particle scattering.

\begin{figure*}[!htp]
\centering
\includegraphics[width=0.45\textwidth]{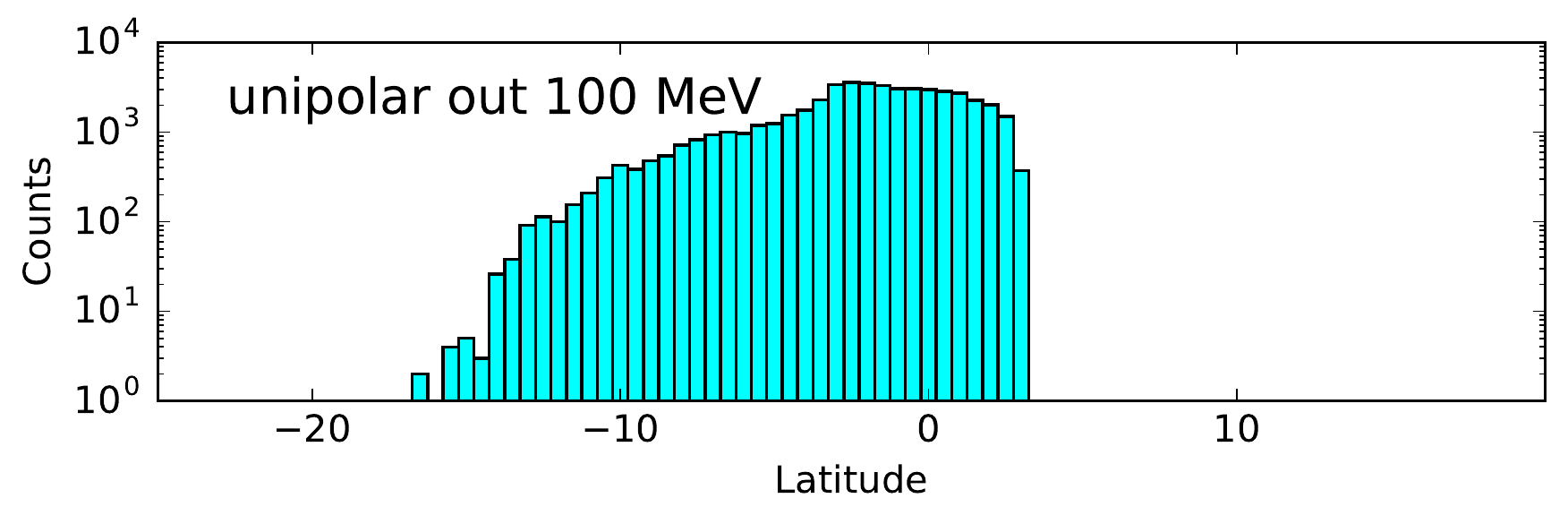}
\includegraphics[width=0.45\textwidth]{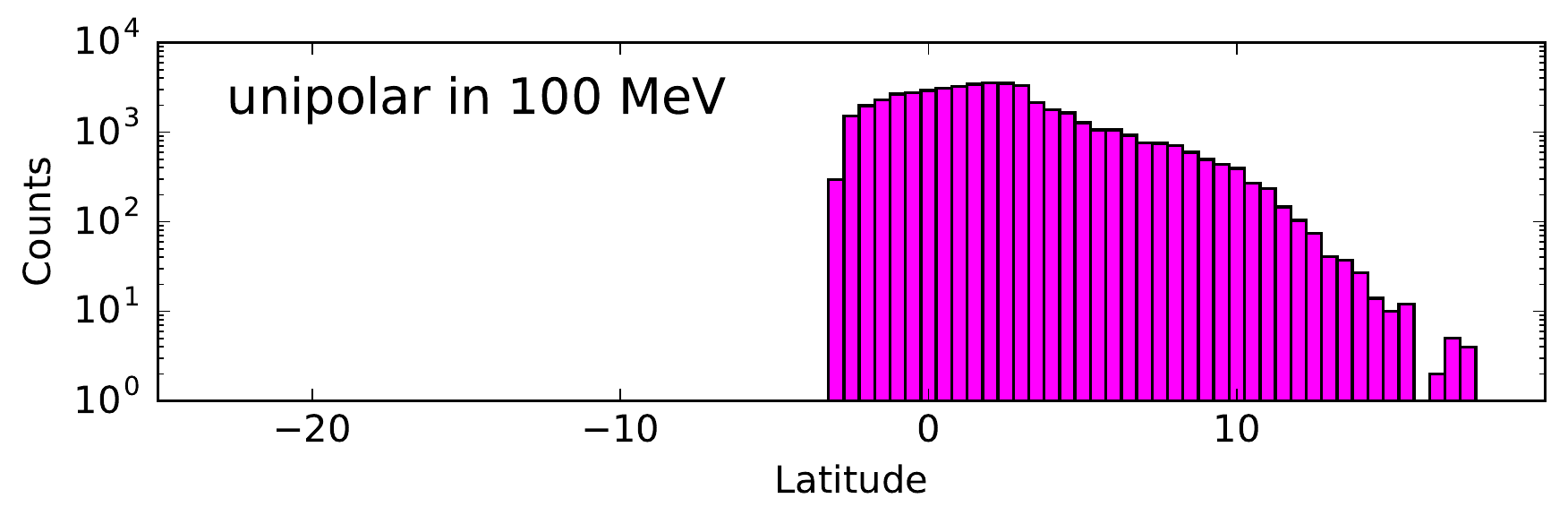}
\includegraphics[width=0.45\textwidth]{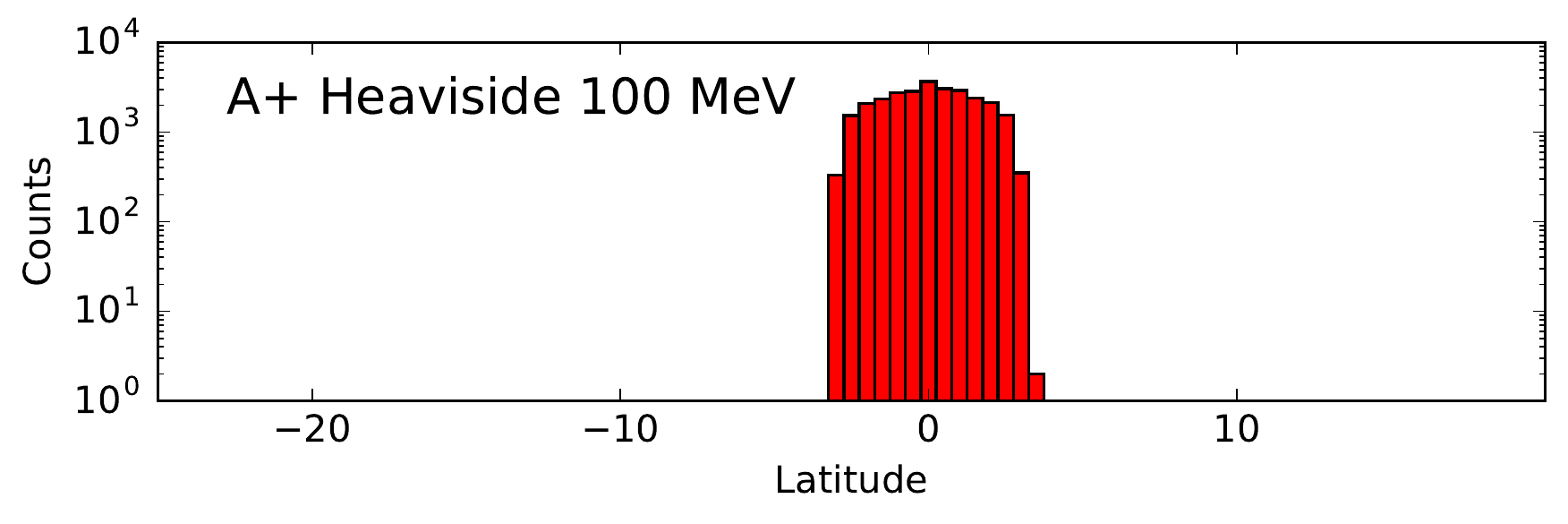}
\includegraphics[width=0.45\textwidth]{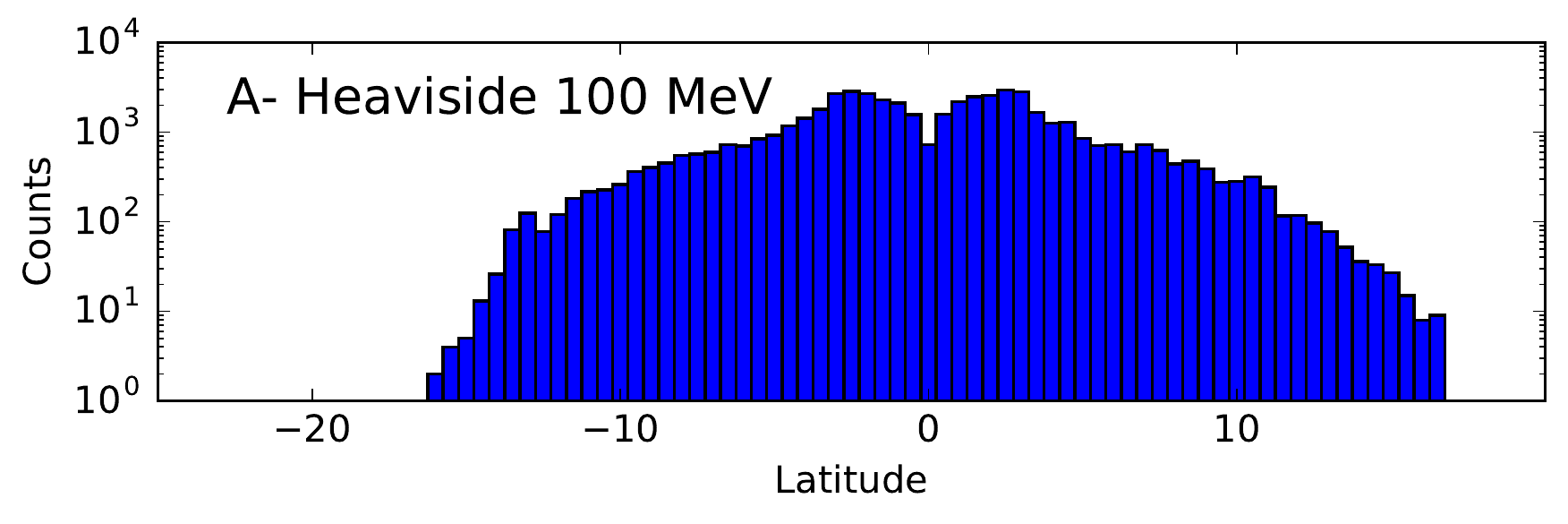}
\includegraphics[width=0.45\textwidth]{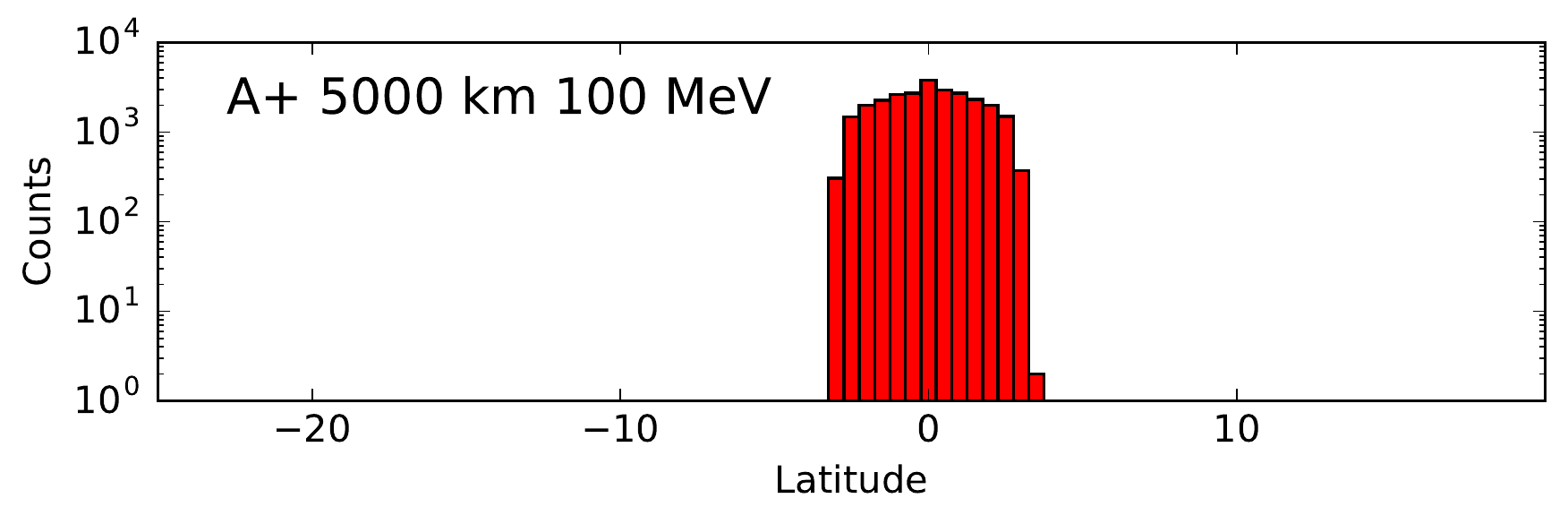}
\includegraphics[width=0.45\textwidth]{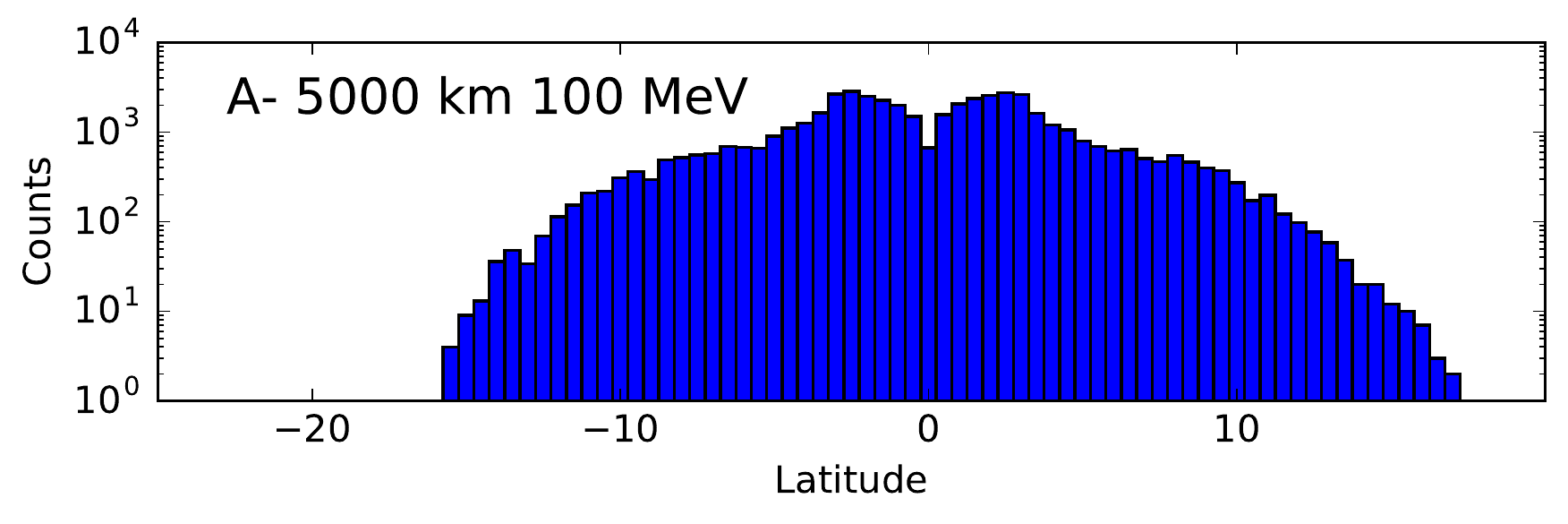}
\includegraphics[width=0.45\textwidth]{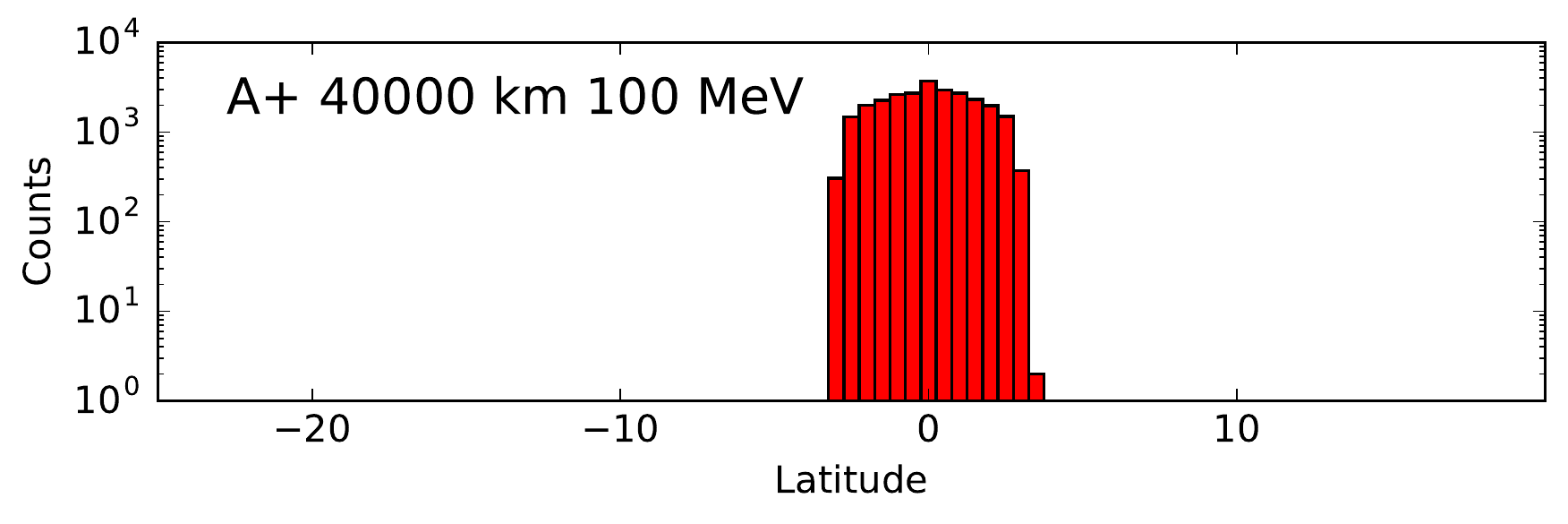}
\includegraphics[width=0.45\textwidth]{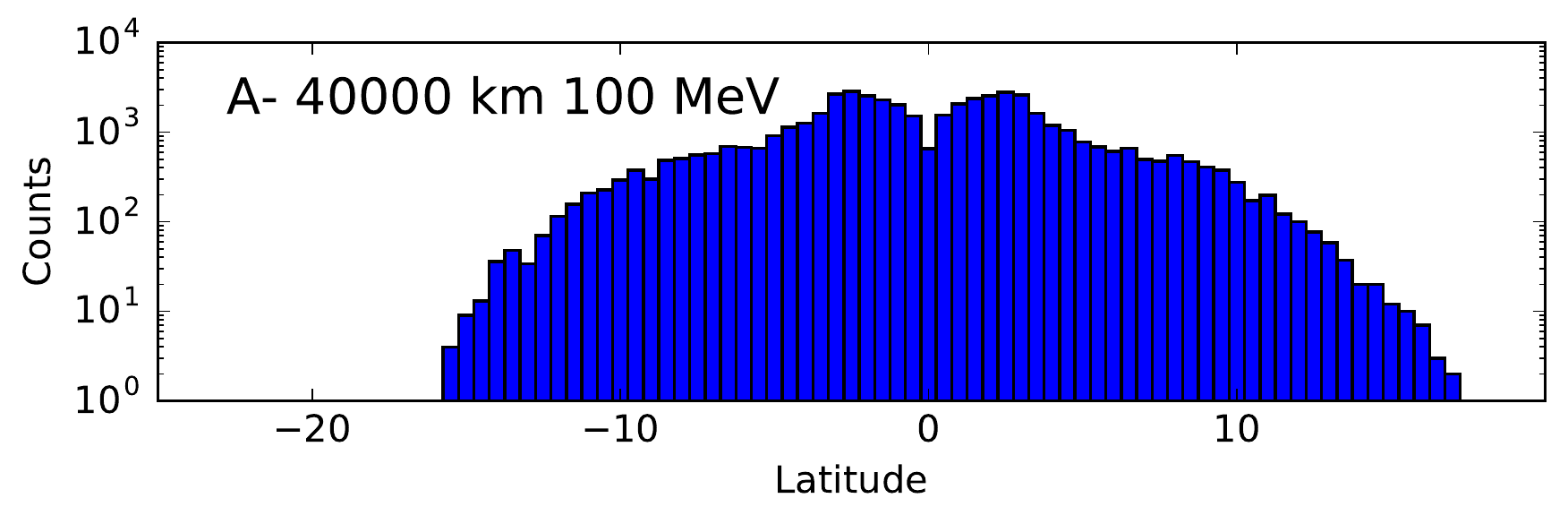}
\caption{Histograms depicting counts of protons, injected at \mbox{100 MeV}, crossing the \mbox{1 au} sphere, as a function of latitude, relative to injection at the equator. Layout as in Figure \ref{fig:maps_grid}.}\label{fig:histograms_grid_lat}
\end{figure*}

Protons experiencing current sheet drift are visible at western heliolongitudes for the $A+$ configuration (left column) and at eastern heliolongitudes for the $A-$ configuration (right column). 
Of particular note for the $A-$ configuration is how, if looking closely at the cells closest to the current sheet at the best-connected field line, intensities are smaller than just above or below it. In Figure \ref{fig:histograms_grid_lat}, we plot histograms of \mbox{100 MeV} proton latitudes at the time they cross the \mbox{1 au} sphere boundary, for eight different IMF configurations. For an $A+$ configuration, protons are preferentially located at the centre of the current sheet, whereas for the $A-$ configuration, two peaks further out are seen. Thus, the depletion at the sheet is shown to be real, not caused by current sheet drift spreading a constant amount of protons over a wider range of longitudes. 

\begin{figure*}[!htp]
\centering
\includegraphics[width=0.45\textwidth]{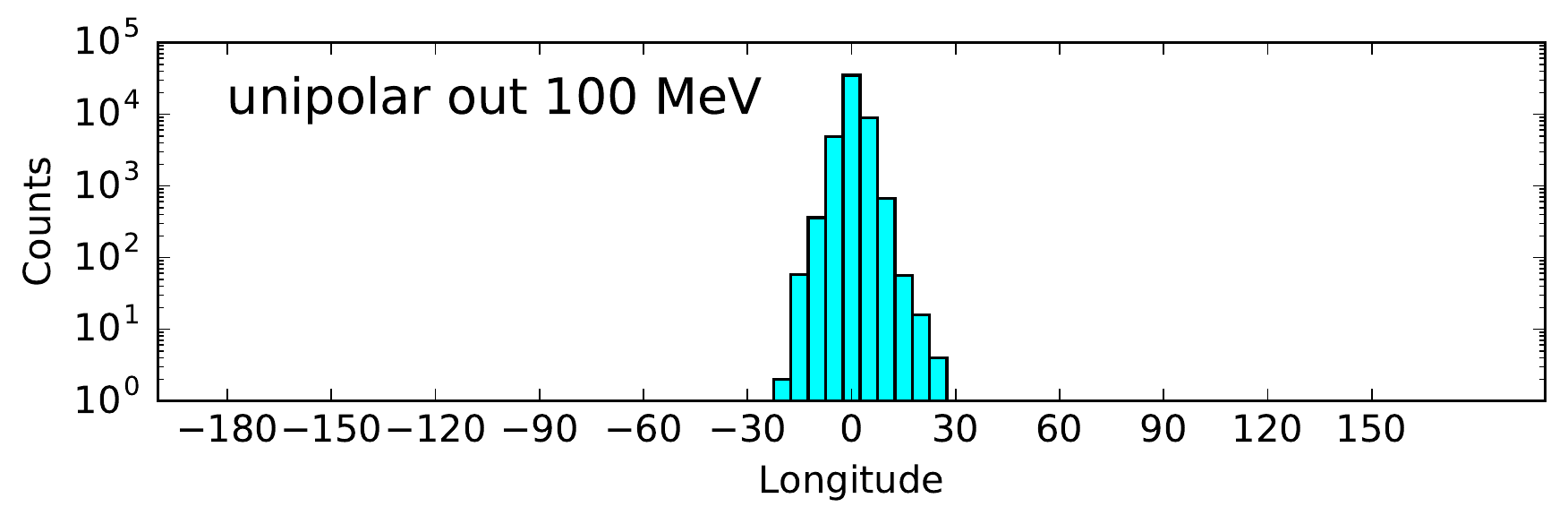}
\includegraphics[width=0.45\textwidth]{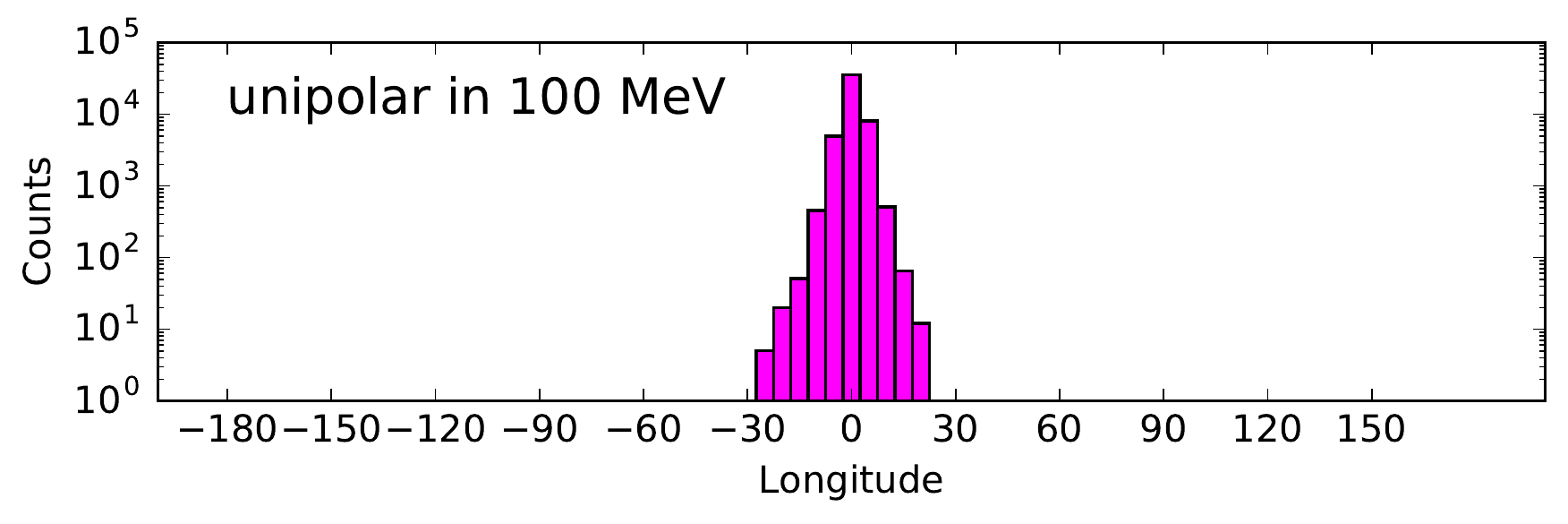}
\includegraphics[width=0.45\textwidth]{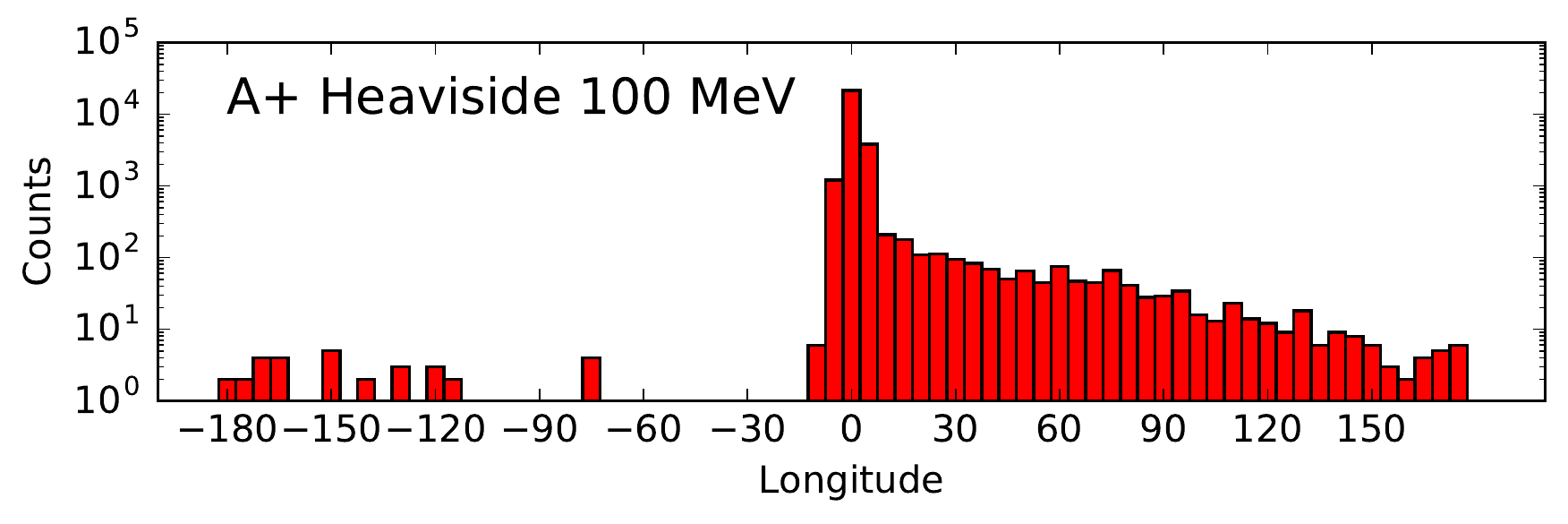}
\includegraphics[width=0.45\textwidth]{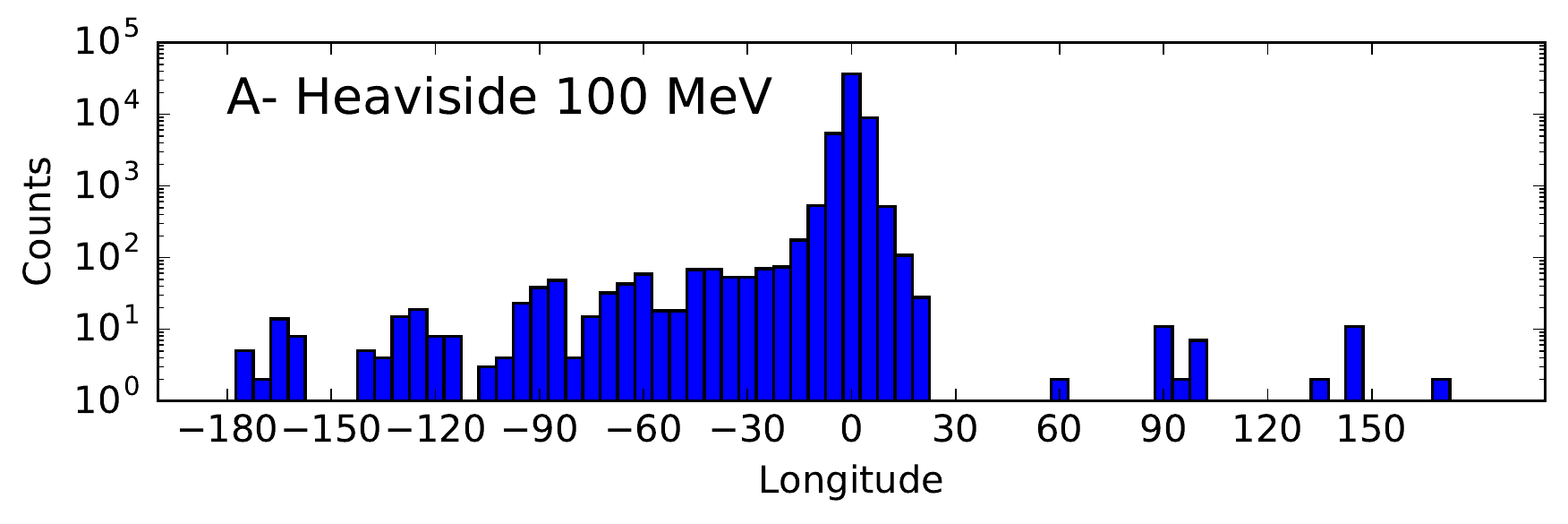}
\includegraphics[width=0.45\textwidth]{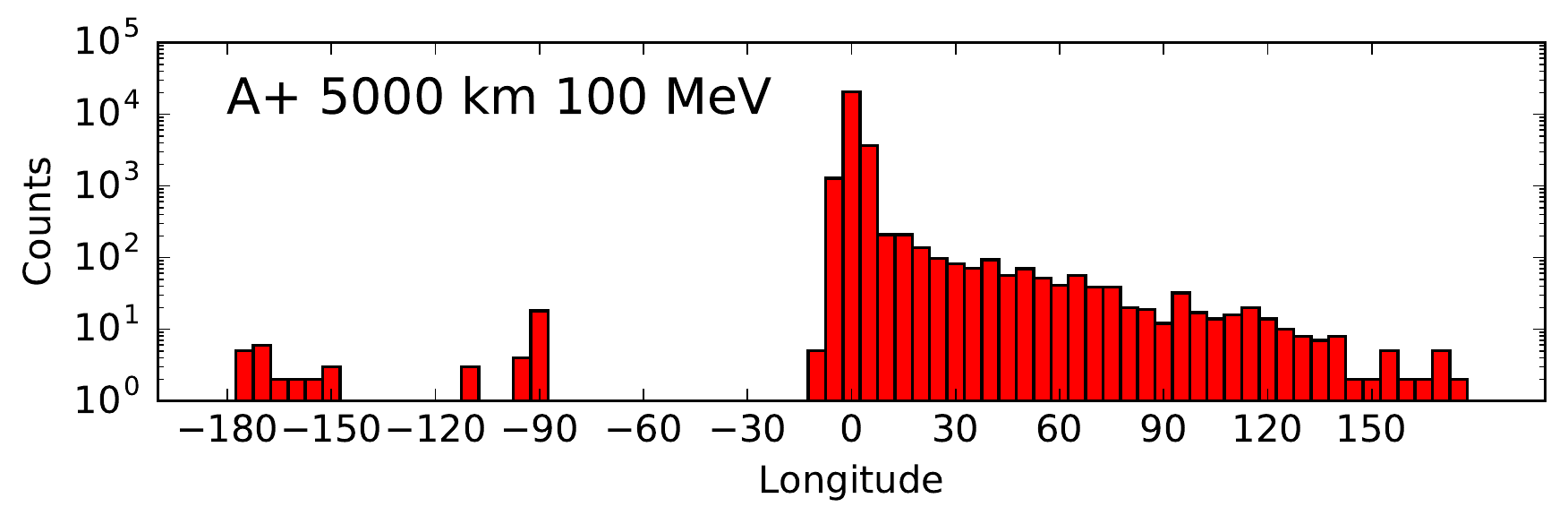}
\includegraphics[width=0.45\textwidth]{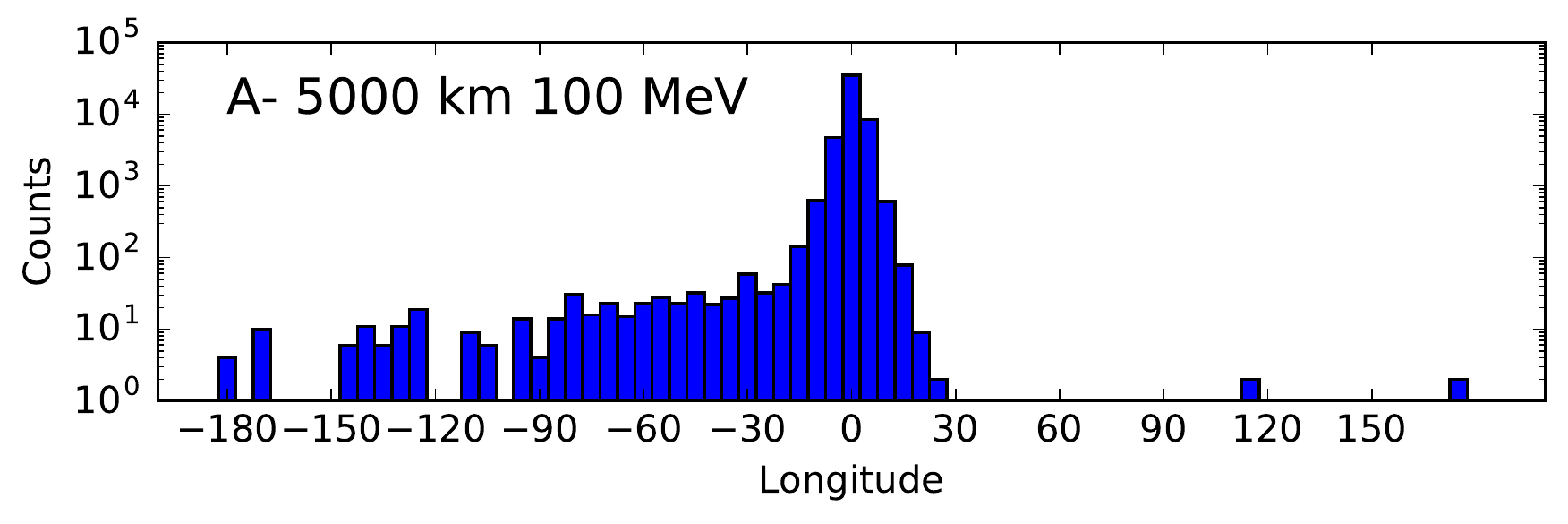}
\includegraphics[width=0.45\textwidth]{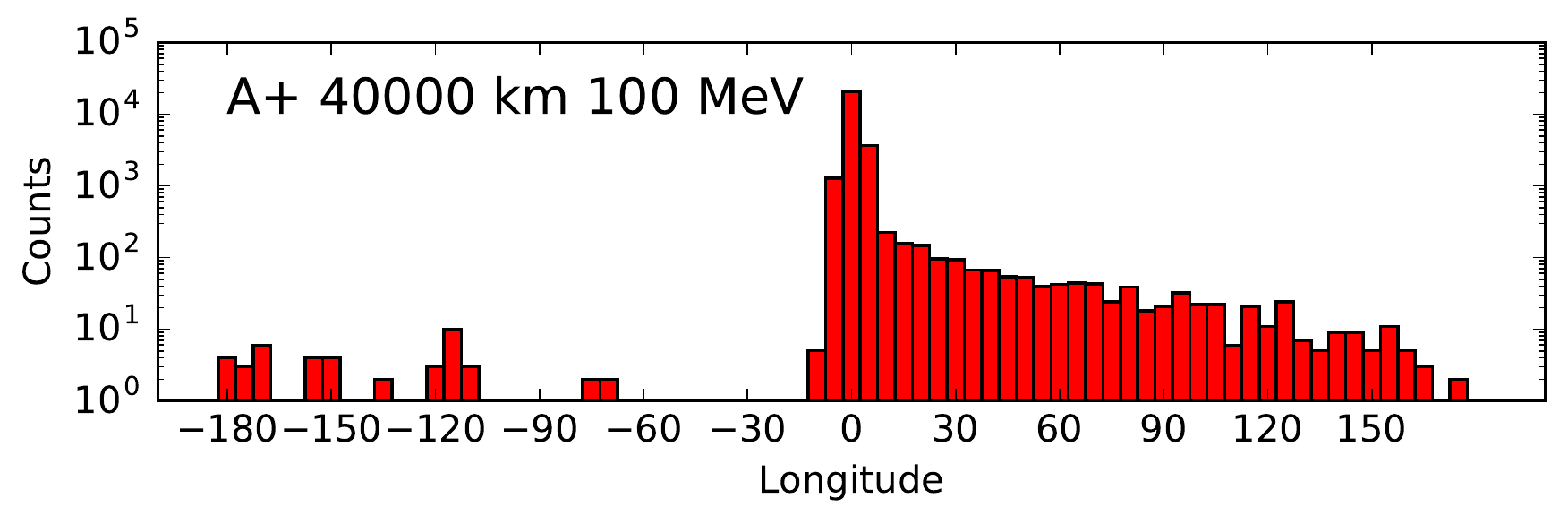}
\includegraphics[width=0.45\textwidth]{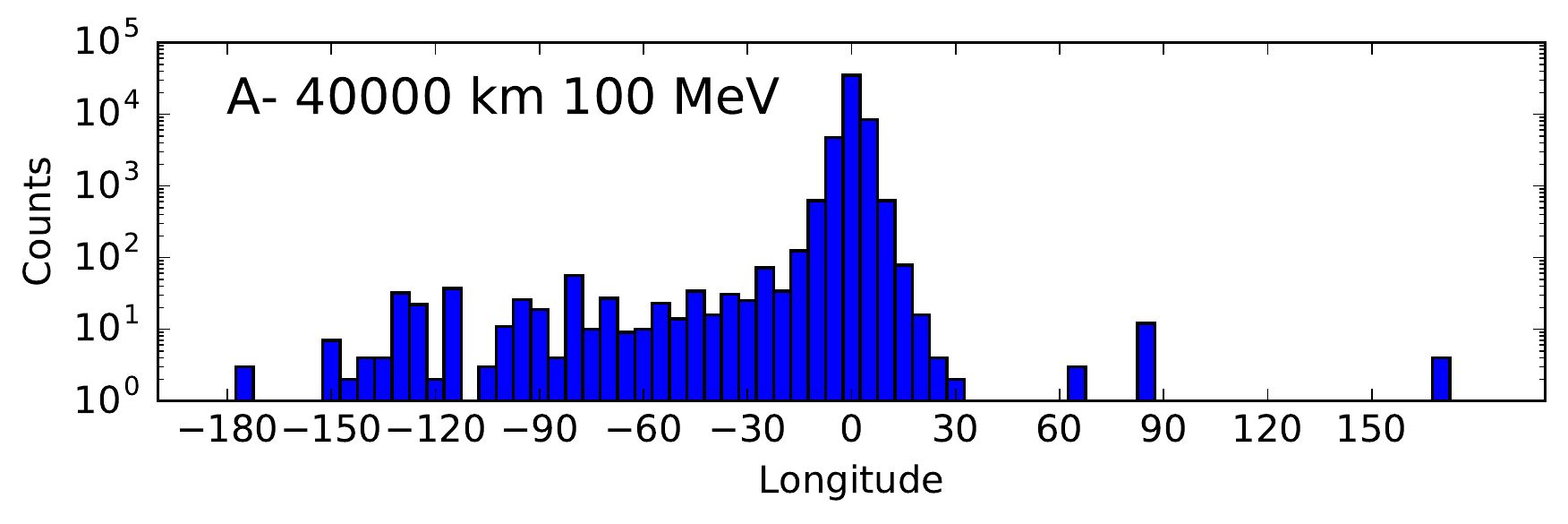}
\caption{Histograms depicting counts of protons, injected at \mbox{100 MeV}, crossing the \mbox{1 au} sphere, as a function of longitude, relative to the best-connected field line. The effect of corotation has been removed. Layout as in Figure \ref{fig:maps_grid}.}\label{fig:histograms_grid_long}
\end{figure*}

In Figure \ref{fig:histograms_grid_long}, we plot histograms of \mbox{100 MeV} proton longitudes at the time they cross the \mbox{1 au} sphere boundary, for eight different IMF configurations. The current sheet drift is seen to have a significant effect, allowing protons to wrap at least 180 degrees around the Sun. An $A+$ configuration is seen to have slightly stronger current sheet drift, which is in agreement with the equator being a stable position in $A+$, and a labile position in $A-$.

\begin{figure*}[!htp]
\centering
\includegraphics[width=0.45\textwidth]{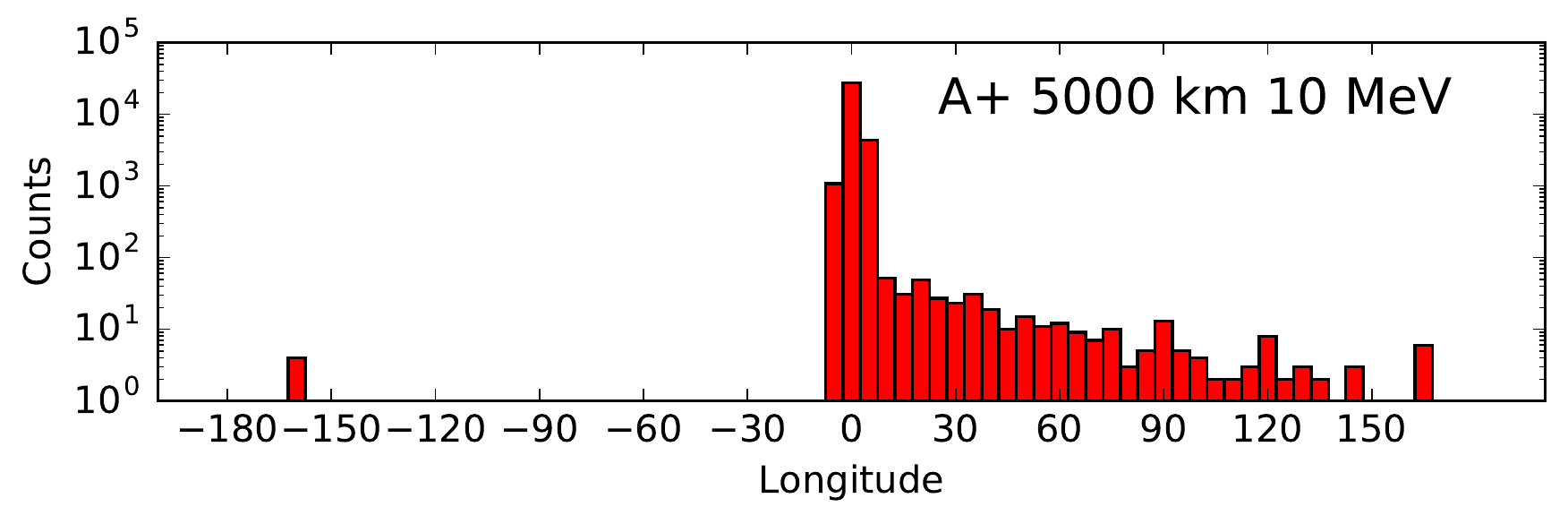}
\includegraphics[width=0.45\textwidth]{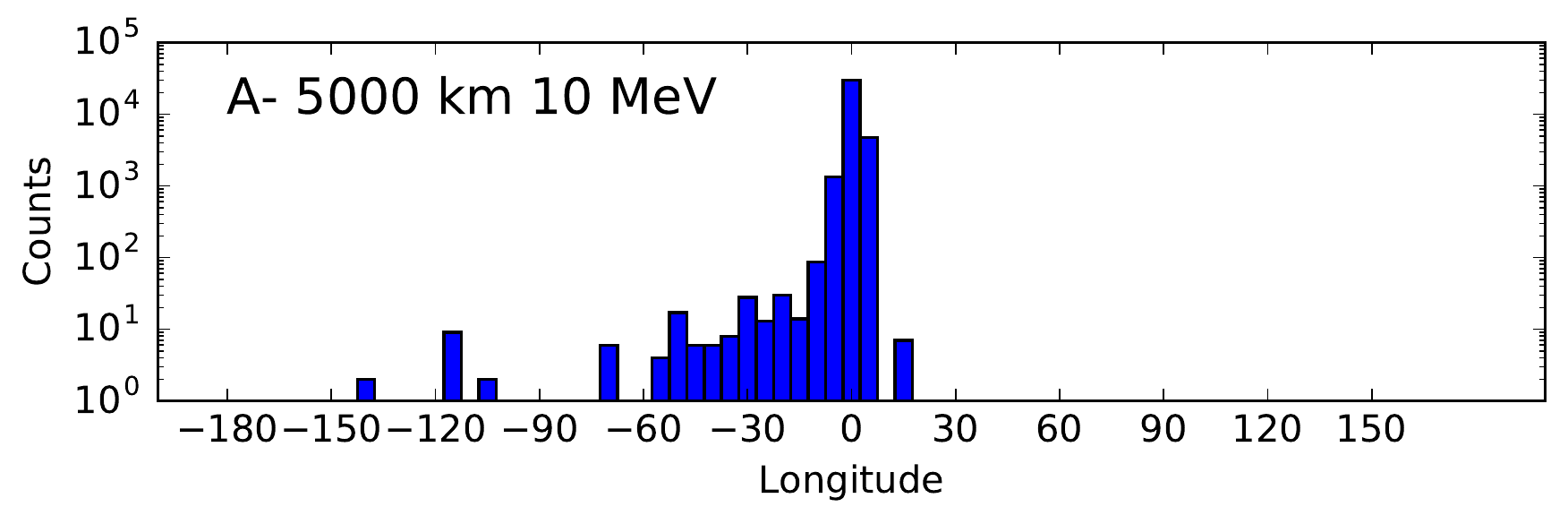}
\includegraphics[width=0.45\textwidth]{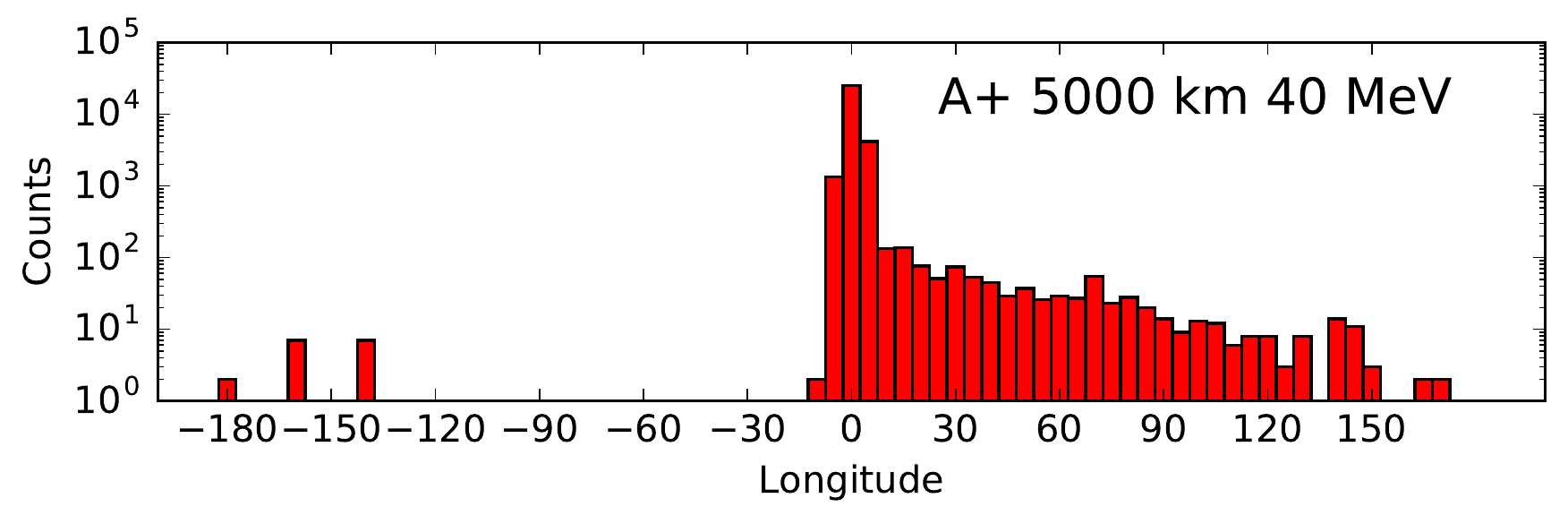}
\includegraphics[width=0.45\textwidth]{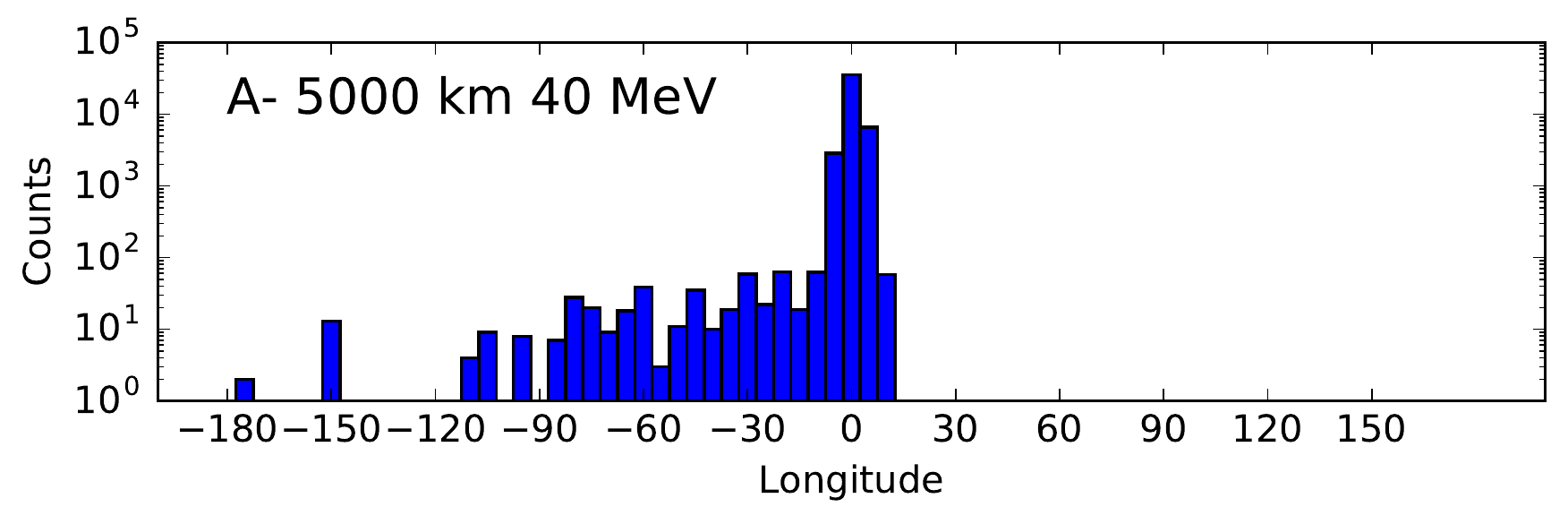}
\includegraphics[width=0.45\textwidth]{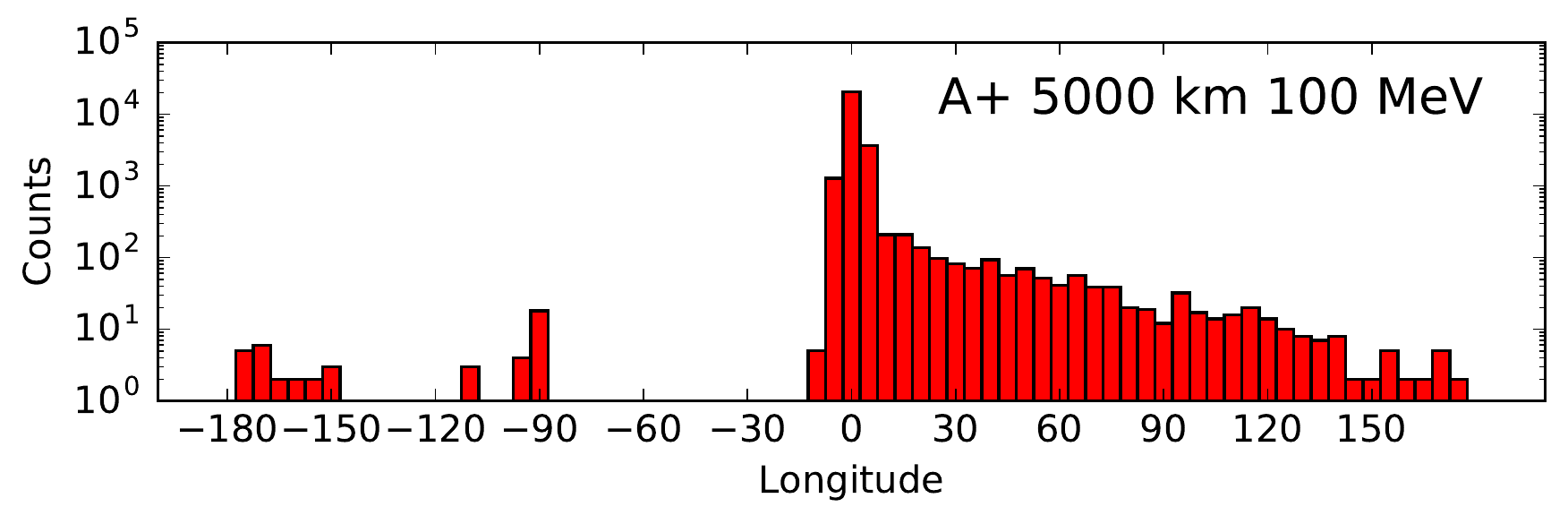}
\includegraphics[width=0.45\textwidth]{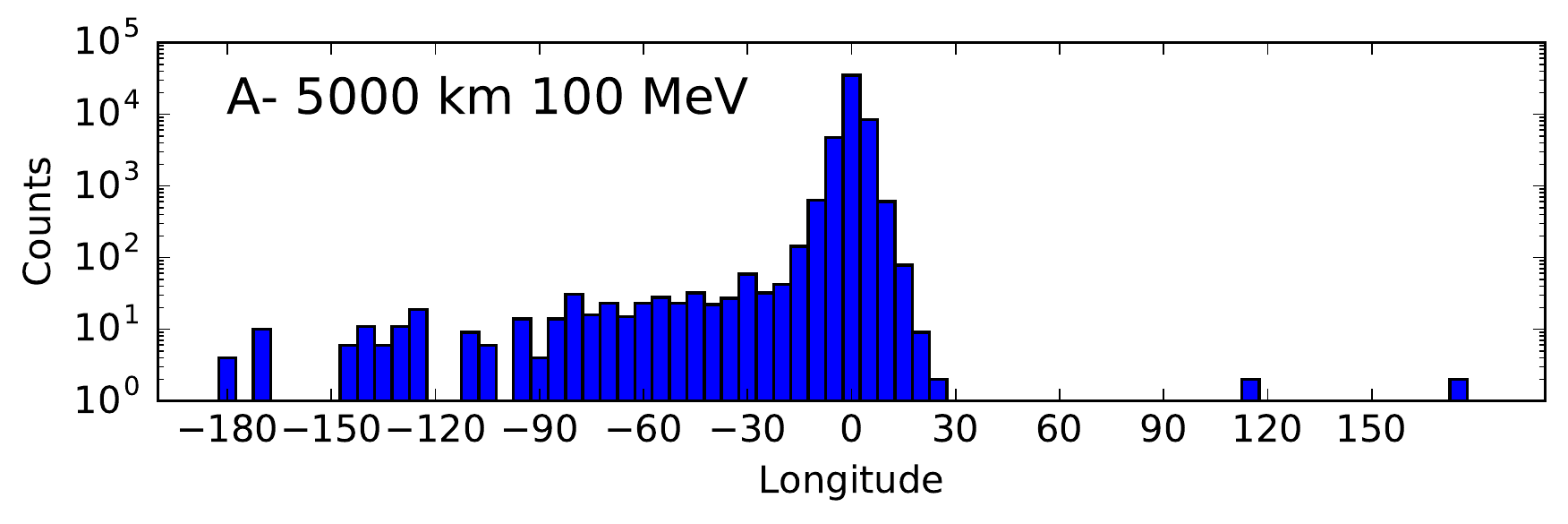}
\includegraphics[width=0.45\textwidth]{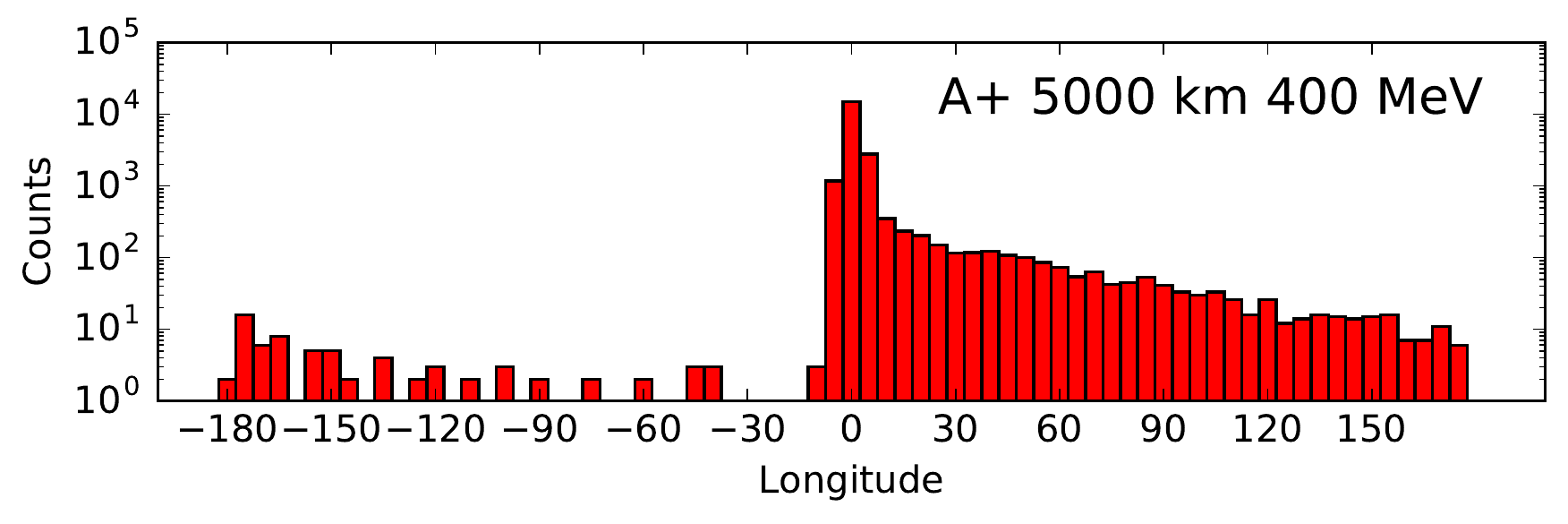}
\includegraphics[width=0.45\textwidth]{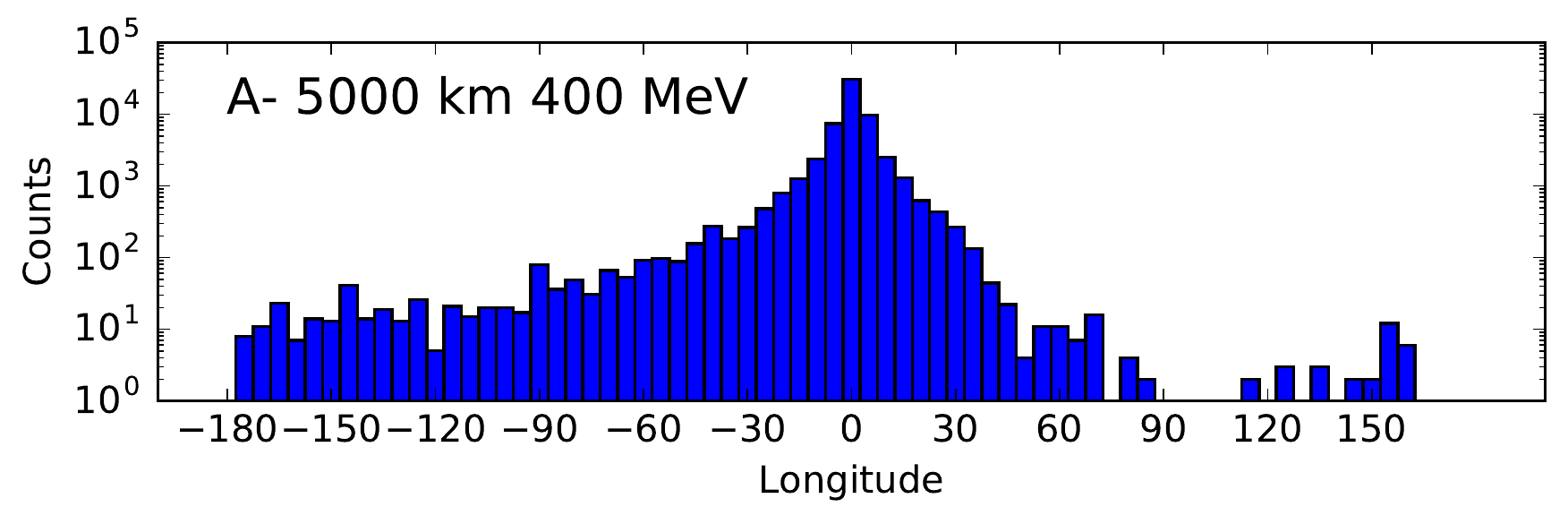}
\caption{Histograms depicting counts of protons, crossing the \mbox{1 au} sphere, as a function of longitude, relative to the best-connected field line. The effect of corotation has been removed. The left column depicts result for an IMF with an $A+$ configuration, the right column for $A-$. HCS thickness was scaled to \mbox{5000 km} at \mbox{1 au}. From top to bottom row, histograms are shown for injection energies of \mbox{10 MeV}, \mbox{40 MeV}, \mbox{100 MeV}, and \mbox{400 MeV}. At high energies, the current sheet drift of protons is seen to wrap around past 180 degrees.}\label{fig:histograms_long_energies}
\end{figure*}

In Figure \ref{fig:histograms_long_energies}, we display histograms of proton longitudes at the time they cross the \mbox{1 au} sphere boundary, for energies of \mbox{10 MeV}, \mbox{40 MeV}, \mbox{100 MeV}, and \mbox{400 MeV}. The left column shows results for an IMF with an $A+$ configuration, the right column for one with an $A-$ configuration, with HCS thickness set to \mbox{5000 km} at \mbox{1 au}. Both the maximum amount drifted and the count of protons at each drifting distance are found to increase with energy. This is as expected, as faster protons are able to sweep across the current sheet from a wider region due to a larger gyroradius, and also due to average Speiser motion being linked with energy \citep{Burger1985}. The $A-$ configuration displays a much stronger energy dependence for current sheet drift, as the reach of particle gyromotion plays a critical role in sampling of the magnetic field reversal, due to lateral drifts transporting protons away from the HCS. The main peak for $A-$, however, does spread out, as longitudinal drifts outside the current sheet can also cause protons to spread westward.

\begin{figure*}[!htp]
\centering
\includegraphics[width=\textwidth]{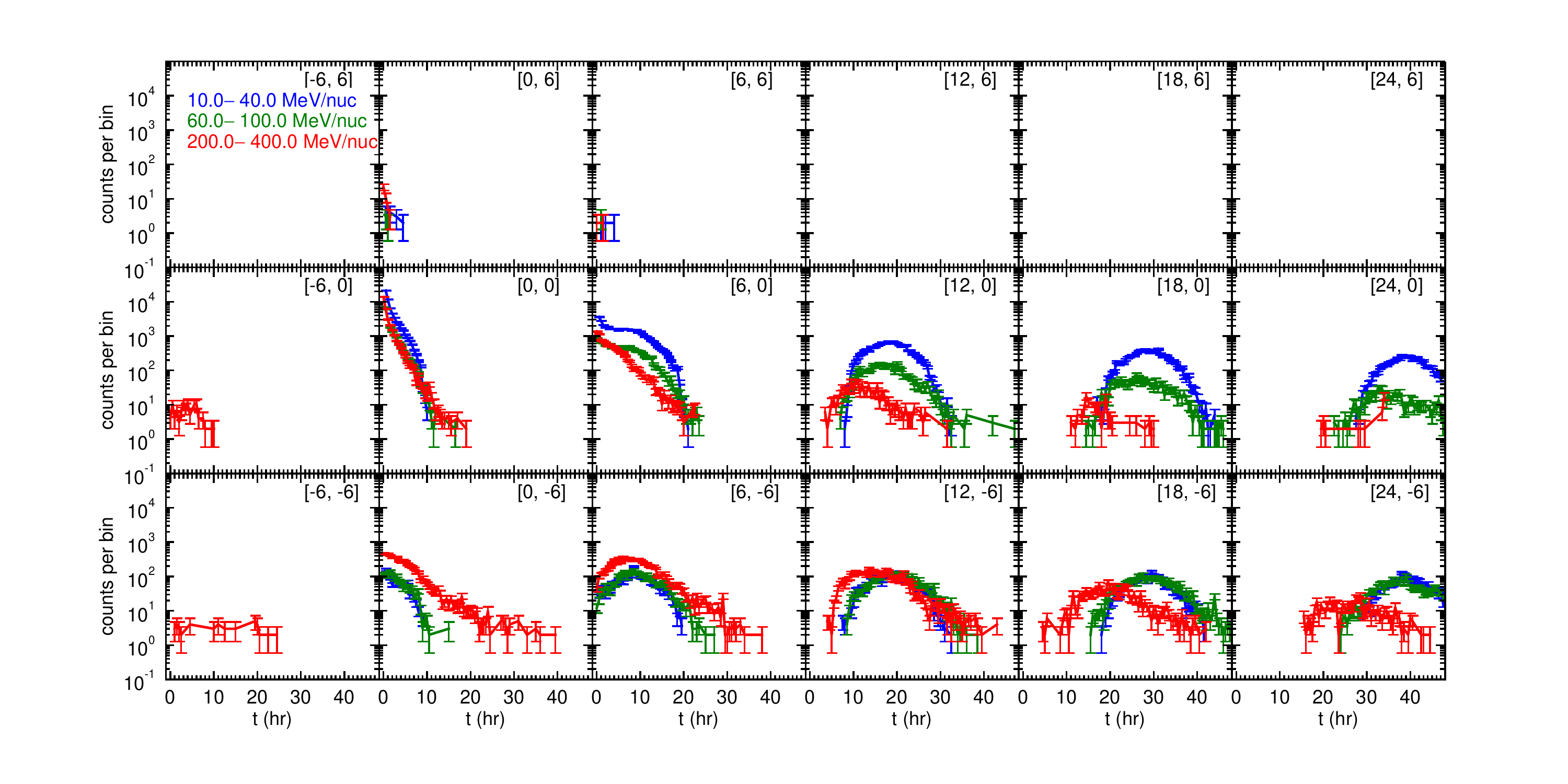}
\caption{Time profiles of protons, where the IMF is an unipolar outwards-pointing field. Each panel shows $4\pi$ steradian and $6^\circ \times 6^\circ$ angular extent virtual observers at \mbox{1 au}, with the captions indicating the \mbox{[lon,lat]} offset in degrees from the position of the best-connected fieldline. Time profiles were generated over an extent of 48 hours, with 30 minute time binning. Injection was at the heliographic equator, with a power law of $\gamma=-1.1$ and an injection energy range spanning \mbox{$10-400$ MeV}. Curvature and gradient drifts cause virtual observers at lower latitudes to see some counts.}\label{fig:timeprofile_unipolar}
\end{figure*}

\begin{figure*}[!htp]
\centering
\includegraphics[width=\textwidth]{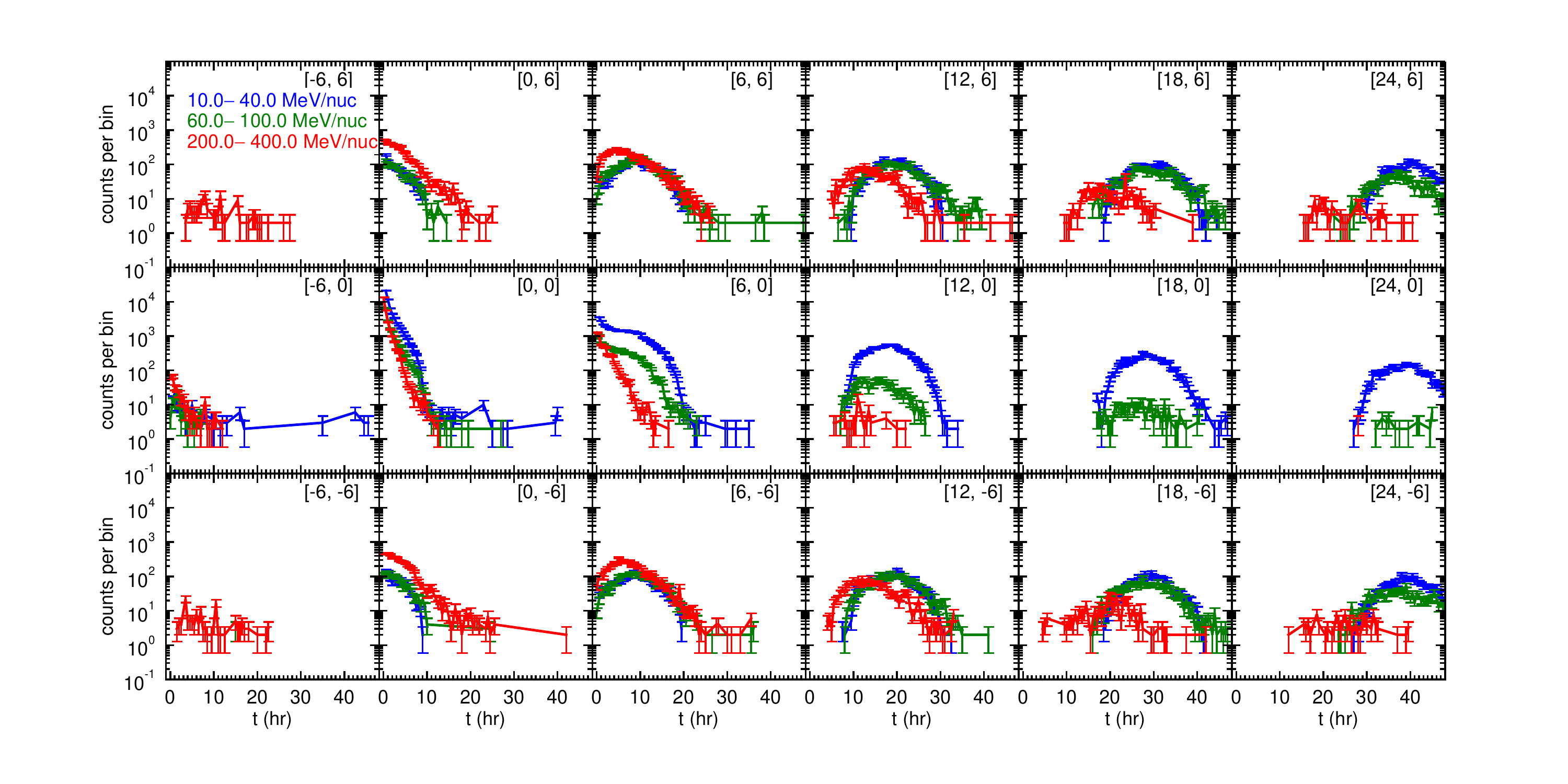}
\caption{Time profiles of protons, where the IMF has an $A-$ configuration, with HCS thickness scaled to \mbox{5 000 km} at \mbox{1 au}. Other properties as in Figure \ref{fig:timeprofile_unipolar}. Curvature and gradient drifts cause virtual observers at both lower and higher latitudes to also see some flux.}\label{fig:timeprofile_a-}
\end{figure*}

\begin{figure*}[!htp]
\centering
\includegraphics[width=\textwidth]{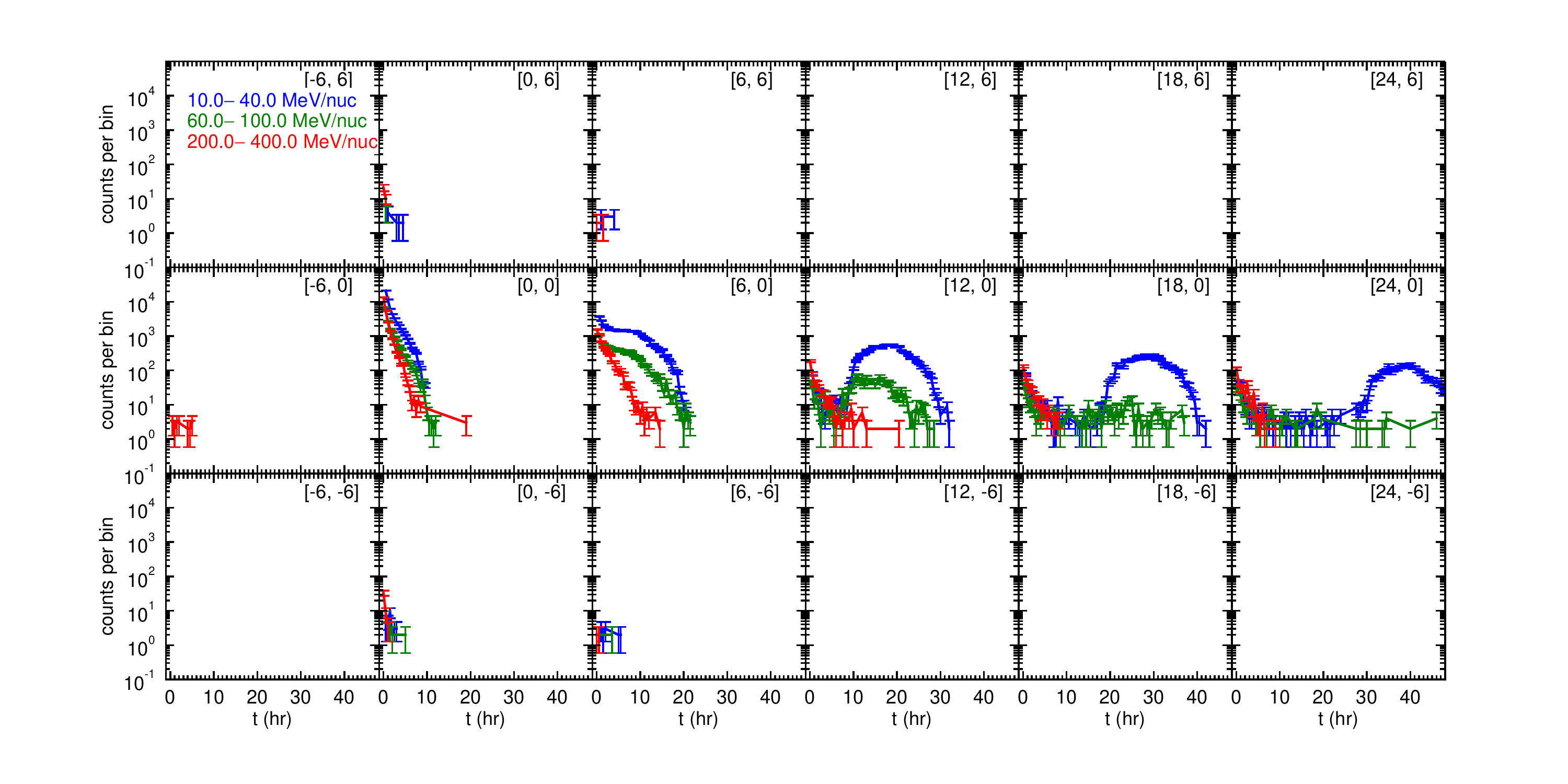}
\caption{Time profiles of protons, where the IMF has an $A+$ configuration, with HCS thickness scaled to \mbox{5 000 km} at \mbox{1 au}. Other properties as in Figure \ref{fig:timeprofile_unipolar}. Curvature and gradient drifts constrain proton flux to the best-connected latitude. Of particular note is the two-component time profile seen at western locations, with an impulsive peak close to injection time in addition to the delayed gradual increase seen with other IMF configurations.}\label{fig:timeprofile_a+}
\end{figure*}

The maps and histograms presented in Figures \ref{fig:map_correction} through \ref{fig:histograms_long_energies} do not explicitly display the time profiles of proton crossings at \mbox{1 au}. In order to allow comparisons with real-world observations, we simulated virtual observers at \mbox{1 au}, by collecting proton counts over neighbouring regions of \mbox{$6^\circ \times 6^\circ$} extent on the surface of the \mbox{1 au} sphere. For this analysis, we performed simulations using a proton injection distribution given by a power-law with $\gamma=-1.1$, extending from \mbox{10 MeV} to \mbox{400 MeV}. We inject $N=10^6$ particles in order to attain better statistics. For gathering of time profiles, we introduced energy channels spanning the extents of \mbox{$10.0-40.0$ MeV}, \mbox{$60.0-100.0$ MeV}, and \mbox{$200.0-400.0$ MeV}. In Figures \ref{fig:timeprofile_unipolar}, \ref{fig:timeprofile_a-}, and \ref{fig:timeprofile_a+}, we display time profiles for an outwards-pointing unipolar field, an $A-$ IMF configuration, and an $A+$ IMF configuration, respectively.


The unipolar field depicted in Figure \ref{fig:timeprofile_unipolar} shows how a single injection event can cause different kinds of observations, depending on virtual observer location, similar to the findings of \cite{Marsh2015}. At the best-connected field line, the proton flux increases abruptly and then decays exponentially. At eastern longitudes, flux is nearly non-existant. With increasing longitudinal separation to the west, the onset is delayed and the shape of the profile becomes more gradual. As flux at western longitudes is influenced by corotation, high energy protons are less numerous, due to propagating rapidly out of the inner heliosphere. At negative heliolatitudes, where in our setup all counts are due to drifting effects, we also find an abrupt rise in flux at connected longitudes and slower rises at western longitudes. However, due to latitudinal drifts being energy-dependent, these time profiles emphasise high energy protons. Thus, if the proton flux of a solar event at an observer is strongly influenced by latitudinal drifts, the observed spectrum can appear much harder than that of the source population. We also note that separation between the observer and the well-connected field line increases the onset time difference between different energies.

With the $A-$ IMF configuration, shown in Figure \ref{fig:timeprofile_a-}, we see a case very similar to the unipolar one, with rapid or prolonged rise phases of intensity, depending on longitude. For this case, however, intensities extend to both positive and negative heliolatitudes. Again, the process of latitudinal proton drifts causes apparent hardening of observed proton spectra north and south of the injection region. We also note that a relatively small abrupt component is seen at the equator, at eastern longitudes, due to protons experiencing current sheet drift. Due to the combined effect of current sheet drift and latitudinal drifts, high energy protons are much less abundant at western longitudes than for the unipolar case.

With the $A+$ IMF configuration, shown in Figure \ref{fig:timeprofile_a+}, we find that the gradient and curvature drifts prevent any significant proton flux from extending to positive or negative heliolatitudes. Protons at lower energies display the same longitudinal characteristic of more prolonged event rise with increasing longitude. Of particular interest is the abrupt current sheet drift associated component at early phases of the simulation, as can be seen by comparing Figure \ref{fig:timeprofile_a+} with Figure \ref{fig:timeprofile_unipolar}. This additional component is found at western observers, causing the time profiles to exhibit two distinct components. Thus, a single injection event could, with a suitable IMF configuration, be observed as two particle events. 

Comparisons of different HCS thickness parameters did not result in noticeable variation in the characteristics of proton time profiles. The additional plots have thus been omitted.

\begin{figure*}[!htp]
\centering
\includegraphics[width=\textwidth]{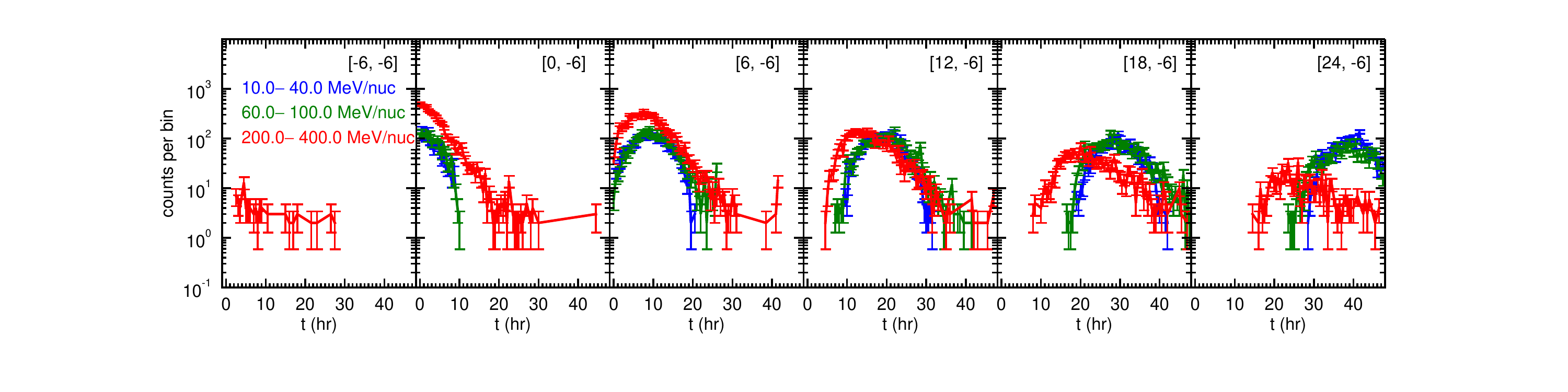}\\
\includegraphics[width=\textwidth]{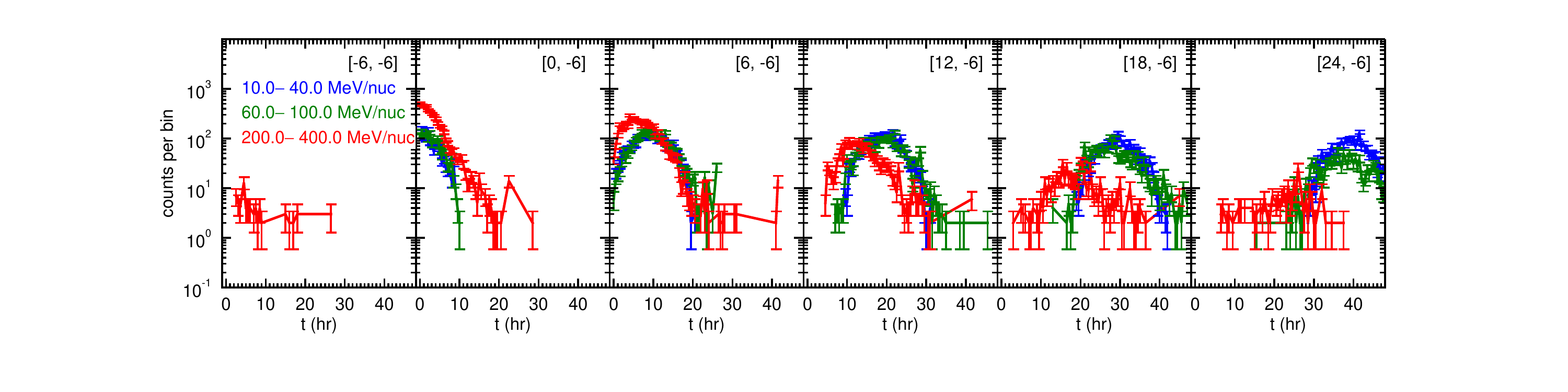}
\caption{\added{Time profiles of protons, where the IMF is an unipolar outwards-pointing field (top row) or where the IMF has an $A+$ configuration with HCS thickness scaled to \mbox{5 000 km} at \mbox{1 au} (bottom row). Each panel shows $4\pi$ steradian and $6^\circ \times 6^\circ$ angular extent virtual observers at \mbox{1 au}, with the captions indicating the \mbox{[lon,lat]} offset in degrees from the position of the best-connected fieldline. Injection was at a heliographic latitude of $+6^\circ$, and thus, all observers are centered on the solar equatorial plane. Injection was with a power law of $\gamma=-1.1$ and an injection energy range spanning \mbox{$10-400$ MeV}. Time profiles were generated over an extent of 48 hours, with 30 minute time binning.}}\label{fig:timeprofile_lat6}
\end{figure*}

\added{Solar active regions are usually associated with sunspots above or below the solar equator. The results presented in Figures \ref{fig:timeprofile_unipolar} to \ref{fig:timeprofile_a+} are applicable if the acceleration region, for example a coronal shock front, spans all the way to the equator. If the injection location is above the HCS, it will take time for particles to reach it and feel its effects. In Figure \ref{fig:timeprofile_lat6}, we display virtual time profiles for an observer at the heliographic equator, when the injection region of $6^\circ \times 6^\circ$ was centered at a latitude of $+6$ degrees. 
The top row shows profiles for an unipolar outwards-pointing IMF, and the bottom row for an IMF with an $A+$ HCS configuration. For high energy protons, the HCS facilitates arrival at the observer earlier than for an unipolar field. However, as the HCS spreads protons across a wide range of longitudes, the achieved peak flux is lower than without the HCS. 
At low energies the difference between the two cases is insignificant, possibly due to protons drifting close to the equator but not quite reaching the current sheet.}

\begin{figure}[htp]
\centering
\includegraphics[width=0.45\textwidth]{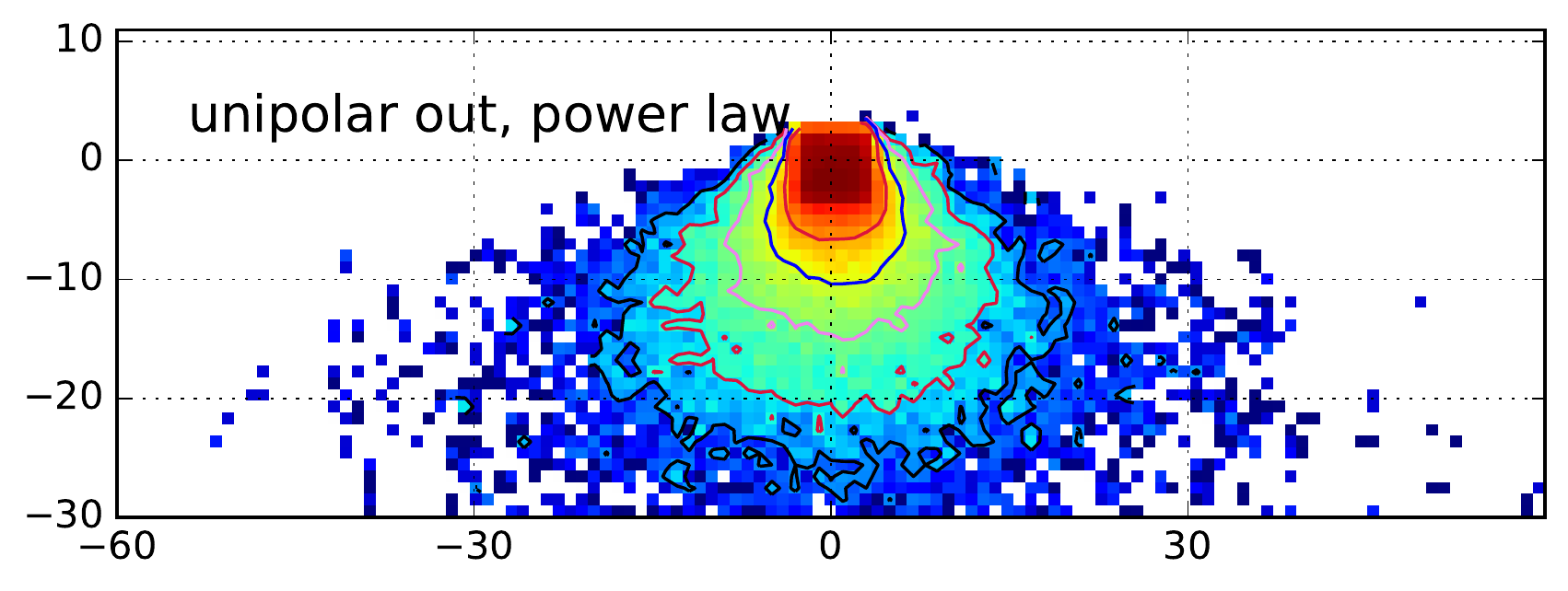}\\
\includegraphics[width=0.45\textwidth]{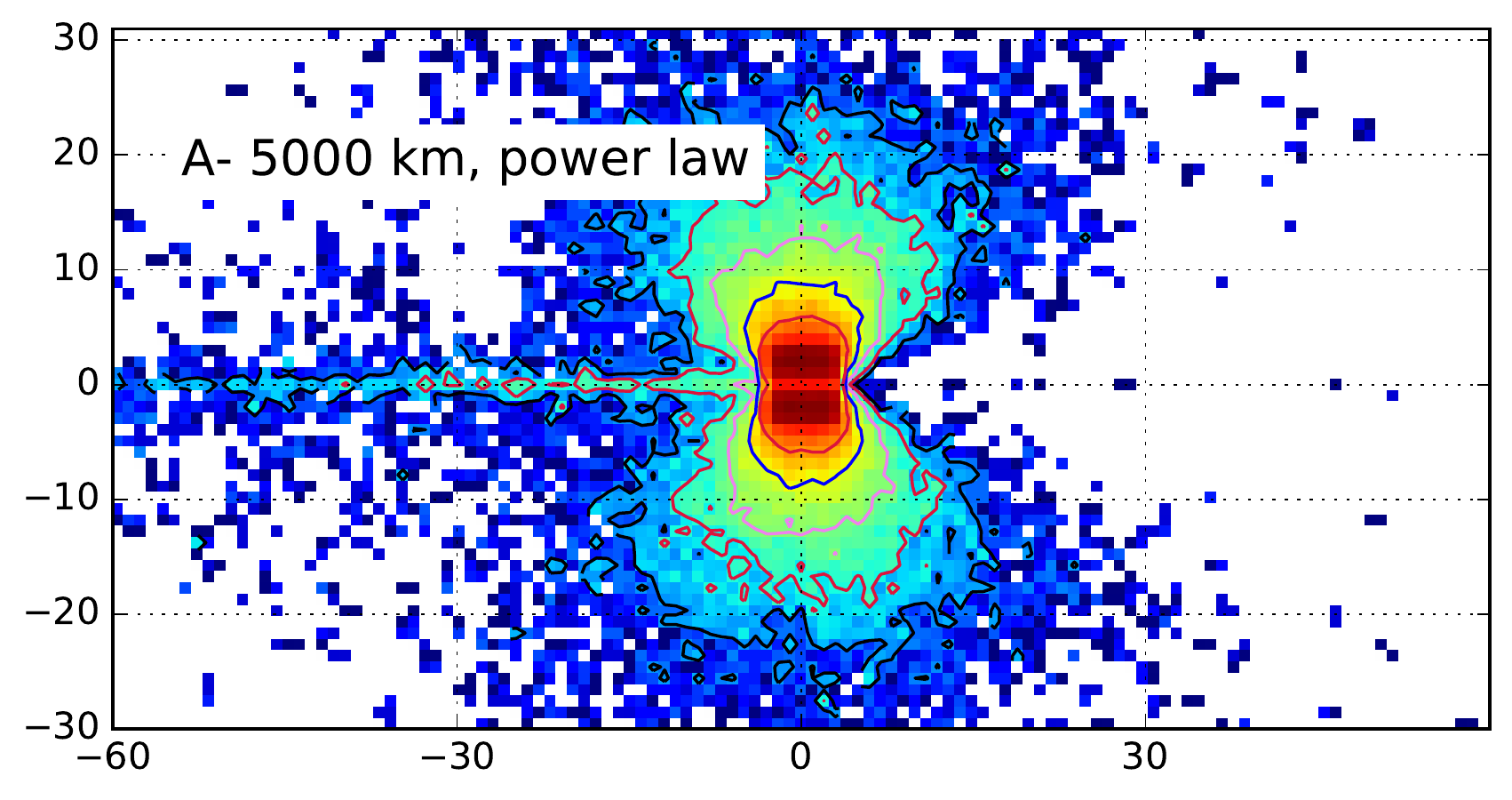}\\
\includegraphics[width=0.45\textwidth]{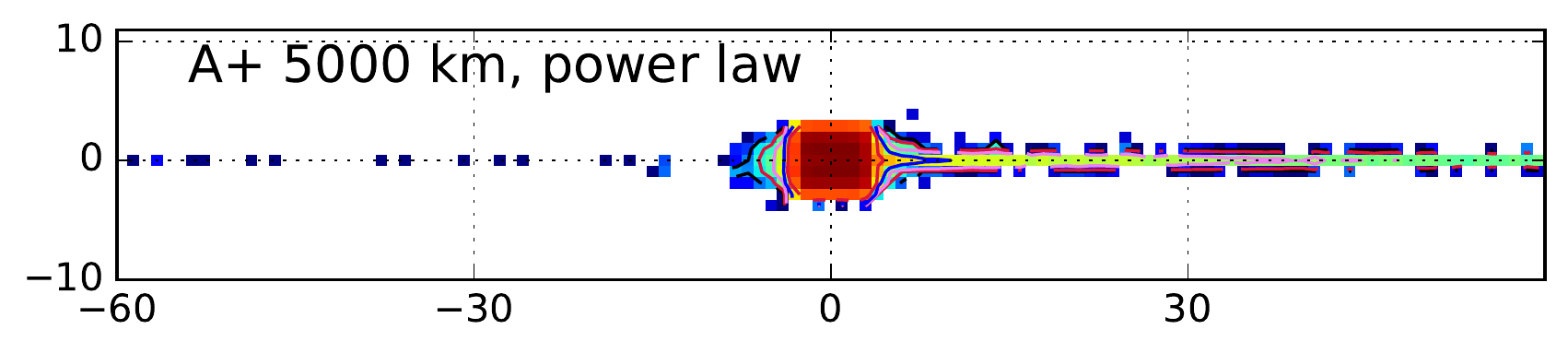}
\caption{Map of crossings of protons, injected from a power-law with $\gamma=-1.1$, spanning the energy range from 10 to 400 MeV, across the \mbox{1 au} sphere, over a time of \mbox{100 hr}, relative to best-connected fieldline, with the effects of corotation removed. Fluence colours are on a logarithmic scale, with two contours per decade. Top: Unipolar field, pointing outwards. Middle: HCS thickness scaled to \mbox{5000 km} at \mbox{1 au}, with an $A-$ field configuration. Bottom: HCS thickness scaled to \mbox{5000 km} at \mbox{1 au}, with an $A+$ field configuration.}\label{fig:crossings_maps_powlaws}
\end{figure}

In Figure \ref{fig:crossings_maps_powlaws} we show fluence maps for the same simulations as seen in Figures \ref{fig:timeprofile_unipolar} through \ref{fig:timeprofile_a+}. Drifts extend protons for significant distances in latitude and longitude. The relative spread, compared with \ref{fig:maps_grid}, is not drastically different, as the number of injected particles for the power-law runs was increased tenfold. One should note that contours are spaced two per decade. 


\begin{figure}[htp]
\centering
\includegraphics[width=0.45\textwidth]{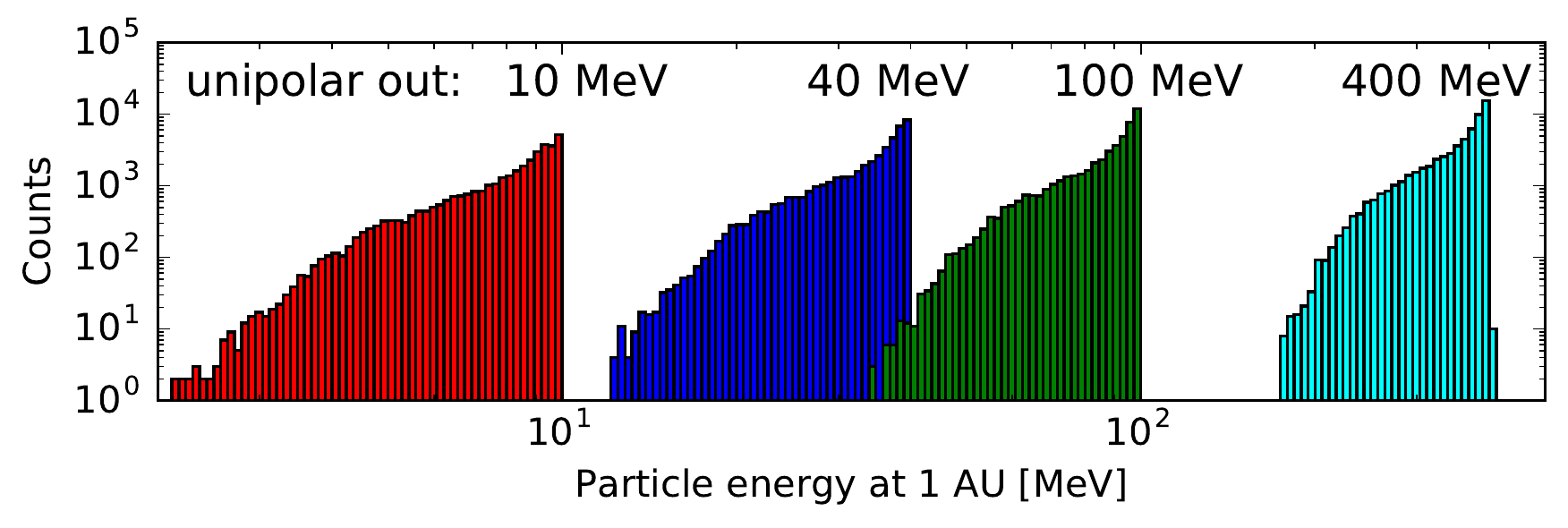}\\
\includegraphics[width=0.45\textwidth]{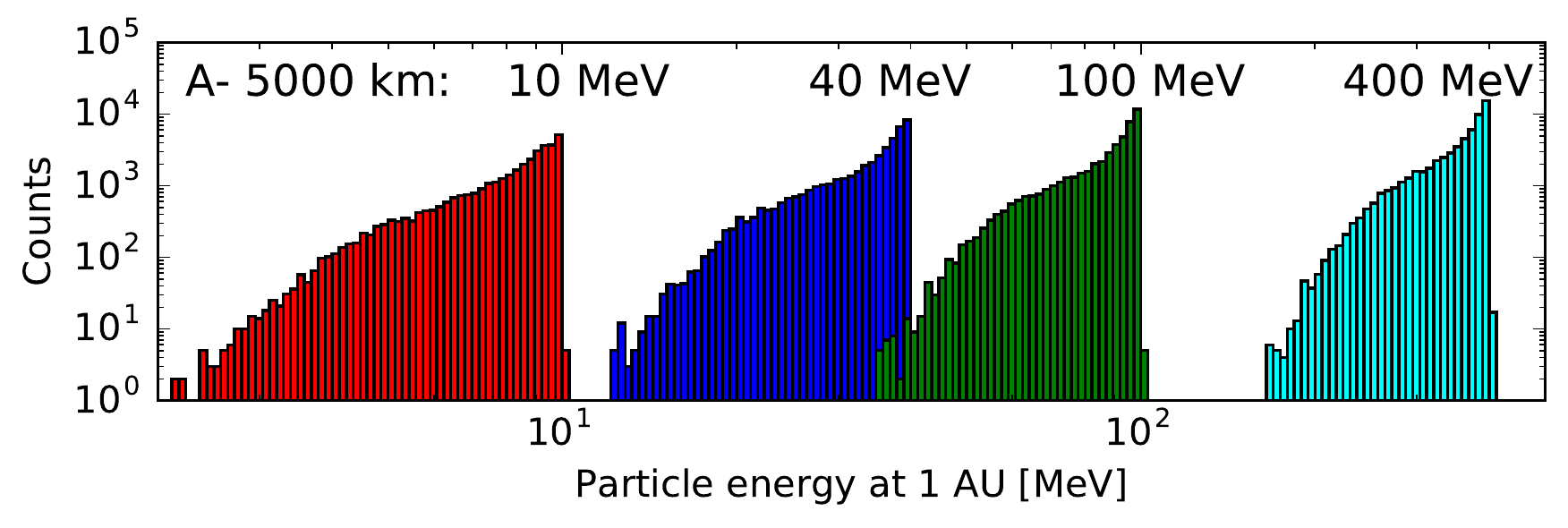}\\
\includegraphics[width=0.45\textwidth]{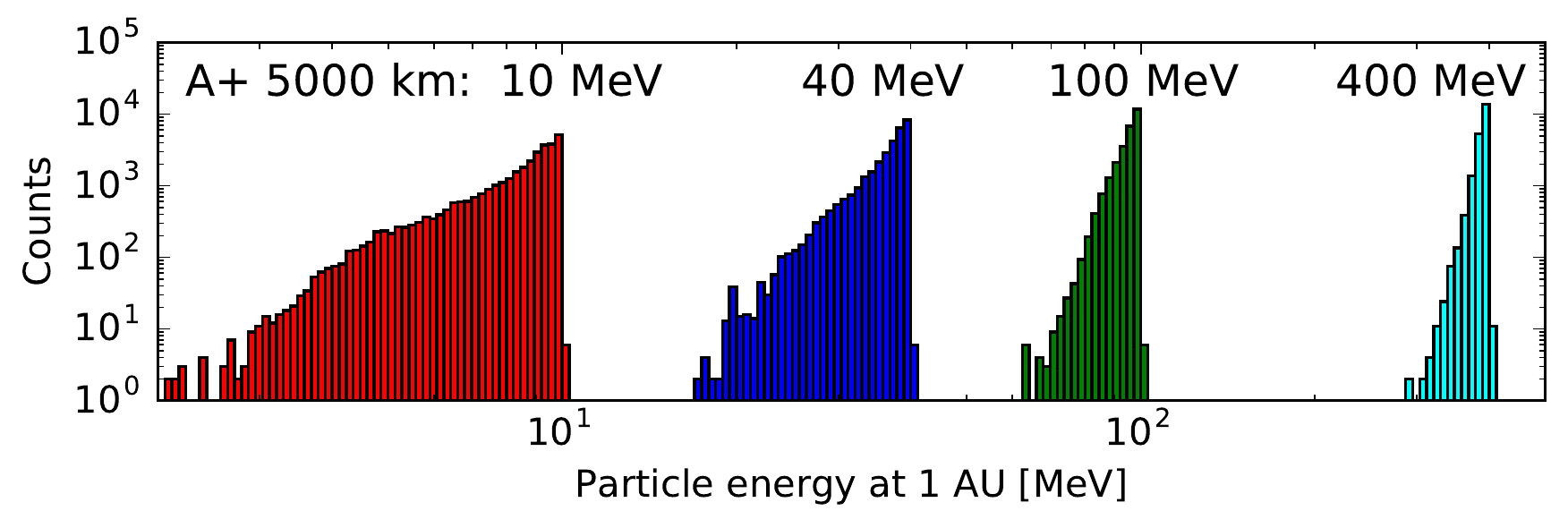}
\caption{Histograms of measured proton energies at the times of \mbox{1 au} crossings. Colours depict protons injected at \mbox{10 MeV} (red), \mbox{40 MeV} (blue), \mbox{100 MeV} (green), and \mbox{400 MeV} (cyan). Top: Unipolar field, pointing outwards. Middle: HCS thickness scaled to \mbox{5000 km} at \mbox{1 au}, with an $A-$ field configuration. Bottom: HCS thickness scaled to \mbox{5000 km} at \mbox{1 au}, with an $A+$ field configuration. Deceleration is seen at all energies, but the $A+$ configuration supresses drifts, and thus, deceleration}\label{fig:histogram_energies}
\end{figure}

As described in \cite{Dalla2015}, SEPs experience deceleration during propagation through interplanetary space due to adiabatic deceleration and drift effects. In the work presented in this manuscript, protons have been injected into the simulation at the described energies, with deceleration happening by the time they reach \mbox{1 au}. Thus, protons which are detected at \mbox{1 au} as, e.g., \mbox{100 MeV} protons, will have likely been injected at higher energies, and will thus have experienced greater drifts due to the velocity dependencies involved. In Figure \ref{fig:histogram_energies}, we plot histograms of \mbox{1 au} crossing energies, over the durationof the simulation, for protons injected at energies of \mbox{10 MeV}, \mbox{40 MeV}, \mbox{100 MeV}, and \mbox{400 MeV}, for three different magnetic field configurations. For an unipolar outwards-pointing IMF, and for an IMF with an $A-$ configuration, protons can decelerate by as much as over 50\% of their initial energy. However, for an $A+$ IMF configuration, protons are confined to the vicinity of the HCS, and deceleration due to drifts is suppressed to as little as $<25\%$.

\begin{figure*}[!htp]
\centering
\includegraphics[width=0.32\textwidth]{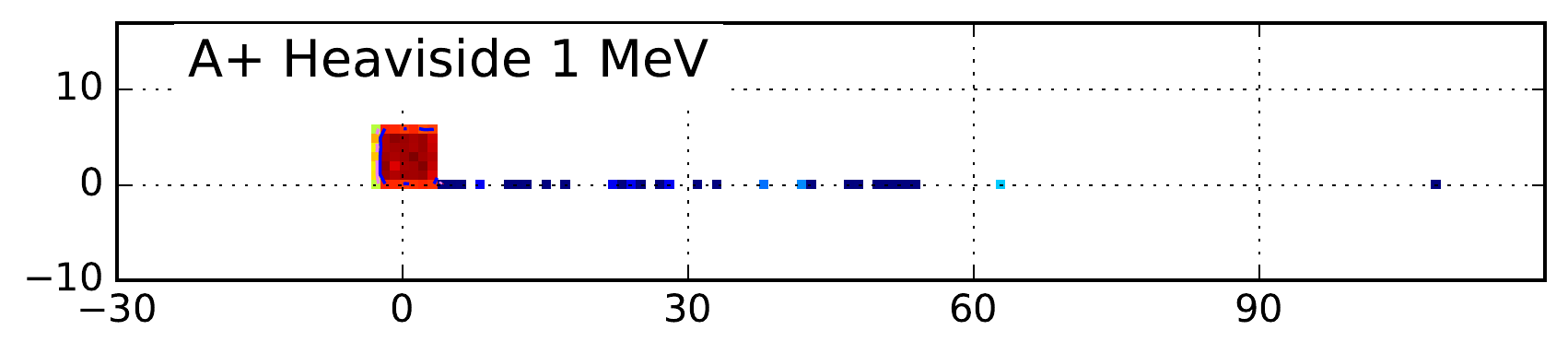}
\includegraphics[width=0.32\textwidth]{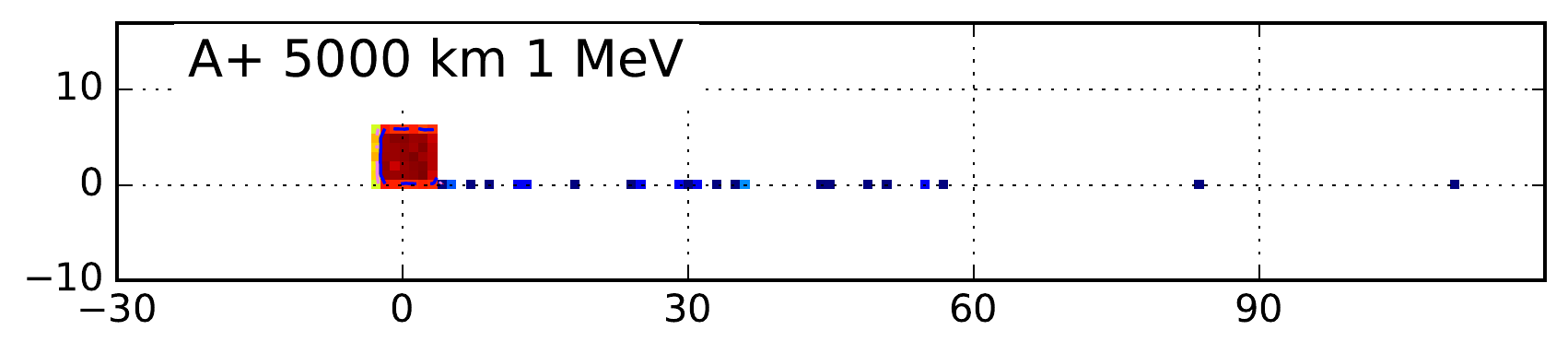}
\includegraphics[width=0.32\textwidth]{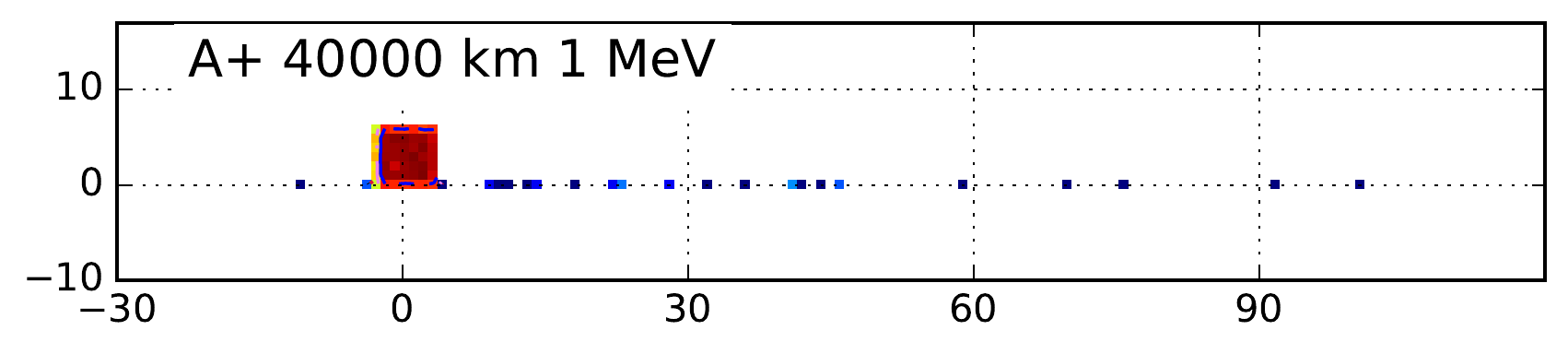}
\includegraphics[width=0.32\textwidth]{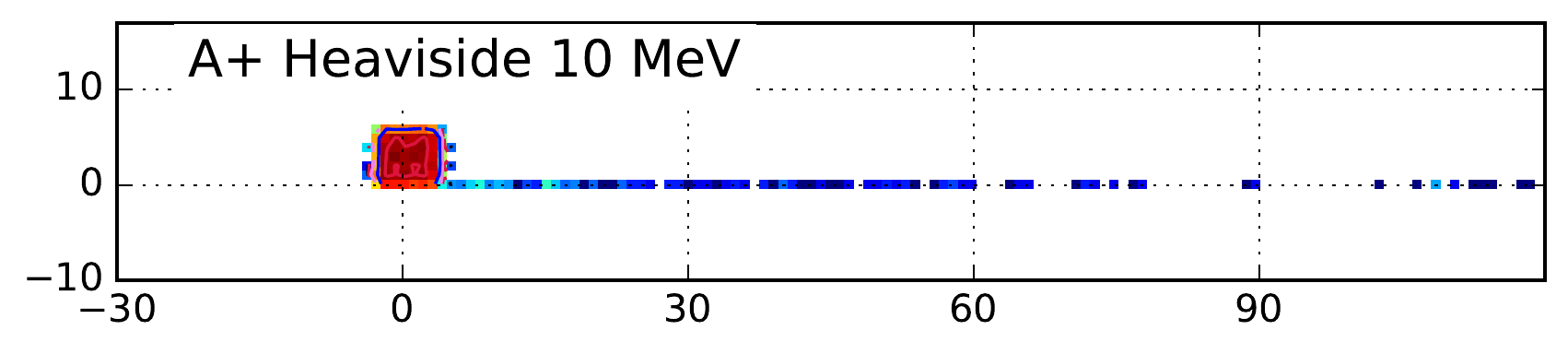}
\includegraphics[width=0.32\textwidth]{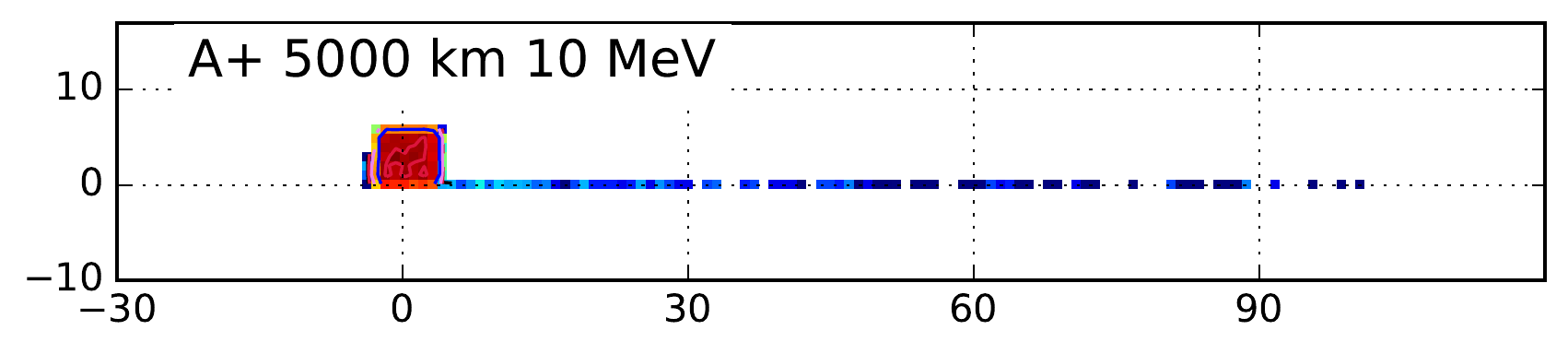}
\includegraphics[width=0.32\textwidth]{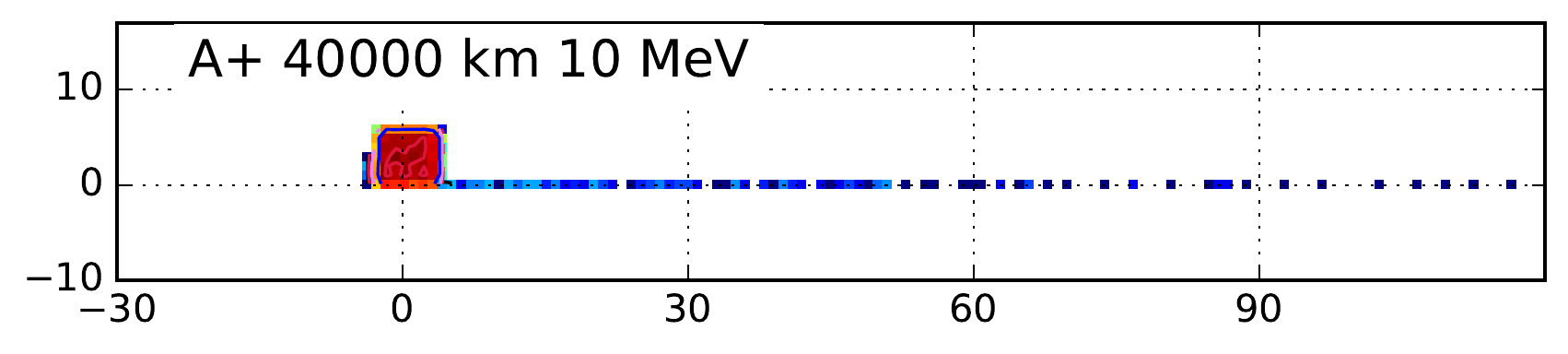}
\includegraphics[width=0.32\textwidth]{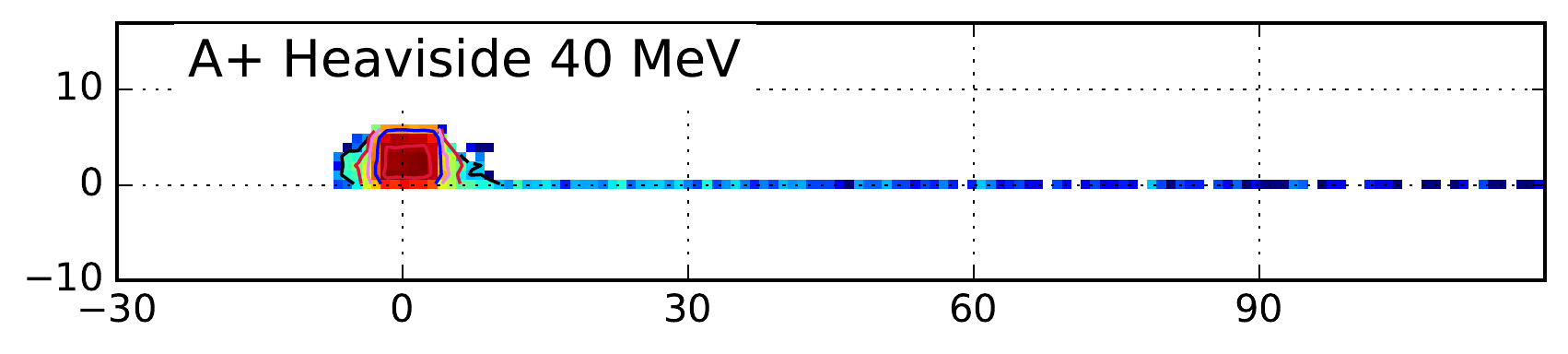}
\includegraphics[width=0.32\textwidth]{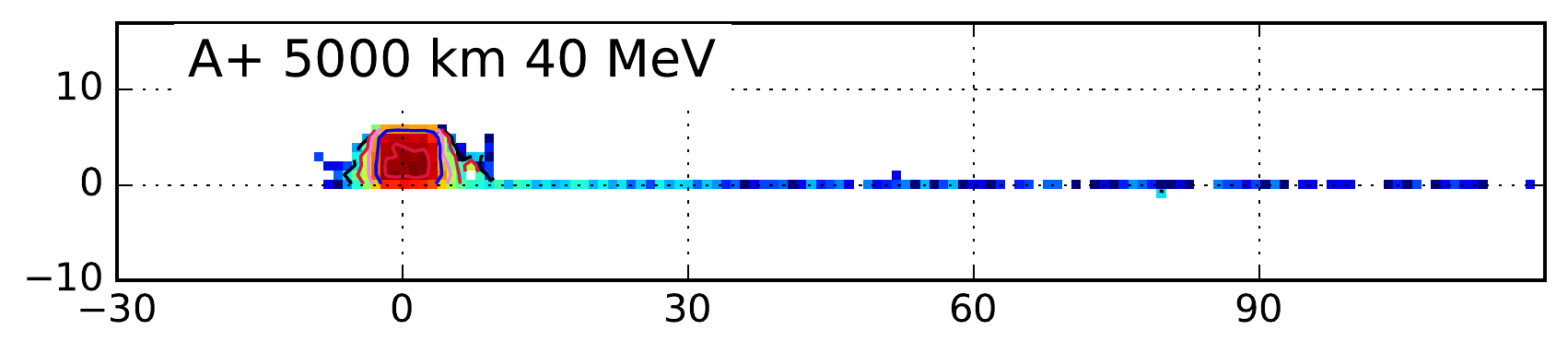}
\includegraphics[width=0.32\textwidth]{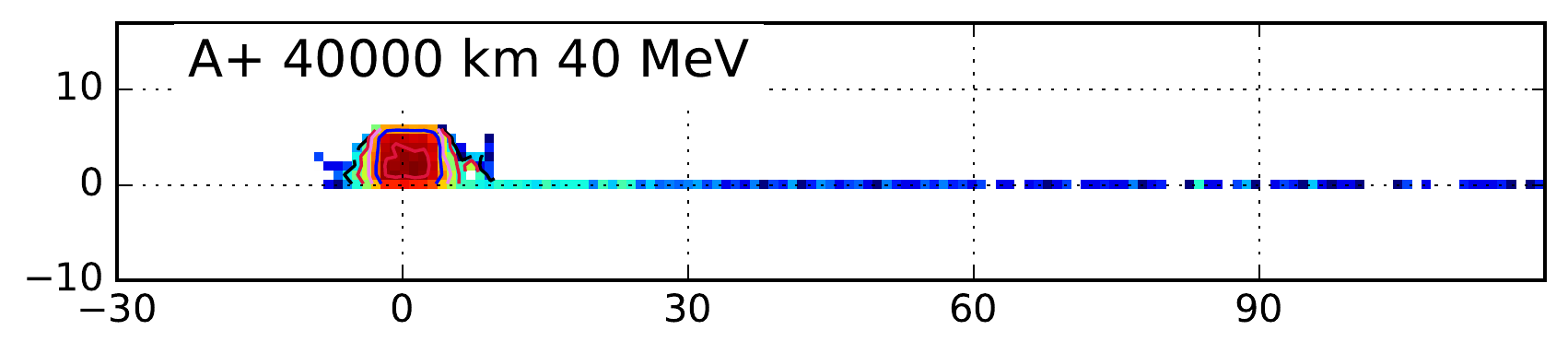}
\includegraphics[width=0.32\textwidth]{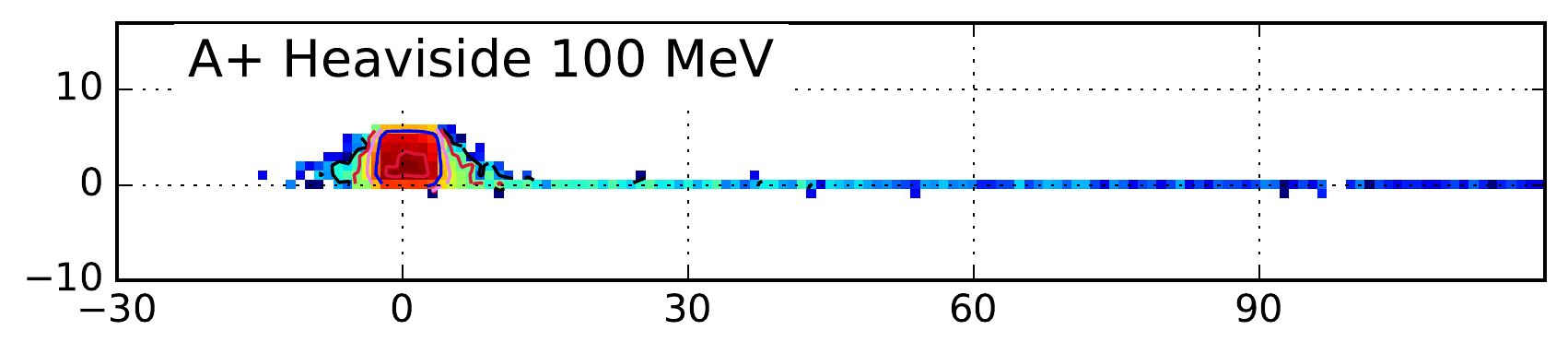}
\includegraphics[width=0.32\textwidth]{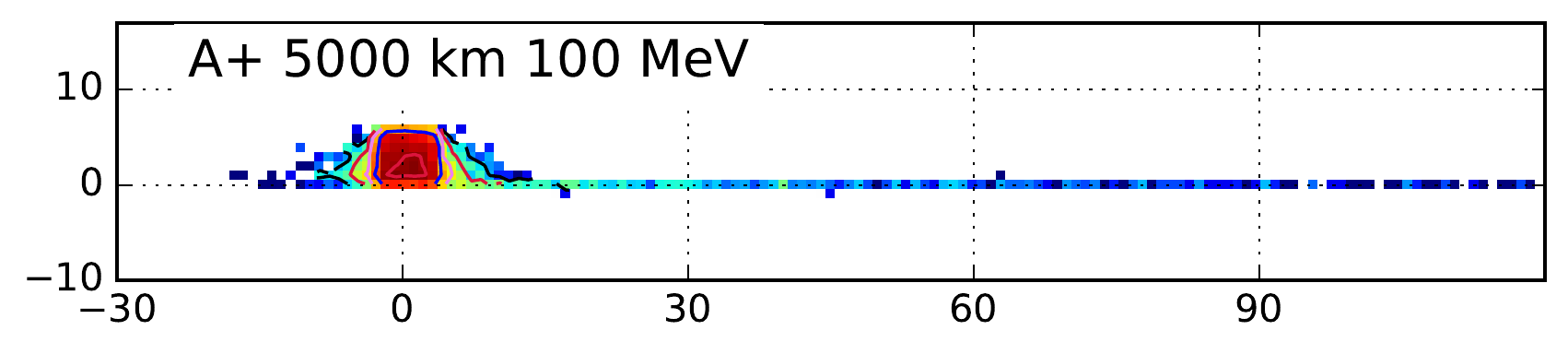}
\includegraphics[width=0.32\textwidth]{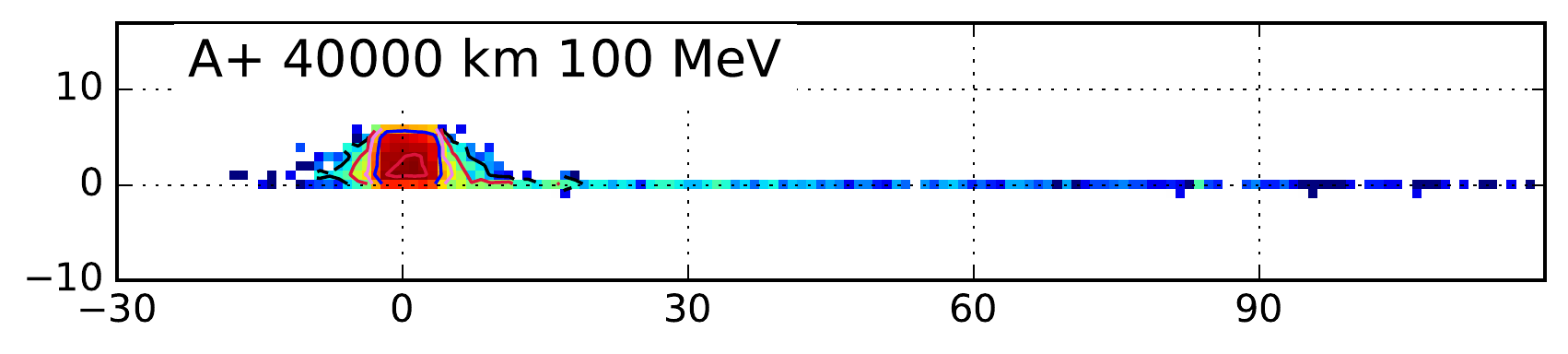}
\includegraphics[width=0.32\textwidth]{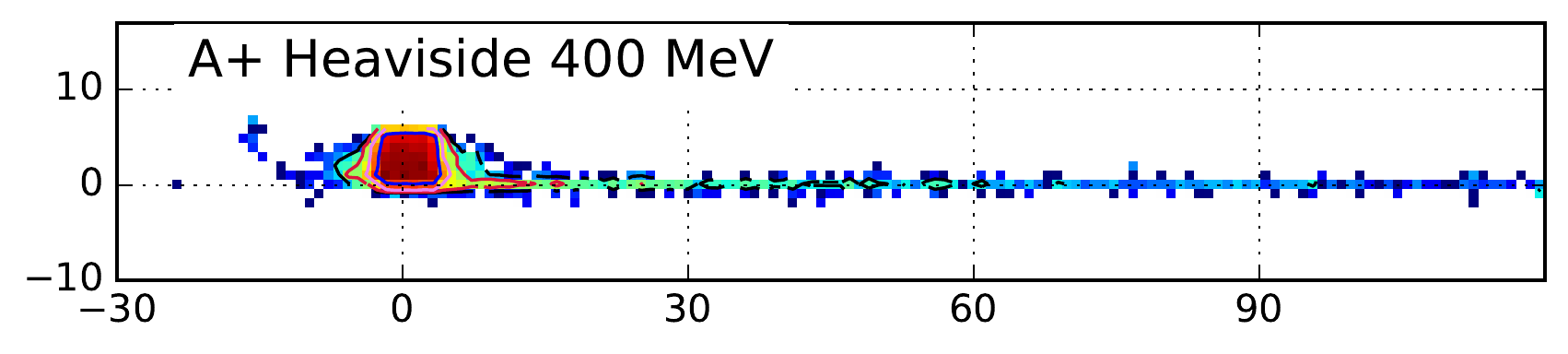}
\includegraphics[width=0.32\textwidth]{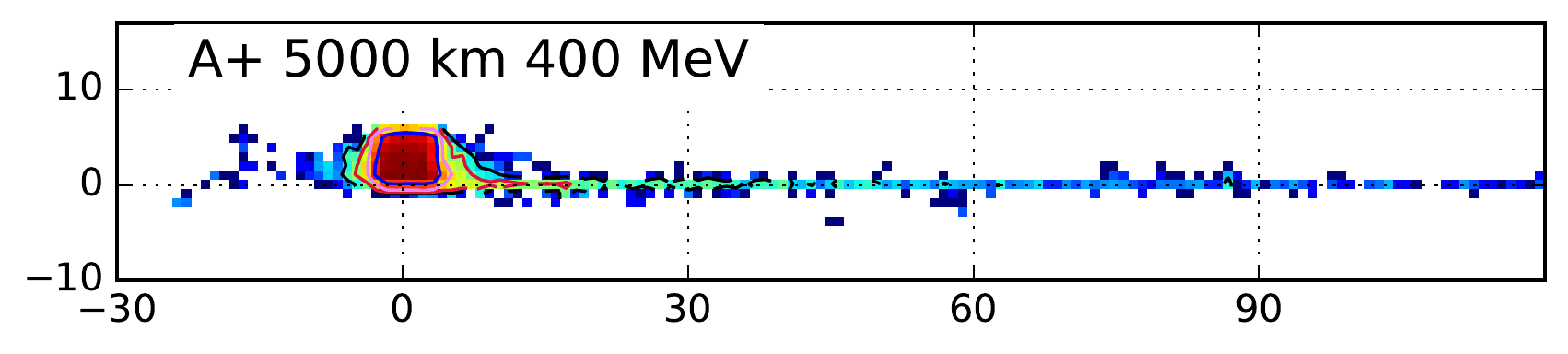}
\includegraphics[width=0.32\textwidth]{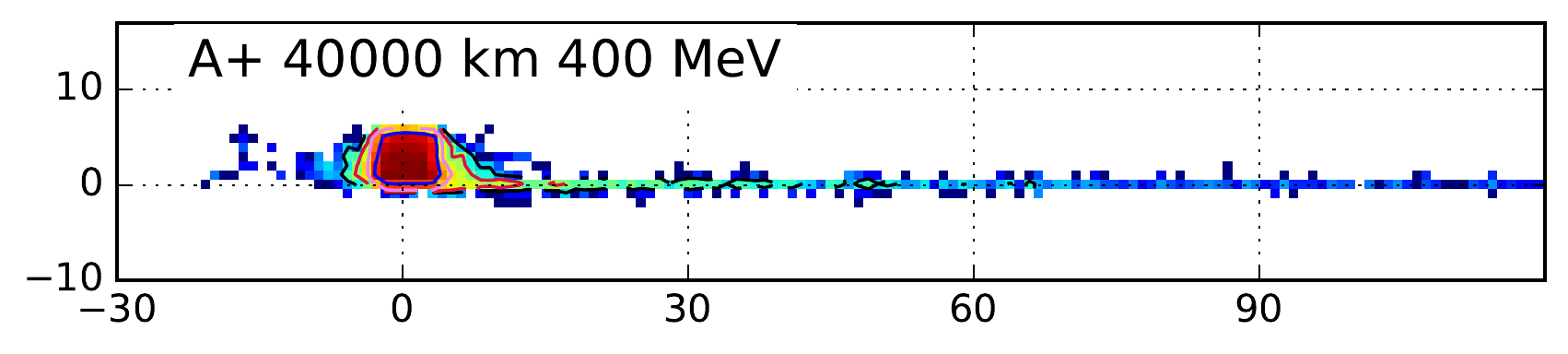}
\includegraphics[width=0.32\textwidth]{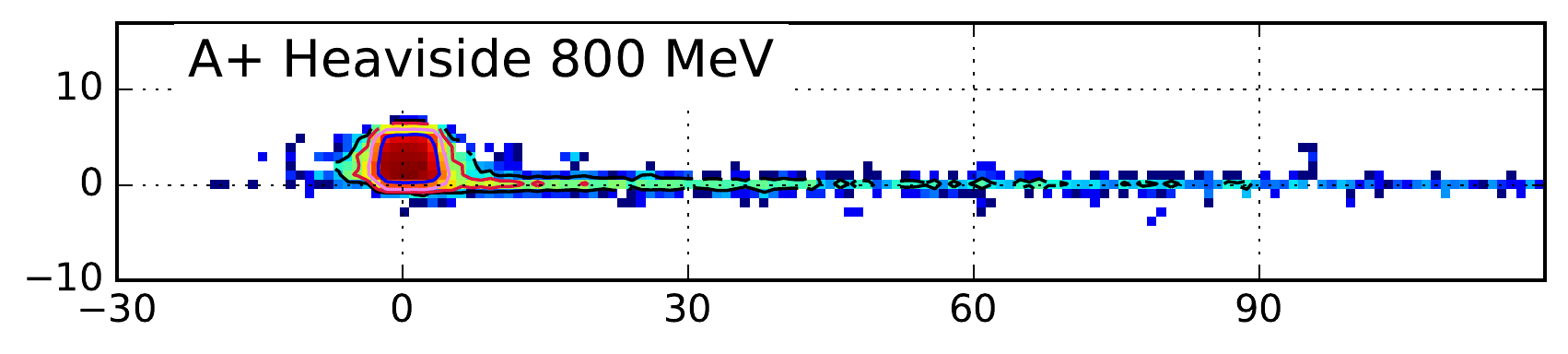}
\includegraphics[width=0.32\textwidth]{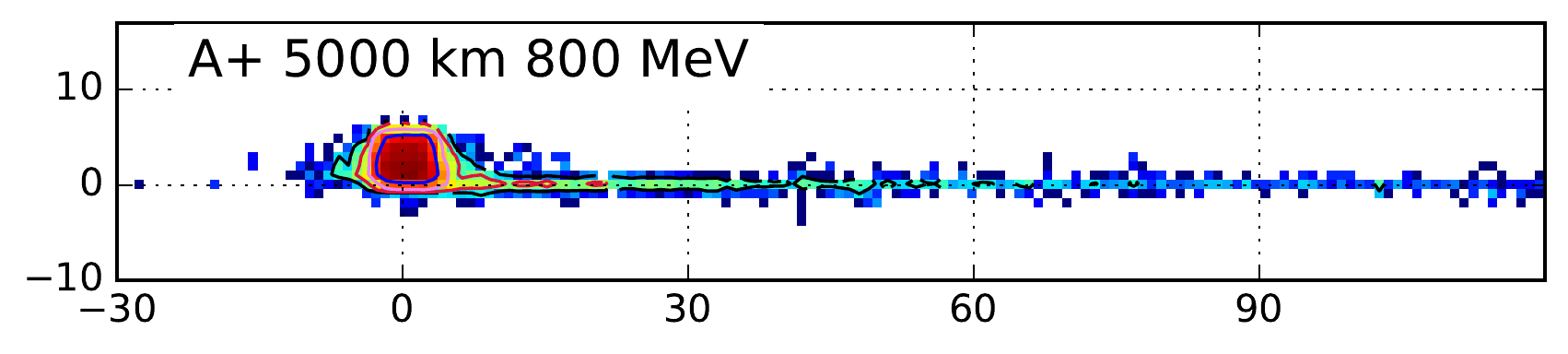}
\includegraphics[width=0.32\textwidth]{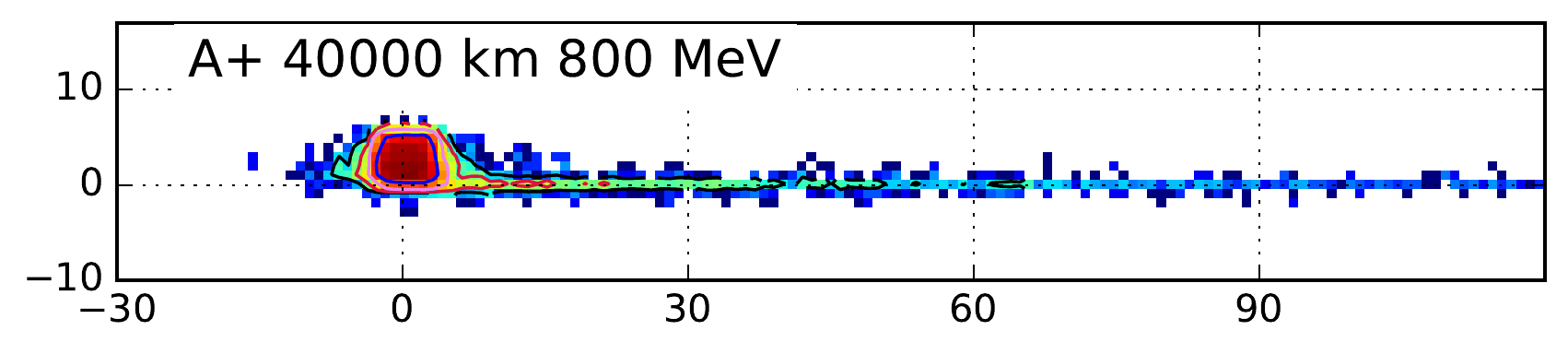}
\caption{Map of crossings of protons, injected at energies ranging from \mbox{1 to 800 MeV}, across the \mbox{1 au} sphere, with the injection site centered at $+3^\circ$  latitude. Fluence colour contours are on a logarithmic scale. The magnetic field has an $A+$ configuration, with HCS thickness as a Heaviside step function (left column), scaled to \mbox{5 000 km} at \mbox{1 au} (centre column), or scaled to \mbox{40 000 km} at \mbox{1 au} (right column). Proton crossing coordinates are shown relative to the best-connected fieldline at injection time with the effects of corotation removed. From top row to bottom row, proton injection energies of \mbox{1 MeV}, \mbox{10 MeV}, \mbox{40 MeV}, \mbox{100 MeV}, \mbox{400 MeV} and \mbox{800 MeV}. Drift of protons across the HCS is non-existant at energies below \mbox{400 MeV}.}
\label{fig:crossings_maps_aplus}%
\end{figure*}

The crossing of SEPs from one IMF polarity to another, across sector boundaries\added{ caused by a wavy HCS}, is a complex question which we can not fully analyse within the scope of this work. A first step, however, is to assess the efficiency of particle drifts and scattering in transporting SEPs across a flat HCS. In order to analyze this, we injected protons of six different energies (\mbox{1 MeV}, \mbox{10 MeV}, \mbox{40 MeV}, \mbox{100 MeV}, \mbox{400 MeV}, and \mbox{800 MeV}) from a $6^\circ \times 6^\circ$ angular injection window, centered at $+3^\circ$ within an $A+$ configuration. In Figure \ref{fig:crossings_maps_aplus}, we plot these results with the effects of corotation removed, for three different current sheet thicknesses. At small energies, only the current sheet drift spreads particles outside the well-connected region, but at energies above \mbox{40 MeV}, some drifts in both latitude and longitude are visible. However, proton energies need to exceed \mbox{100 MeV} in order to be ejected from the current sheet to the southern hemisphere. For comparison, the Larmor radius of \mbox{400 MeV} protons at \mbox{1 au}, assuming, for example, a pitch-angle of $\alpha \approx 5^\circ$, is of the order of \mbox{40000 km}.

\begin{figure*}[!htp]
\centering
\includegraphics[width=0.32\textwidth]{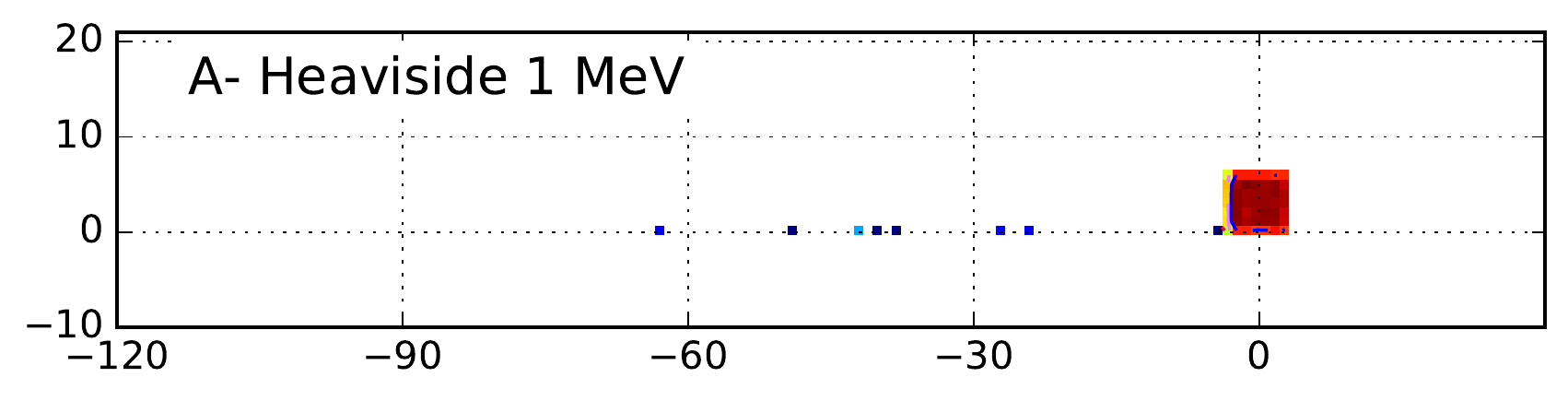}
\includegraphics[width=0.32\textwidth]{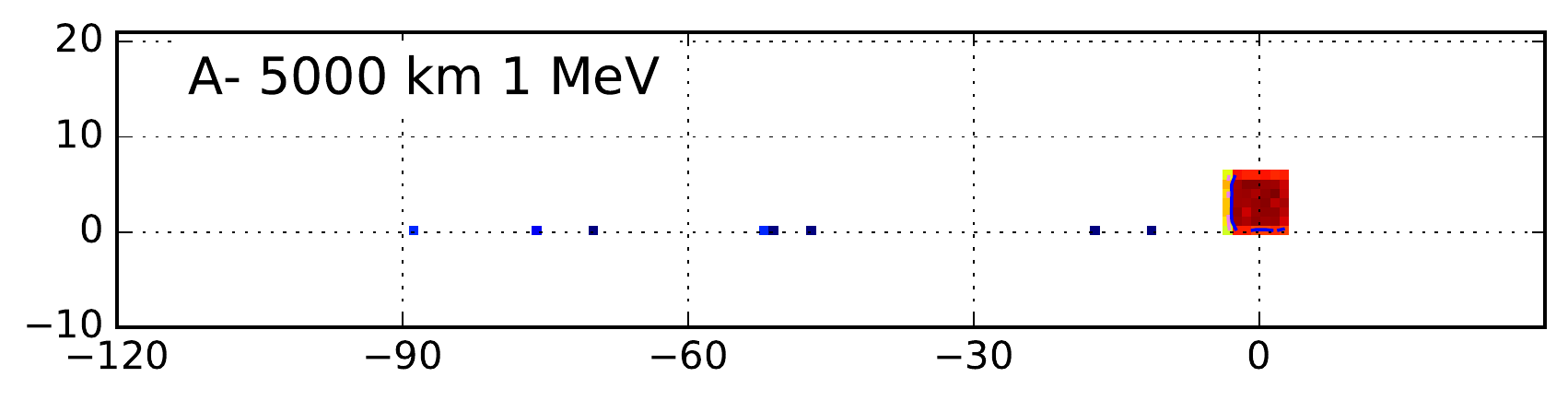}
\includegraphics[width=0.32\textwidth]{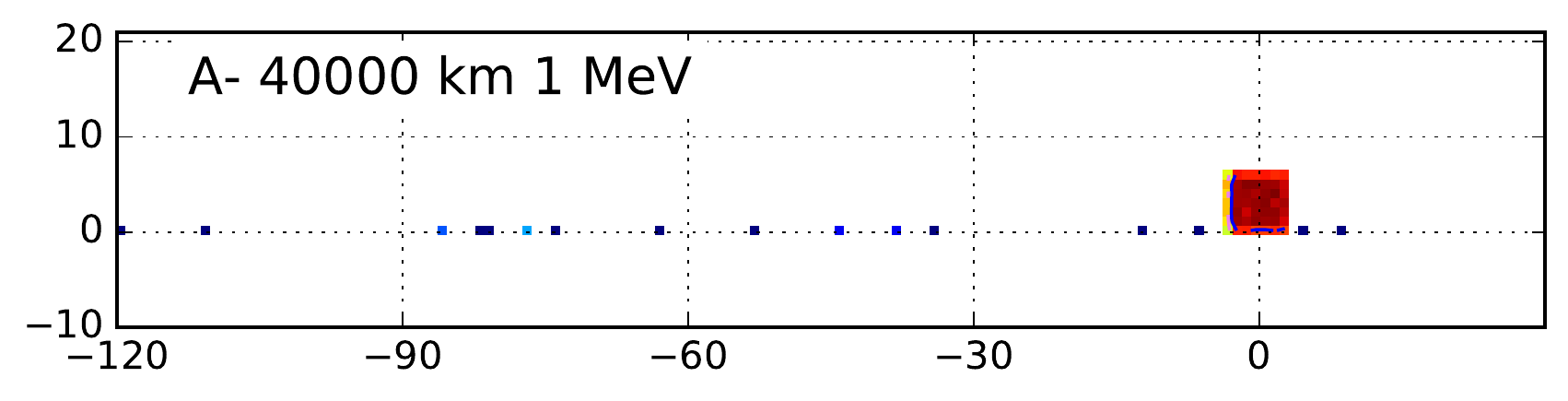}
\includegraphics[width=0.32\textwidth]{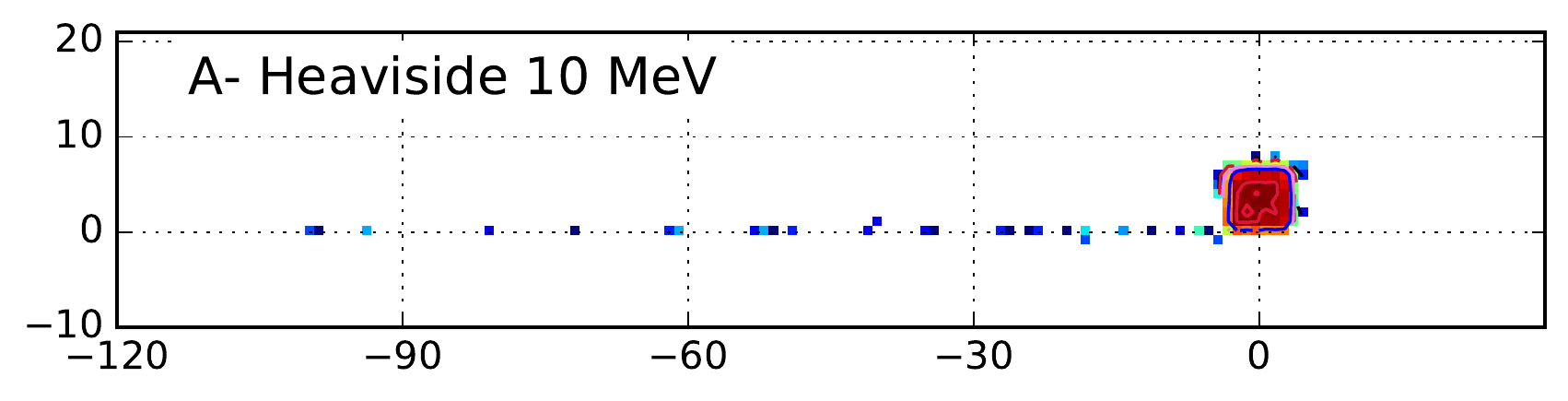}
\includegraphics[width=0.32\textwidth]{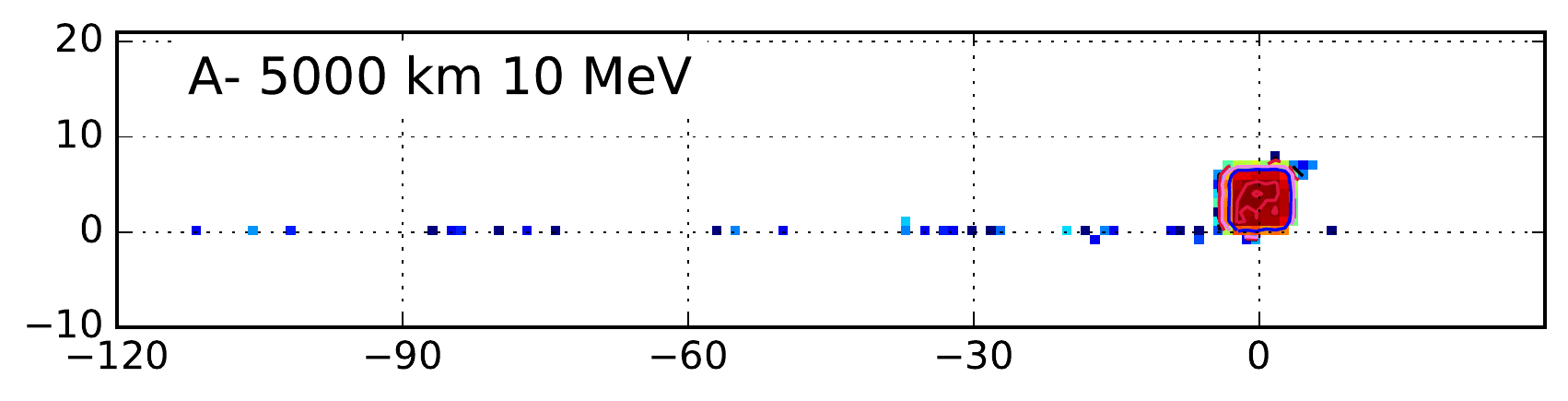}
\includegraphics[width=0.32\textwidth]{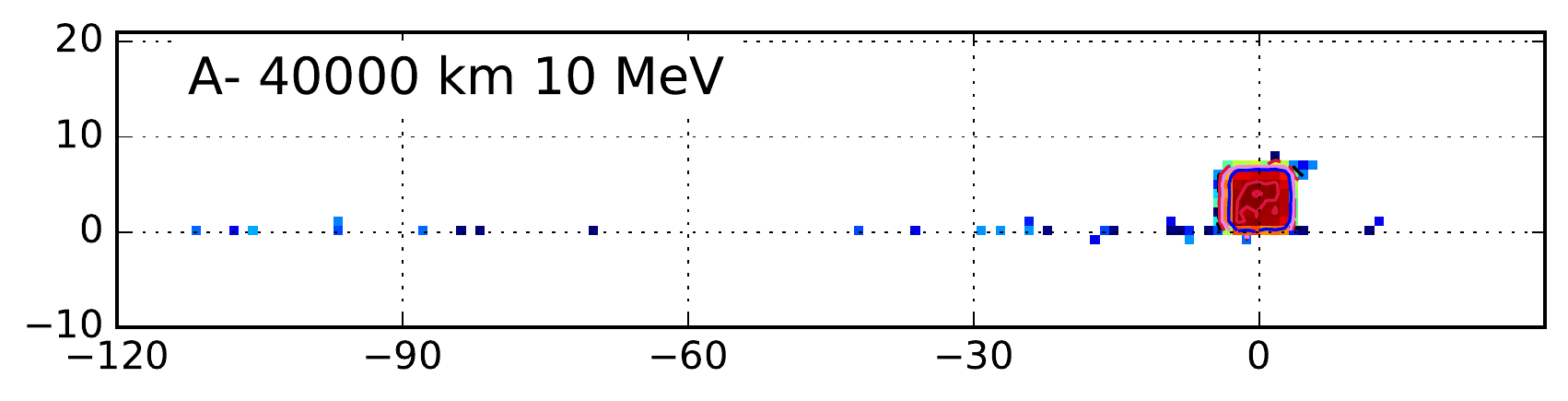}
\includegraphics[width=0.32\textwidth]{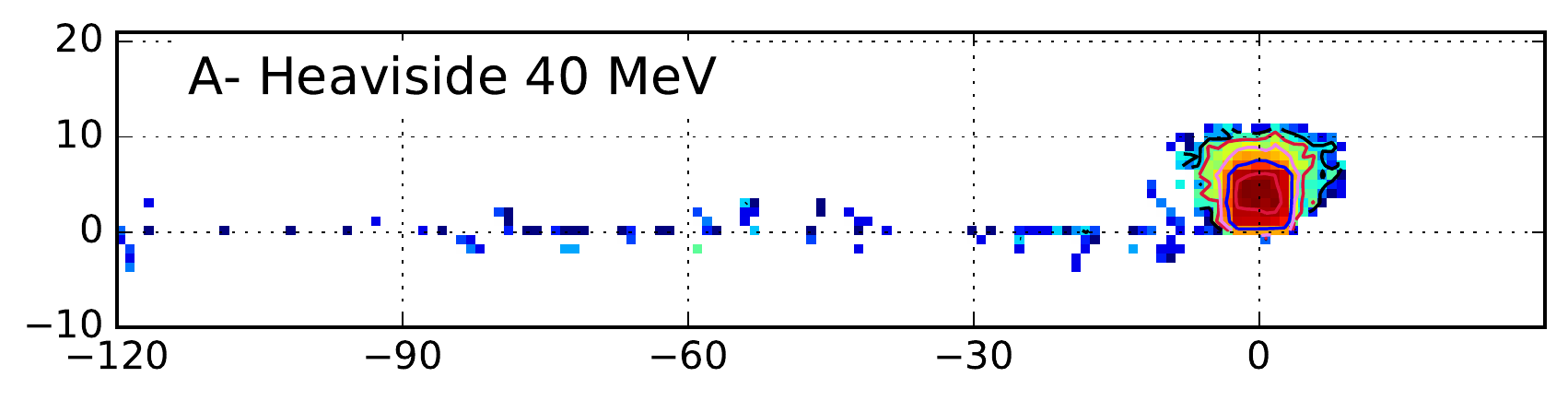}
\includegraphics[width=0.32\textwidth]{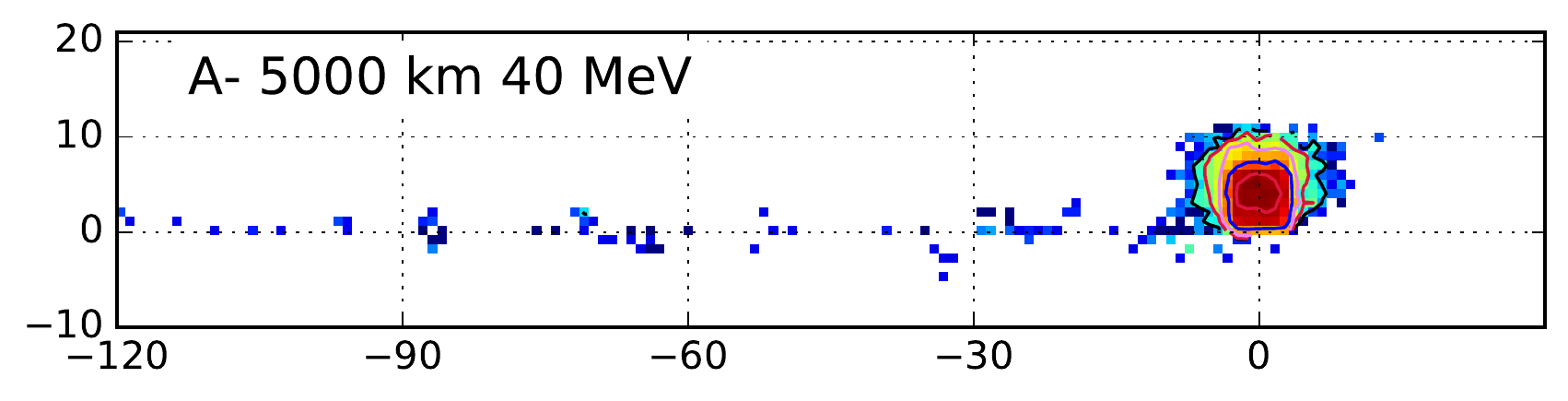}
\includegraphics[width=0.32\textwidth]{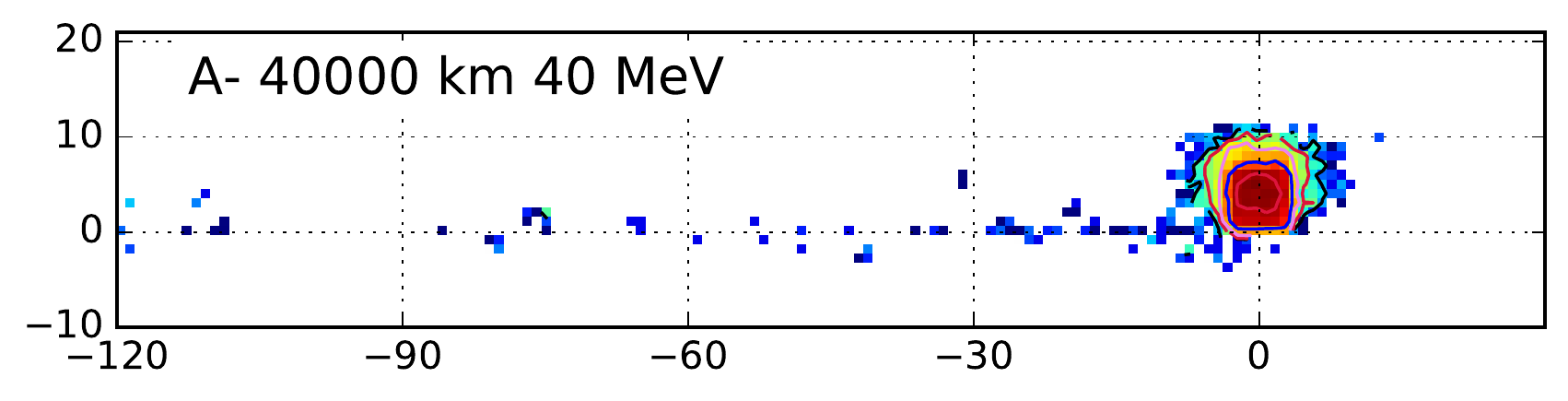}
\includegraphics[width=0.32\textwidth]{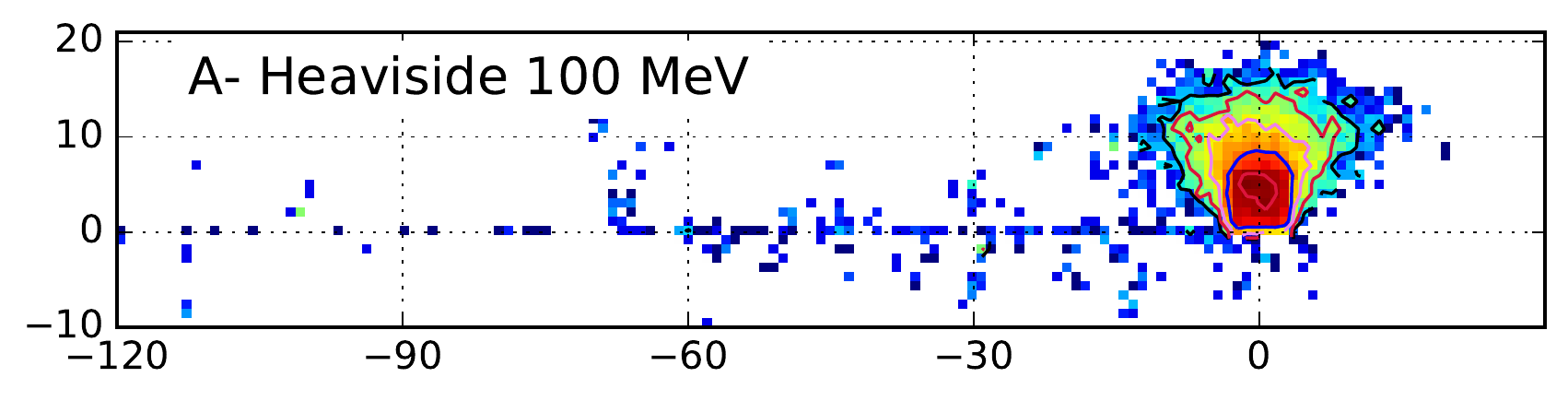}
\includegraphics[width=0.32\textwidth]{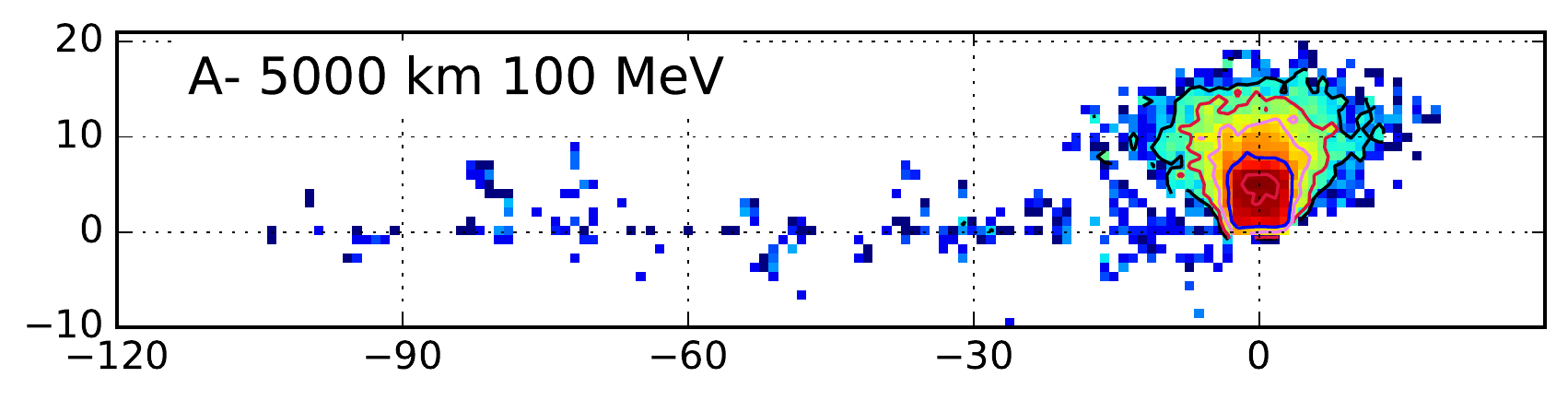}
\includegraphics[width=0.32\textwidth]{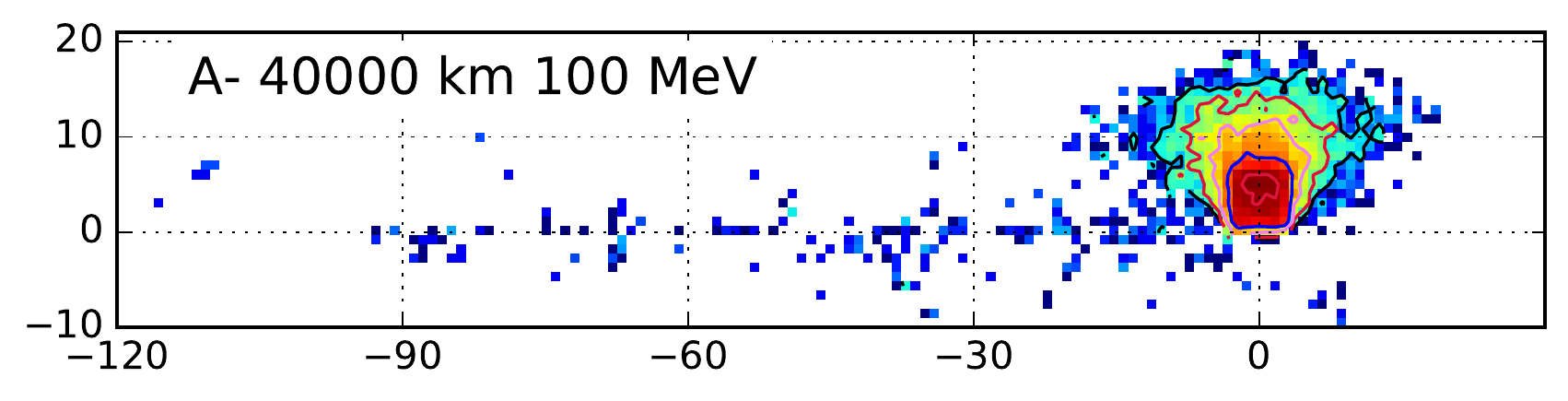}
\includegraphics[width=0.32\textwidth]{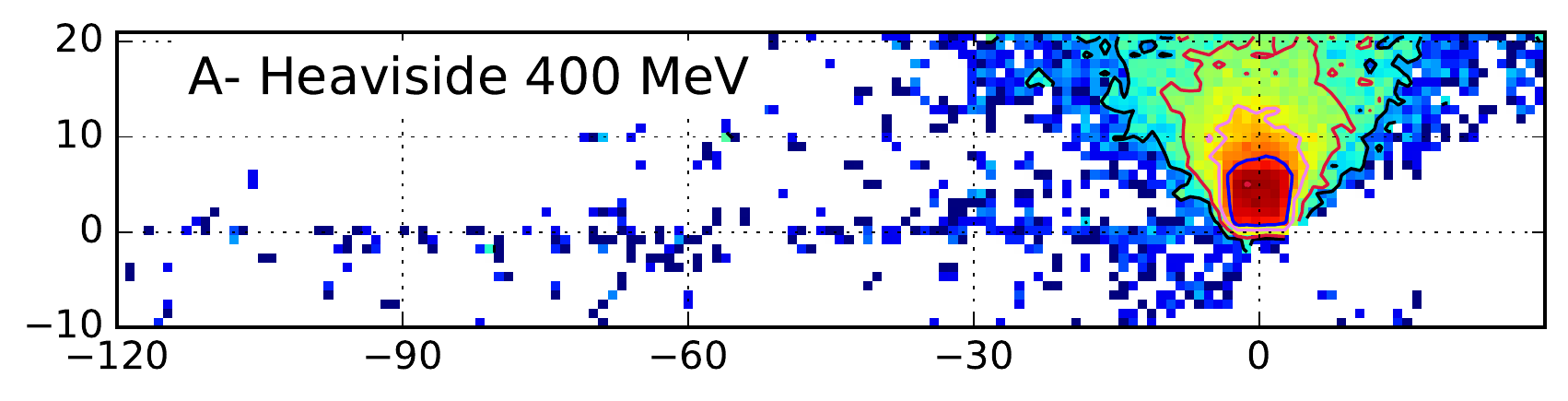}
\includegraphics[width=0.32\textwidth]{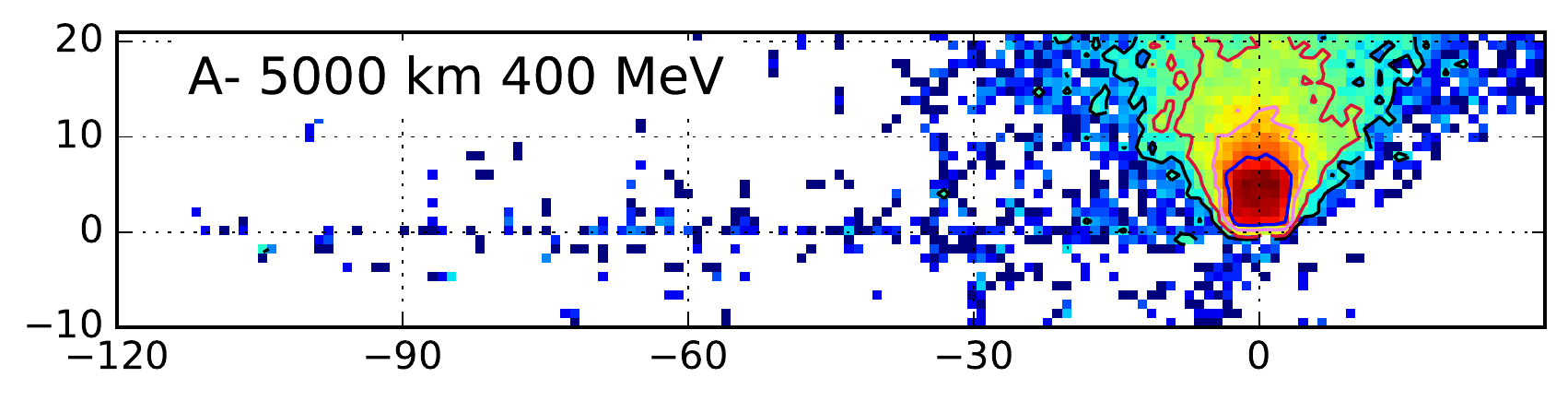}
\includegraphics[width=0.32\textwidth]{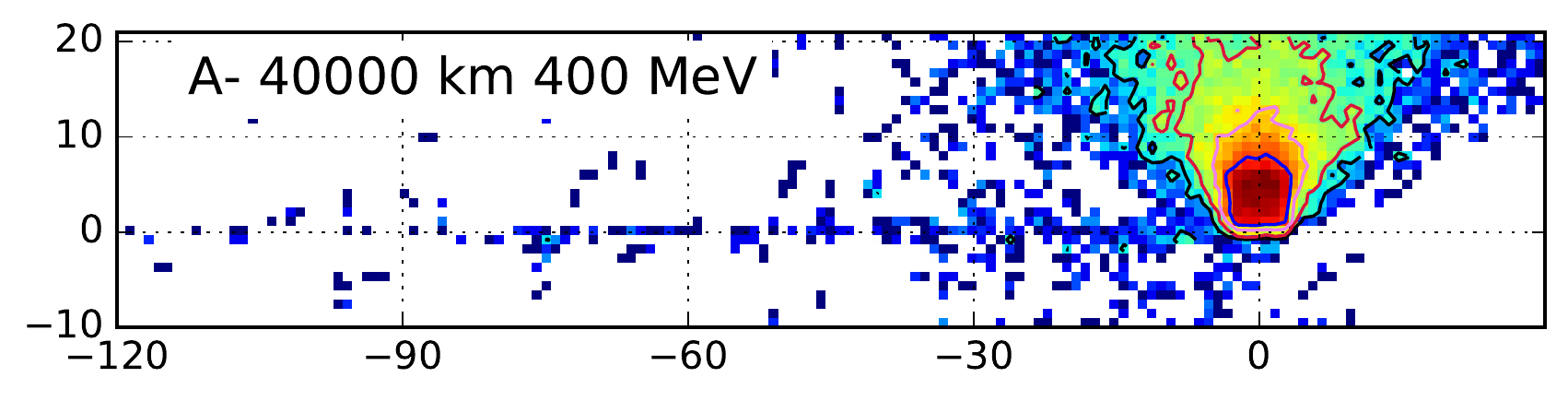}
\includegraphics[width=0.32\textwidth]{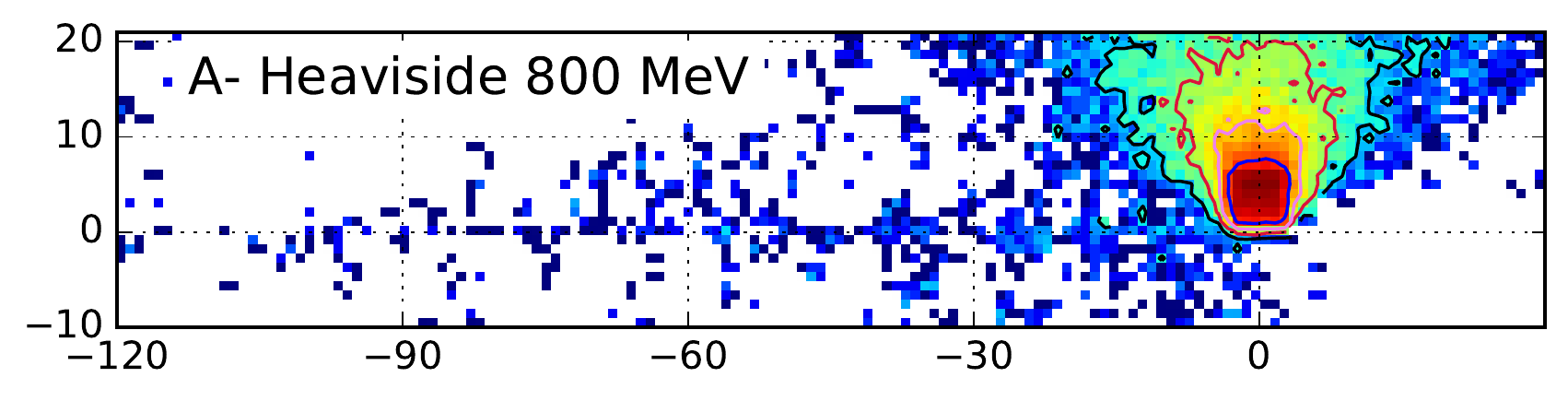}
\includegraphics[width=0.32\textwidth]{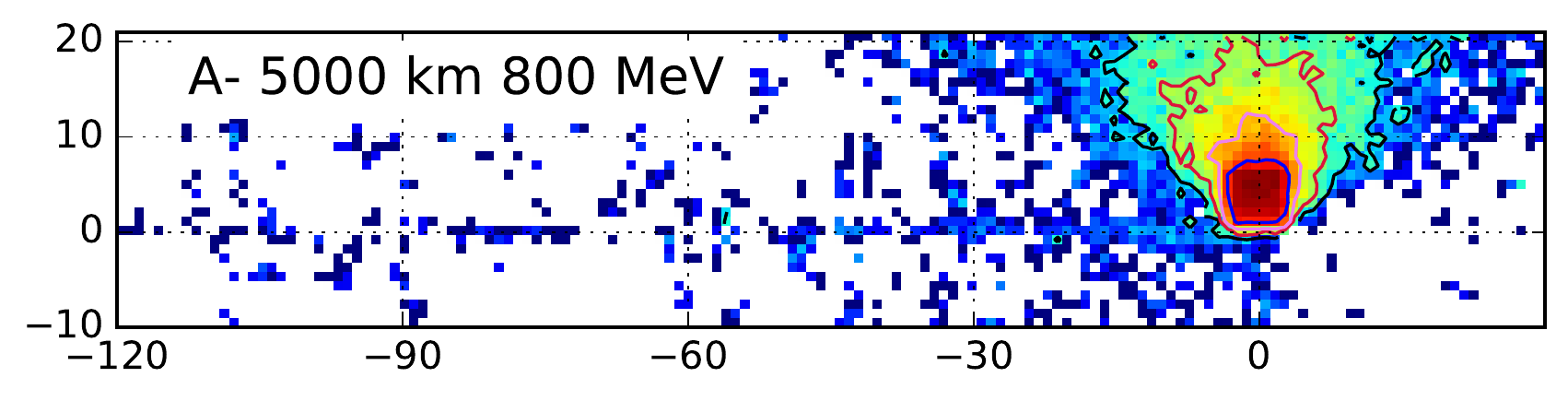}
\includegraphics[width=0.32\textwidth]{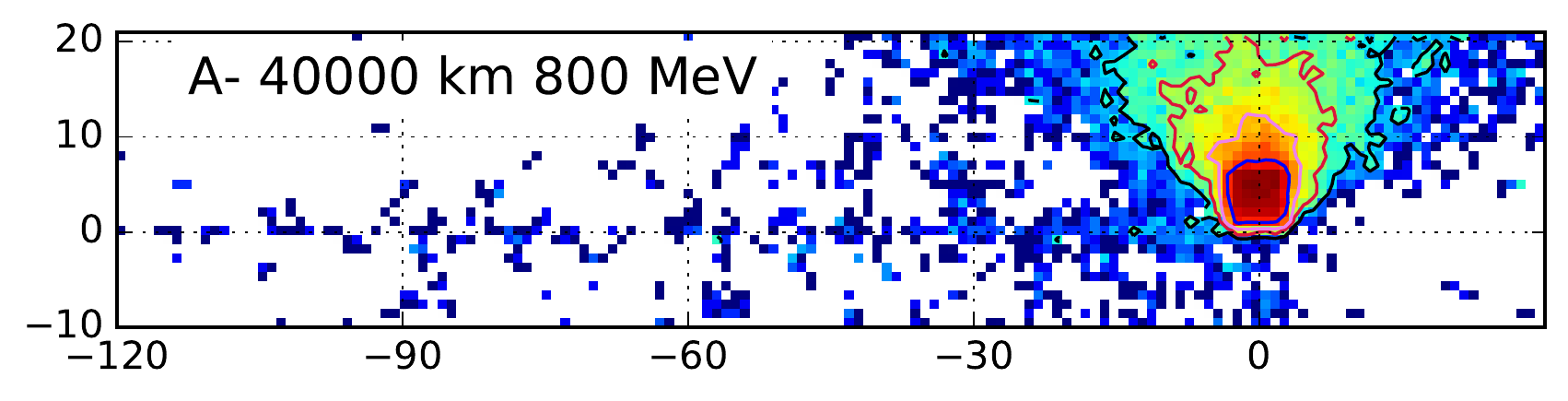}
\caption{Map of crossings of protons, injected at energies ranging from \mbox{1 to 800 MeV}, across the \mbox{1 au} sphere, as in Figure \ref{fig:crossings_maps_aplus}, but for a magnetic field with an $A-$ configuration. Protons at injection energies as low as \mbox{40 MeV} are able to cross the HCS, as even a small scattering across the sheet will bring them to a region where they drift further away from the sheet. Crossing counts, however, remain low.}
\label{fig:crossings_maps_aminus}%
\end{figure*}

In Figure \ref{fig:crossings_maps_aminus}, we plot the same crossings as in Figure \ref{fig:crossings_maps_aplus}, but for an IMF with a $A-$ configuration. Again, at low energies, the current sheet drift is the primary way particles spread outside the well-connected region. However, as general drift directions are away from the current sheet, any particles which are transported along the current sheet and then scatter away from it can easily propagate further away from it. Thus, at energies as low as \mbox{40 MeV}, protons are seen to scatter into the southern hemisphere. We note, however, that if the injection region of protons does not coincide with the current sheet, protons within an $A-$ configuration are unlikely to reach the current sheet, and thus, unlikely to scatter across it.

Thus, we conclude that an injection event constrained to one hemisphere can, due to lateral drifts and the heliospheric current sheet, remain undetectable in the opposite hemisphere. 

\section{Conclusions}

We simulated the propagation of solar energetic protons with energies ranging from \mbox{$1$ to $800$ MeV} within multiple different IMF conditions, assessing the effects a flat heliospheric current sheet located at the heliographic equator has on proton drifts and propagation. We show that, in the presence of a flat HCS, drifts along the current sheet are significant, allowing high-energy protons to drift over 180 degrees in longitude. We show that both $A+$ and $A-$ configurations of the IMF allow for significant current sheet drift which helps protons reach regions far from the injection longitude. 

We assessed the effects of current sheet thickness on proton propagation, and found there to be negligible difference between simulations using realistic parameters or a step function. Gradient drifts due to sheet thickness profiles are found to be non-existant. Thus, we conclude that using a step-mode current sheet is an adequate tool in numerical simulations.

We placed virtual observers at a distance of \mbox{1 au} from the Sun and generated time profiles mimicking space observations. The IMF configuration was confirmed to significantly affect the qualitative shape of time profiles at different observer locations. For an injection location centered at the HCS, the $A+$ configuration confined protons to the vicinity of the HCS, whereas the $A-$ configuration caused observers at both northern and southern latitudes to observe particle fluxes. Latitudes separated from the injection region exhibited harder power-laws in particle flux compared with latitudes with injection, due to energy dependence of drifts. The current sheet drift of protons was detectable for an observer at the solar equator, causing an additional abrupt peak early in the event at eastern longitudes for the $A-$ configuration and western longitudes for the $A+$ configuration. In the $A+$ case, this generated dual-component time profiles, which could be misinterpreted as being accelerated by two distinct events.

We find that the IMF configuration can affect the extent of deceleration experienced by SEPs, with high energy particles in an $A+$ configuration retaining a significant portion of their energy due to constraints upon latitudinal drifting extents. When propagating within an $A-$ IMF configuration, latitudinal drifts are not constrained as they are towards the poles, and particles experience deceleration in agreement with the unipolar case. As drift effects are dependent on particle energy, deceleration plays an important role in all of our other results as well. Protons detected at \mbox{1 au} with a given energy will have been injected at sometimes significantly higher energies, and thus, will have been able to drift greater distances during their interplanetary propagation.

Finally, we assess the motion of particles across a flat HCS, and find that only scattering allows for protons to cross into the opposite hemisphere. The amount of scattering across the HCS is found to be small. Thus, within our model, if a current sheet exists between and injection site and an observer, fluences can be strongly suppressed.

 We note that all presented results involve a relatively small injection region, perhaps associated best with a flare without an associated CME. For injection at wider fronts, 
a tile-based approach like the one presented in \cite{Marsh2015} can be applied.

The configuration of the HCS and its associated drifts have thus been show to play an important role in the longitudinal and latitudinal transport of SEPs, the generation of complex time profiles, hardened proton spectra, and also the absence of flux at field lines close to the best-connected one. The presented results are applicable to eruptions during the solar minimum, when the solar magnetic field resembles a dipole. To model active phases of the Sun, when the HCS is tilted and wavy, resulting in a more complex IMF, additional investigations are required, and these will be the subject of future work.
%

\acknowledgements

 The authors wish to thank the Leverhulme Trust for providing funding for this research through grant number RPG-2015-094.
 SD acknowledges support from the UK Science and Technology Facilities Council (STFC) (grant ST/M00760X/1).

%


\begin{thebibliography}{}
\expandafter\ifx\csname natexlab\endcsname\relax\def\natexlab#1{#1}\fi
\providecommand{\url}[1]{\href{#1}{#1}}

\bibitem[{Aran {et~al.}(2005)Aran, Sanahuja, \& Lario}]{Aran2005}
Aran, A., Sanahuja, B., \& Lario, D. 2005, AdSpR, 36, 2333

\bibitem[{Boris(1970)}]{Boris1970}
Boris, J. 1970, {Acceleration calculation from a scalar potential.}, Tech.
  rep., Princeton Univ., N. J. Plasma Physics Lab., MATT--769

\bibitem[{Burger(2012)}]{Burger2012}
Burger, R.~A. 2012, \apj, 760, 60

\bibitem[{Burger {et~al.}(1985)Burger, Moraal, \& Webb}]{Burger1985}
Burger, R.~A., Moraal, H., \& Webb, G.~M. 1985, \apss, 116, 107

\bibitem[{Chollet {et~al.}(2010)Chollet, Giacalone, \& Mewaldt}]{Chollet2010}
Chollet, E.~E., Giacalone, J., \& Mewaldt, R.~A. 2010, JGRA, 115, 6101

\bibitem[{Dalla \& Browning(2005)}]{Dalla2005}
Dalla, S., \& Browning, P.~K. 2005, \aap, 436, 1103

\bibitem[{Dalla {et~al.}(2017)Dalla, Marsh, \& Battarbee}]{Dalla2017}
Dalla, S., Marsh, M.~S., \& Battarbee, M. 2017, \apj, 834 (2), p. 167

\bibitem[{Dalla {et~al.}(2013)Dalla, Marsh, Kelly, \& Laitinen}]{Dalla2013}
Dalla, S., Marsh, M.~S., Kelly, J., \& Laitinen, T. 2013, JGRA, 118, 5979

\bibitem[{Dalla {et~al.}(2015)Dalla, Marsh, \& Laitinen}]{Dalla2015}
Dalla, S., Marsh, M.~S., \& Laitinen, T. 2015, \apj, 808, 62

\bibitem[{Eastwood {et~al.}(2002)}]{Eastwood2002}
Eastwood, J.~P., Balogh, A., Dunlop, M.~W., Smith, C.~W. 2002, JGRA, 107, 1365

\bibitem[{Ebert(2003)}]{ebert2003texturing}
Ebert, D. 2003, Texturing \& Modeling: A Procedural Approach, Morgan Kaufmann
  series in computer graphics and geometric modeling.

\bibitem[{Guo \& Florinski(2014)}]{Guo2014}
Guo, X., \& Florinski, V. 2014, JGRA, 119, 2411

\bibitem[{He {et~al.}(2011)He, Qin, \& Zhang}]{He2011}
He, H.-Q., Qin, G., \& Zhang, M. 2011, \apj, 734, 74

\bibitem[{Jokipii \& Levy(1977)}]{Jokipii1977}
Jokipii, J.~R., \& Levy, E.~H. 1977, \apj, 213, L85

\bibitem[{Kelly {et~al.}(2012)Kelly, Dalla, \& Laitinen}]{Kelly2012}
Kelly, J., Dalla, S., \& Laitinen, T. 2012, \apj, 750, 47

\bibitem[{K{\'{o}}ta \& Jokipii(2001)}]{Kota2001}
K{\'{o}}ta, J., \& Jokipii, J. 2001, AdSpR, 27, 607

\bibitem[{Kubo {et~al.}(2009)}]{KuboYukiNagatsumaTsutomu2009}
Kubo, Y., Nagatsuma, T., \& Akioka, M. 2009, Journal of the National
  Institute of Information and Communications Technology, 56, 17.

\bibitem[{Laitinen {et~al.}(2016)Laitinen, Kopp, Effenberger, Dalla, \&
  Marsh}]{Laitinen2016}
Laitinen, T., Kopp, A., Effenberger, F., Dalla, S., \& Marsh, M.~S. 2016, \aap, 591, A18

\bibitem[{Luhmann {et~al.}(2007)Luhmann, Ledvina, Krauss-Varban, Odstrcil, \&
  Riley}]{Luhmann2007}
Luhmann, J., Ledvina, S., Krauss-Varban, D., Odstrcil, D., \& Riley, P. 2007, AdSpR, 40, 295

\bibitem[{Marsh {et~al.}(2015)Marsh, Dalla, Dierckxsens, Laitinen, \&
  Crosby}]{Marsh2015}
Marsh, M.~S., Dalla, S., Dierckxsens, M., Laitinen, T., \& Crosby, N.~B. 2015, SpWea, 13, 386

\bibitem[{Marsh {et~al.}(2013)Marsh, Dalla, Kelly, \& Laitinen}]{Marsh2013}
Marsh, M.~S., Dalla, S., Kelly, J., \& Laitinen, T. 2013, \apj, 774, 4

\bibitem[{Parker(1958)}]{Parker1958}
Parker, E.~N. 1958, \apj, 128, 664

\bibitem[{Pei {et~al.}(2012)Pei, Bieber, Burger, \& Clem}]{Pei2012}
Pei, C., Bieber, J.~W., Burger, R.~A., \& Clem, J. 2012, \apj, 744, 170

\bibitem[{Press(1996)}]{press1996numerical}
Press, W. 1996, Numerical Recipes in Fortran 90: Volume 2 (Cambridge University Press)

\bibitem[{Roelof(1969)}]{Roelof1969}
Roelof, E. 1969, in Lectures in High-Energy Astrophysics, ed. H.~{\"{O}}gelman \& J.~Wayland, 111

\bibitem[{Speiser(1965)}]{Speiser1965}
Speiser, T.~W. 1965, \jgr, 70, 4219

\bibitem[{Strauss {et~al.}(2012)Strauss, Potgieter, B{\"{u}}sching, \&
  Kopp}]{Strauss2012}
Strauss, R.~D., Potgieter, M.~S., B{\"{u}}sching, I., \& Kopp, A. 2012, \apss, 339, 223

\bibitem[{Turner(2000)}]{Turner2000}
Turner, R. 2000, IEEE Transactions on Plasma Science, 28

\bibitem[{Winterhalter {et~al.}(1994)Winterhalter, Smith, Burton, Murphy, \&
  McComas}]{Winterhalter1994}
Winterhalter, D., Smith, E.~J., Burton, M.~E., Murphy, N., \& McComas, D.~J.
  1994, \jgr, 99, 6667

\bibitem[{Zhang {et~al.}(2003)Zhang, Jokipii, \& McKibben}]{Zhang2003}
Zhang, M., Jokipii, J.~R., \& McKibben, R.~B. 2003, \apj, 595, 493

\end{thebibliography}

\end{document}